\documentclass[a4paper,12pt]{report}
\usepackage[english,greek]{babel}
\usepackage[iso-8859-7]{inputenc}
\usepackage{mathrsfs}
\usepackage{amsmath}
\usepackage{amsfonts}
\usepackage{amssymb}
\usepackage{amsthm}
\usepackage{graphics}
\usepackage{subfig}
\usepackage{makeidx}
\usepackage{float}
\usepackage{enumerate}
\usepackage[top=2.2cm, bottom=2.5cm, left=2.2cm, right=2.2cm]{geometry}
\usepackage{color}
\usepackage{soul}
\usepackage{fancyhdr}
\usepackage[pdftex]{graphicx}
\usepackage{slashed}
\usepackage[toc,page]{appendix}
\usepackage{titlepic}
\usepackage{afterpage}
\usepackage{epsfig}
\usepackage{epstopdf}

\newcommand{\sen}{\selectlanguage{english}}

\newcommand{\beq}{\begin{equation}}
\newcommand{\eeq}{\end{equation}}
\newcommand{\HRule}{\rule{\linewidth}{0.5mm}}

\begin{document}
\setlength{\parindent}{2em}
\sen

\begin{titlepage}
\begin{center}
\includegraphics[width=1.5cm,height=3cm]{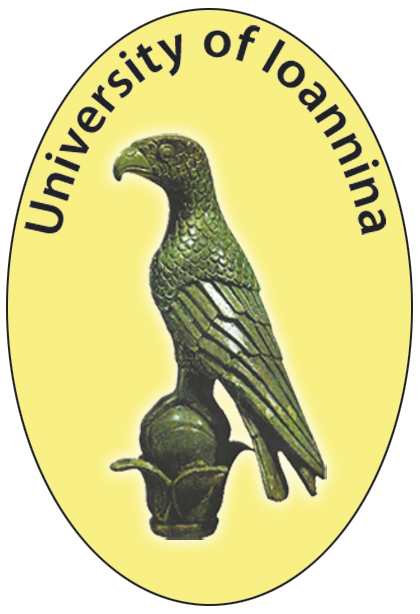}\\[1cm]
\sen
\textsc{\LARGE University of Ioannina}\\[0.5cm]
\textsc{\Large Physics Department}\\[0.5cm]
\textsc{\ Section of Astrogeophysics}\\[0.4cm]
\vspace{20mm}
\HRule \\[0.4cm]
{ \Huge \bfseries Axisymmetric Equilibria with Pressure Anisotropy and Plasma Flow \\[0.6cm] }

\HRule \\[1.5cm]
\end{center}
\vspace{15mm}

	\begin{center} \LARGE
		\emph{Achilleas Evangelias}
		\end{center}
	\vspace{5mm}
	\begin{center}
		\includegraphics[width=3in]{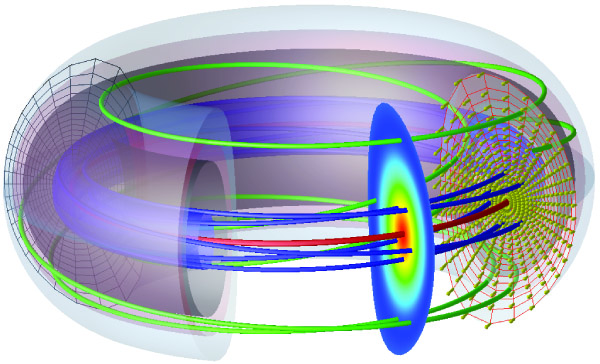}
		\end{center}

\vfill
\begin{center}
\large Ioannina, Greece \\
  June 16, 2015
  \end{center}
\end{titlepage}
\newpage\thispagestyle{empty}
\sen
\begin{center}
{ \huge \bfseries Axisymmetric Equilibria with Pressure Anisotropy and Plasma Flow \\[0.4cm] }

\vspace{5mm}
By
\vspace{5mm}

{\Large \emph{Achilleas Evangelias}}

\vspace{15mm}

{A thesis submitted in partial fulfillment of the requirements\\
for the degree of Master in Physics\\
of the University of Ioannina, Physics Department}

\includegraphics[width=1.3in]{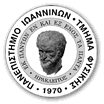}
\end{center}

\vspace{15mm}
\begin{flushleft}
	Approved by the Examining Committee:
	\vspace{15mm}
	\begin{itemize}
\item	\underline{George Throumoulopoulos}, Associate Professor, Supervisor
	
\item	\underline{George Leontaris}, Professor, Member
	
	\item	\underline{Alexander Nindos}, Associate Professor, Member

\end{itemize}
\end{flushleft}
\vspace{40mm}

\begin{center}
       University of Ioannina
       \end{center}
       \begin{center}
      Ioannina, Greece, June 16, 2015
      
      \end{center}  
   \newpage\thispagestyle{empty}    
   \sen
   \vspace{28mm}
       \begin{center}
        [This page was intentionally left blank]
        \end{center}
\newpage\thispagestyle{empty}
\sen

\begin{minipage}{0.8\textwidth}
\vspace{110mm} 
\emph{``In communist society, where nobody has one exclusive sphere of activity but each can become accomplished in any branch he wishes, society regulates the general production and thus makes it possible for me to do one thing today and another tomorrow, to hunt in the morning, fish in the afternoon, rear cattle in the evening, criticise after dinner, just as I have a mind, without ever becoming hunter, fisherman, herdsman or critic.''}\\

\emph{Karl Marx, German Ideology (1845)}
\end{minipage}
\newpage\null\thispagestyle{empty}\newpage

\setcounter{page}{4}
\begin{center}
\large \bfseries ABSTRACT
\end{center}

\vspace{40mm}

In this Master thesis we investigate the influence of pressure anisotropy and incompressible flow of arbitrary direction on the equilibrium properties of magnetically confined, axisymmetric toroidal plasmas. The main novel contribution is the derivation of a pertinent generalised Grad-Shafranov equation. This equation includes six free surface functions and recovers known Grad-Shafranov-like equations in the literature as well as the usual static, isotropic one. The form of the generalised equation indicates that pressure anisotropy and flow act additively on equilibrium.\par
In addition, two sets of analytical solutions, an extended Solovev one with  a plasma reaching the separatrix  and an extended Hernegger-Maschke one for a plasma surrounded by a fixed boundary possessing an X-point, are constructed, particularly in relevance to the ITER and NSTX tokamaks. Furthermore, the impacts both of pressure anisotropy, through an anisotropy function $ \sigma _d$ assumed to be uniform on the magnetic surfaces, and plasma flow, via the variation of an Alfv\'enic Mach function, on these equilibria are examined.\par 
 It turns out that depending on the maximum value and the shape of an anisotropy function, the anisotropy can act either paramagnetically or diamagnetically. Also, in most of the cases considered both the anisotropy and the flow have stronger effects on NSTX equilibria than on ITER ones. We conjecture that these effects may have an influence on plasma stability and transport, and play a role in the transitions to the improved confinement regimes in tokamaks.
\newpage\null\newpage

\begin{center}
\large \bfseries ACKNOWLEDGEMENTS
\end{center}

\vspace{40mm}
I am using this opportunity to express my gratitude to my supervisor, Associate Professor George Throumoulopoulos, who supported me throughout the course of this Master programme and of the present thesis with his patience and knowledge whilst allowing me the room to work in my own way. I am thankful for his aspiring guidance, invaluably constructive criticism and friendly advice during the project work. I am sincerely grateful to him for sharing his truthful and illuminating views on a number of issues related to the project. I attribute the level of my Masters degree to his encouragement and effort and without him this thesis, too, would not have been completed or written. One simply could not wish for a better or friendlier supervisor.
\par
Also, i would like to thank the members of the examination committee: Professor George Leontaris and Associate Professor Alexander Nindos, for dedicating time on this project.
\par
I would also like to thank my friends in University of Ioannina: Athanasios Kondaxis, Athanasios Karozas, Panagiota Grigoriadou and Katerina Ferentinou, for the stimulating discussions, and for all the fun we have had in the last two years.
\par
Last but not the least, I would like to thank my parents and my sister for the financial and moral support they provided me, giving me the opportunity to complete this Master programme.

\newpage\null\newpage

\tableofcontents

\newpage

\chapter{Brief Introduction to Plasma Physics and Controlled Fusion}

\section{What is plasma?}

\hspace{2em}A plasma is quasi-neutral gas of charged and neutral particles which exhibits collective behaviour and constitutes the fourth fundamental state of matter. When a gas is heated enough that the atoms collide with each other and knock their electrons off in the process, a plasma is formed. In a plasma, charge separation between ions and electrons gives rise to electric fields and charged particle flows give rise to currents and magnetic fields. These fields result in `action at a distance' and a range of different phenomena. Plasma is the most abundant form of ordinary matter in the visible Universe, as it makes up nearly 100\% of the Sun and all stars and of the interplanetary, interstellar and intergalactic medium. Also, the Earth's ionosphere is plasma. On Earth, bolt lightning, electric sparks and the aurora borealis are everyday examples of phenomena associated with plasmas. Also,  neon lights could more accurately be called ``plasma lights'', because the light comes from the plasma inside of them.

\vspace{10mm}

\begin{flushleft}
{\bfseries {Quasineutrality - Debye Shielding}}
\end{flushleft}

A fundamental characteristic of the behaviour of a plasma is its ability to shield out electric potentials that are applied to it. Suppose we put a charged particle with $q>0$ inside a plasma in thermal equilibrium. This would create an electric field so would attract the negative charged particles and repel the positive ones.  As a result there would exist a charge density around $q$

\[\rho =(n_i-n_e)e \]
where $n_i$ and $n_e$ are the ions and electrons densities. Due to Boltzmann's Law the number density $n$ of a species of particle in thermal equilibrium subject to a potential $\phi$ , can be expressed as

\[n_{\alpha}(r)=n_{0}e^{-\frac{q_{\alpha}\phi(r)}{kT_{\alpha}}}\]
where $\alpha =i,e $. In the quasi-neutral situation, the number of ions $n_i$ and electrons $n_e$ are approximately equal, $ n_i \approx n_e \approx n_0$. However, the distribution of these different kinds of particles are not necessarily uniform in space in the quasineutral situation, and so charge density gradients and fields can still exist within the plasma. As soon as the distribution starts to evolve, the change in the density $n$ changes the potential $\phi $. To find the potential, we need Poisson's equation,

\[ \nabla ^2 \phi (r)=-\frac{\rho (r)}{\epsilon _0}\]
In the one-dimensional plane geometry, Poisson's equation with the inclusion of the test charge, is

\[-\frac{\partial ^2 \phi (x)}{\partial x^2}=\frac{n_0 e}{\epsilon _0}\left[e^{-\frac{e\phi(x)}{kT_i}}-e^{\frac{e\phi(x)}{kT_e}}\right] +q \] 
Finding an analytic solution to this equation is non trivial. The physical
solution is to make an approximation and solve the equation in the resulting
limit. In the region where $\frac{e\phi}{kT_\alpha}\ll 1$, the exponents can be Taylor expanded

\[e^{-\frac{e\phi(x)}{kT_i}}\approx 1-\frac{e\phi(x)}{kT_i}\]

\[ e^{\frac{e\phi(x)}{kT_e}}\approx 1+\frac{e\phi(x)}{kT_e}\]
This approximation corresponds to $kT_\alpha \gg e\phi$, meaning that most particles in the plasma are ``free-streaming'', unaffected by the potential $\phi$.
No simplification is possible for small x, where $\frac{e\phi}{kT_\alpha}$ may be large. Fortunately, this region does not contribute much to the thickness of the sheath, because the potential falls very rapidly there. The linearised form of Poisson's equation then is

\[-\frac{\partial ^2 \phi (x)}{\partial x^2}=\frac{n_0 e^2}{\epsilon _0}\left(\frac{1}{kT_i}+\frac{1}{kT_e}\right)\phi+q\]
The coefficient of the $\phi$ term is a constant quantity. By comparing
with the left side, it must have units of inverse length squared. The Debye length is defined as the inverse square root of this constant. We can simplify the expression further by defining an effective temperature,

\[\frac{1}{T_{eff}}=\frac{1}{T_e}+\frac{1}{T_i}\]
The Debye length can then be defined as

\beq \label{Debye}\lambda _D = \sqrt{\frac{\epsilon _{0}kT_{eff}}{n_0 e^2}}\eeq
and so the solution for the potential can be written in the form

\[ \phi =\phi _{0}e^{-\frac{x}{\lambda _D}}\]
In position space, $\phi (x)$ decays exponentially away from the sheet of
charge. The scale length of the exponential decay is $\lambda _D$. Hence the Debye
length is the scale length to which a charge in the plasma is shielded.
Debye length is inversely proportional to $n_ 0$, and hence the shielding length
decreases as the density increases. This results because there are more
electrons to cancel an existing charge distribution, and hence a shorter
length scale to reach neutrality in the plasma, as the density of the
plasma increases. Furthermore, $\lambda_D$ increases with increasing $kT_{eff}$ and one observes that it is determined by the temperature of the colder species. 
A criterion for an ionized gas to be a plasma is that it be dense enough or/and cold enough that $\lambda _D$ is much smaller than the dimensions L of the system

\beq \label{lambda} \lambda _D \ll L\eeq
This is related crucially to the quasineutrality, because whenever local concentrations of charge arise or external potentials are introduced into the system, they are shielded out in a distance short compared with $L$, leaving the bulk of the plasma free of large electric potentials or fields, and so, outside of the sheath $n_i$ is equal to $n_e$. That is, the plasma is neutral enough so that one can take $ n_i \approx n_e \approx n_0$, where $n_0$ is a common density called the plasma density, but not so neutral that all the interesting electromagnetic forces vanish. The Debye length is the smallest macroscopic natural scale in the plasma. This is because every particle in the plasma is effectively shielding
every other particle on the Debye scale.
\vspace{10mm}

\begin{flushleft}
{\bfseries {Collective Behaviour}}
\end{flushleft}

Shielding is the first example we have seen of collective behaviour in a plasma. If an electric field is introduced in the plasma, the particles of the plasma respond and rearrange themselves to maintain charge neutrality. Recall that exhibiting collective behaviour was one of the characteristics listed
in the definition of a plasma. All of the previous work was based on the assumption that the Boltzmann distribution is applicable. The system cannot exhibit collective behaviour if the number of particles in the system is too small and the Boltzmann
statistics break down. Clearly, if there are one or two particles in the sheath region, Debye shielding would not be a statistically valid concept. If $N_D$ represents the number of particles in a sphere with radius equal to the Debye length, then

\[N_D=n_0 \left(\frac{4}{3}\pi \lambda ^3_D\right)\]
Thus, in addition to $\lambda_D \ll L$, collective behaviour requires
\beq \label{ND} N_D\gg 1\eeq
We have given two conditions that an ionized gas must satisfy to be called a plasma. A third condition has to do with collisions. We introduce a frequency involving the electron charge as

\[\omega _p^2=\frac{n_0 e^2}{\epsilon _0 m}\]
called the plasma frequency, which sets the most fundamental time-scale of plasma physics. There is evidently a plasma frequency for each species, but the relatively fast electron plasma frequency is more important, and thus, it references to `the plasma frequency' refer implicitly to the electron version. It can be seen that $\omega _p$ corresponds to the electrostatic oscillation in response to small charge separations, but of course plasma oscillation will be observed only if the plasma system is studied over time periods $\tau$ longer than $1/\omega _p$. The period $\tau$ represents the mean time between collisions with neutral atoms, so we require 

\beq \label{omega} \omega _p \tau >1\eeq
for a gas to behave like a plasma rather than a neutral gas.\par
In conclusion, the three conditions that a plasma must satisfy in order to be a plasma are:

\[ 1. \, \, \lambda_D \ll \L \]
\[ 2. \, \, N_D \gg 1 \]
\[ 3. \, \, \omega _p \tau >1 \]

\section{The Ideal Magnetohydrodynamics (MHD) model}
\hspace{2em}It has been established in a number of fusion devices
that sheared flow both zonal and mean (equilibrium)
play a role in the transitions to improved confinement
regimes as the L-H transition and the Internal Transport
Barriers \cite{zonal}, \cite{flow}. These flows can be driven externally
in connection with electromagnetic power and
neutral beam injection for plasma heating and current
drive or can be created spontaneously (zonal flow).
The MHD model is a single-fluid model which is concerned with the mutual interaction of fluid flow and magnetic fields. The fluids in question must be electrically conducting and that is why this model is used to describe the effects of magnetic geometry on the macroscopic equilibrium and stability properties of fusion plasmas.
Since the MHD equations describe the motion of a conducting fluid interacting with a magnetic field, we need to combine Maxwell's equations with the equations of gas dynamics and provide equations describing the interaction. A brief presentation of the ideal MHD model equations is given below.
\par
Maxwell's equations describe the evolution of electric field $\vec{E}(\vec{r},t)$ and magnetic field  $\vec{B}(\vec{r},t)$ in response to current density  $\vec{J}(\vec{r},t)$ and space charge  $s(\vec{r},t)$:
\beq \label{Max1} \vec{\nabla} \times \vec{E}=-\frac{\partial \vec{B}}{\partial t}\eeq
\beq \label{Max2} \vec{\nabla} \times \vec{B}=\mu _0\vec{J}+\frac{1}{c^2}\frac{\partial \vec{E}}{\partial t}\eeq
\beq \label{Max3} \vec{\nabla} \cdot \vec{E}=\frac{s}{\epsilon _0}\eeq
\beq \label{Max4} \vec{\nabla} \cdot \vec{B}=0\eeq
Gas dynamics equations describe evolution of density $\rho (\vec{r},t)$ and pressure $p(\vec{r},t)$:
\beq \label{Gas1} \frac{D\rho}{Dt}+\rho \vec{\nabla} \cdot \vec{v}=0\eeq
\beq \label{Gas2} \frac{Dp}{Dt}+\gamma p \vec{\nabla} \cdot \vec{v}=0\eeq
where $\frac{D}{Dt}\equiv \frac{\partial}{\partial t}+\vec{v}\cdot \vec{\nabla}$  is the Lagrangian time derivative (moving with the fluid element).
Eq. (\ref{Gas1}) expresses mass conservation and (\ref{Gas2}) relates to adiabatic energy conservation.
Coupling between the system described by $[\vec{E} ,\vec{B}]$ and the system described by $[\rho ,p]$ comes about through two equations involving the velocity $\vec{v}(\vec{r},t)$ of the fluid.
The first one is the equation that describes the acceleration of a fluid element by pressure gradient, gravity, and electromagnetic contributions
\beq \label{un1} \rho \frac{D\vec{v}}{Dt}=-\vec{\nabla}p+\rho \vec{g}+\vec{J}\times \vec{B}+s \vec{E} \eeq
indicating the momentum conservation (here $\vec{g}$ is the gravitational acceleration which has been included for completeness, but though important for certain astrophysical plasmas is negligible for laboratory ones).
The second is the Ohm's law (for a perfectly conducting moving fluid) which expresses the fact that the electric field $\vec{E'}$ in the  frame of reference co-moving with the fluid element, vanishes
\beq \label{un2} \vec{E'} \equiv \vec{E}+\vec{v} \times \vec{B}=0\eeq
In fact the more general form of the Ohm's law, $ \vec{E}+\vec{v} \times \vec{B}=\eta \vec{J}$, involves the electric resistivity of the fluid. In the Ideal MHD model $\eta$ is considered to be zero, an approximation which is good for fusion plasmas. This is because according to the Spitzer's law the electrical resistivity depends on the temperature as  $ \eta \sim T^{-3/2}$, and in fusion experiments the prevailing temperatures are of the order of $T\sim 10^8 K$.
\par
Equations (\ref{Max1}) - (\ref{un2}) are complete, and, on the side, become simpler in the non relativistic regime, $v\ll c$.
Considering the pre-Maxwell equations we see that Maxwell's displacement current is negligible $(\sim \frac{v^2}{c^2})$

\[ \frac{1}{c^2}|\frac{\partial \vec{E}}{\partial t}|\sim \frac{v^2}{c^2}\frac{B}{l_0}\ll \mu _0|\vec{J}|\sim | \vec{\nabla} \times \vec{B}|\sim \frac{B}{l_0}\]
indicating length scales by $l_0$ and time scales by $t_0$, so that $v\sim \frac{l_0}{t_0}$.
Electrostatic force is also negligible $(\sim \frac{v^2}{c^2})$

\[ s |\vec{E}|\sim \frac{v^2}{c^2}\frac{B^2}{\mu _0 l_0}\ll |\vec{J}\times\vec{B}|\sim \frac{B^2}{\mu _0 l_0}\]
so that space charge effects may be ignored and Poisson's law is not employed. Actually in MHD the Poisson's law is replaced by the quasineutrality condition.
\par
Employing these approximations the basic equations of Ideal MHD are recovered in their usual form. In the present study we are going to construct and investigate the properties of fusion plasmas in equilibrium state, so we are interested in the time-independent form of the  Ideal MHD equations with convective plasma  flow, which are:
\beq \label{MHD1} \vec{\nabla}\cdot (\rho \vec{v})=0\eeq
\beq \label{MHD2} \rho (\vec{v}\cdot \vec{\nabla})\vec{v}=\vec{J}\times \vec{B}-\vec{\nabla}p\eeq
\beq \label{MHD3} \vec{\nabla}\times \vec{B}=\mu _0\vec{J}\eeq
\beq \label{MHD4} \vec{\nabla}\times \vec{E}=0\eeq
\beq \label{MHD5} \vec{\nabla}\cdot \vec{B}=0\eeq
\beq \label{MHD6} \vec{E}+\vec{v}\times \vec{B}=0\eeq
In order for ideal MHD to be valid, three conditions must be satisfied: (1) there must happen frequent collisions, (2) the ion gyro-radius has to be small, and (3) the plasma resistivity must be small.
Conditions (2) and (3) are well satisfied for fusion plasmas since MHD frequencies are much smaller than the ion gyro frequency, and since plasma temperature is so, the resistivity is too small because of the Spitzer law. On the other hand, condition (1) is never satisfied for plasmas of fusion interest, since the the ion-electron Coulomb collision frequency is dependent on the plasma resistivity as

\[ \nu _{ei}=\frac{ne^2}{m_e}\eta\]
and therefore it is also small. However, in a magnetized plasma the magnetic field plays the role of collisions for the motion perpendicular to $\vec{B}$ because of the particle gyro motion, and so the Larmor radius plays the role of effective mean free path. Thus, the ideal MHD model can be applied in fusion plasmas.

\section{Pressure Anisotropy}

\hspace{2em}An additional effect of external heating, depending on the direction of the injected momentum,  is pressure anisotropy.
For tokamaks and stellarators the MHD pressure is usually considered isotropic. However, plasma heating by the neutral beam injection, ion resonance waves and electron cyclotron waves can produce strong plasma anisotropy. Thus, pressure anisotropy is present in strongly magnetized plasmas and may play a role in several magnetic fusion as well as in some astrophysical related problems.
When collision time is considered to be the shortest time in the problem, with the possible exception of the gyration period, a small element of mass of a plasma will relax quickly to a Maxwellian distribution function before it can change its properties, and a local description in terms of the parameters characterizing this Maxwellian is appropriate. But in many important plasmas as the high temperature ones the collision time is so long that collisions can be ignored. It would appear that for such collisionless plasmas a fluid theory is not appropriate. However as already mentioned, for perpendicular motions the magnetic field plays the role of the collisions, thus making a fluid description appropriate.
\par
Macroscopic equations for a collisionless plasma with pressure anisotropy have been derived by Chew, Goldberger, and Low \cite{CGL}. A brief  discussion  on this derivation is made below.  Vlasov equation for the ions is
\beq \label{CGL1} \frac{\partial f}{\partial t}+(\vec{v}\cdot \vec{\nabla}_{\vec{r}})f+\frac{e}{m_i}(\vec{E}+\vec{v}\times \vec{B})\cdot \vec{\nabla}_{\vec{v}}f=0\eeq
where $f$ is the distribution function, depending on position $\vec{r}$, velocity $\vec{v}$ and time t. The ion charge is $e$ and the mass $m_i$.
Expanding the distribution function in powers of $m_i/e$ (which in appropriate units is equivalent to an expansion in powers of the Larmor radius)
\beq \label{CGL2} f=f_0+f_1+... \eeq
 we find that the equations satisfied by the sequence of  functions are
\beq \label{CGL3} (\vec{E}+\vec{v}\times \vec{B})\cdot \vec{\nabla}_{\vec{v}}f_0=0\eeq
\beq \label{CGL4} \frac{\partial f_0}{\partial t}+(\vec{v}\cdot \vec{\nabla}_{\vec{r}})f_0+\frac{e}{m_i}(\vec{E}+\vec{v}\times \vec{B})\cdot \vec{\nabla}_{\vec{v}}f_1=0\eeq
\begin{center}
.....
\end{center}
We observe that (\ref{CGL3}) cannot be fulfilled unless the electric field is perpendicular to the magnetic field, a condition which is satisfied to a very good approximation because of the presence of the electrons. Indeed, owing to their small mass the electrons are highly mobile, so any electric field which develops parallel to the magnetic field will cause violent electron motion and cannot long persist. If we set $\vec{u}_E=\vec{E}\times \vec{B}/B^2$ and assume that the electric field and the magnetic field are perpendicular to each other, it follows that
\beq \label{CGL5} \vec{E}=-\vec{u}_E\times \vec{B}\eeq
and Eq. (\ref{CGL3}) becomes $((\vec{v}-\vec{u}_E)\times \vec{B})\cdot \vec{\nabla}_{\vec{v}}f_0$, so that 
$\vec{\nabla}_{\vec{v}}f_0$, $(\vec{v}-\vec{u}_E)$, and $\vec{B}$ are seen to be coplanar. Therefore the general form of $f_0$ is
\beq \label{CGL6} f_0((\vec{v}-\vec{u}_E)^2,(\vec{v}\cdot \vec{B}),\vec{r},t)\eeq
The equation of continuity is
\beq \label{CGL7} \frac{\partial n_0}{\partial t}+\vec{\nabla}\cdot(n_0\vec{V})=0\eeq
where
$n_0=\int f_0d^3\vec{r}$
and $\vec{V}$ is the average velocity of the ions
\beq \label{CGL8} n_0\vec{V}=\int \vec{V}f_0 d^3\vec{r}\eeq
Next, by multiplying Eq. (\ref{CGL4}) by $\vec{v}$ and integrate, we find
\beq \label{CGL9} \rho _0\left(\frac{\partial}{\partial t}+\vec{V}\cdot \vec{\nabla}\right)V=-\vec{\nabla}\cdot \stackrel{\textstyle\leftrightarrow}{\rm {\mathbb P}
}+\left(e\int (\vec{v}-\vec{V})f_1 d\vec{v}\right)\times\vec{B}\eeq
where $\rho _0=m_i n_0$
and the pressure tensor is defined by
\beq \label{CGL10} \stackrel{\textstyle\leftrightarrow}{\rm {\mathbb P}}=m_i\int (\vec{v}-\vec{V})(\vec{v}-\vec{V})f_0 d\vec{v} \eeq
The restriction (\ref{CGL6}) on the functional form of $f_0$ implies that pressure tensor must be of the form

\beq \label{CGL11} \stackrel{\textstyle\leftrightarrow}{\rm {\mathbb P}}=p_{\|}\vec{b}\vec{b}+p_{\bot}( \stackrel{\textstyle\leftrightarrow}{\rm {\mathbb I}}-\vec{b}\vec{b})\eeq
where $\vec{b}$ is a unit vector pointing along the magnetic field and $\stackrel{\textstyle\leftrightarrow}{\rm {\mathbb I}}$ signifies the unit dyadic. In other words, the pressure tensor is diagonal in a local rectangular coordinate system one of whose axes (i.e $x_3$) points along $\vec{B}$
\beq \label{CGL12} {\mathbb P}_{ij}=\left(\begin{array}{ccc} p_{\bot} & 0 & 0 \\ 0 & p_{\bot} & 0 \\ 0 & 0 & p_{\|} \end{array} \right)\eeq
In the plane perpendicular to $\vec{B}$, associated with two degrees of freedom, the pressure is a scalar of magnitude $p_{\bot}$. The pressure along $\vec{B}$ is $p_{\|}$, which in general need not equal $p_{\bot}$. It is clear that $p_{\bot}$ represents the thermal energy associated with gyration, while $p_{\|}$ measures the thermal motion along the magnetic field.
When the distribution function is isotropic, the pressure tensor is
\beq \label{CGL13} {\mathbb P}_{ij}=p\delta _{ij}\eeq
and $p (= p_{\bot} =p_{\|}) =nT$ for a Maxwellian distribution function.
\par
In this work, we generalise the ideal MHD model to include equilibrium pressure anisotropy in the fluid part of the theory. Thus, the basic equations (\ref{MHD1})-(\ref{MHD6}) take the form

 \beq \label{continuity} \vec{\nabla}\cdot (\rho \vec{v})=0\eeq
\beq \label{momentum1} \rho (\vec{v}\cdot \vec{\nabla})\vec{v}=\vec{J}\times \vec{B}-\vec{\nabla}\cdot \stackrel{\textstyle\leftrightarrow}{\rm {\mathbb P}}\eeq
\beq \label{Ampere} \vec{\nabla}\times \vec{B}=\mu _0\vec{J}\eeq
\beq \label{Faraday} \vec{\nabla}\times \vec{E}=0\eeq
\beq \label{BGauss} \vec{\nabla}\cdot \vec{B}=0\eeq
\beq \label{Ohm} \vec{E}+\vec{v}\times \vec{B}=0\eeq
Eq. (\ref{Faraday}) implies that $\vec{E}=-\vec{\nabla}\Phi $, where $\Phi$ is the electrostatic potential.

\begin{flushleft}
{\bfseries {Pressure Anisotropy Function and the Momentum Equation}}
\end{flushleft}

A more convenient form for Eq. (\ref{CGL11}) is
\beq \label{tensor1} \stackrel{\textstyle\leftrightarrow}{\rm {\mathbb P}}=p_{\bot}\stackrel{\textstyle\leftrightarrow}{\rm {\mathbb I}}+\frac{p_{\|}-p_{\bot}}{|\vec{B}|^2}\vec{B} \vec{B}\eeq
At this point we introduce the function 
\beq \label{sigma} \sigma =\frac{p_{\|}-p_{\bot}}{|\vec{B}|^2} \eeq 
so that the pressure tensor can be written in the form
\beq \label{tensor2} \stackrel{\textstyle\leftrightarrow}{\rm {\mathbb P}}=p_{\bot}\stackrel{\textstyle\leftrightarrow}{\rm {\mathbb I}}+\sigma \vec{B} \vec{B}\eeq
One can see that $\sigma $ is a measurement of the pressure anisotropy and has dimensions of $\frac{1}{\mu _0}$. It is clear that particle collisions, in equilibrating parallel and perpendicular energies, will reduce $\sigma $, and therefore that a collision-dominated plasma is described accurately by a scalar pressure. However, because of the low collision frequency a high-temperature confined plasma remains for long anisotropic, once anisotropy is induced  by  external heating sources.
Using the identity
$\vec{\nabla}\cdot (\sigma \vec{B}\vec{B})=[\vec{\nabla}\cdot(\sigma \vec{B})]\vec{B}+
[(\sigma \vec{B})\cdot \vec{\nabla}]\vec{B}$
and Eq. (\ref{BGauss}), then the divergence of the pressure tensor is put in the form
\beq \label{tensordiv} \vec{\nabla}\cdot \stackrel{\textstyle\leftrightarrow}{\rm {\mathbb P}}=\vec{\nabla}p_{\bot}+(\vec{B}
\cdot \vec{\nabla}\sigma)\vec{B}+\sigma (\vec{B}\cdot \vec{\nabla})\vec{B}\eeq
To make the analysis tractable, following \cite{Cotsaftis,Clement,clst}, we assume that $\sigma $ is a function only of $\psi $, where $\psi $ is the usual poloidal magnetic  flux  function (cf.  section 2.1 of chapter 2). This implies that $\sigma$ is uniform on magnetic surfaces. A similar assumption was adopted recently in \cite{Kuznetsov} to study static mirror structures. Because of the large parallel thermal conductivity a good assumption is that $p_{\|}=p_{\|}(\psi )$, but this is not sufficient to explain that $\sigma =\sigma (\psi)$. However, the hypothesis $\sigma =\sigma (\psi)$, according to Mercier and Cotsaftis \cite{Cotsaftis}, may be the only suitable for satisfying the boundary conditions on a rigid, perfectly conducting wall.
 Taking into account $\sigma =\sigma (\psi )$, the momentum equation (\ref{momentum1}) with the use of (\ref{tensordiv}) can be written
 \beq \label{momentum2} \rho (\vec{v}\cdot \vec{\nabla})\vec{v}+\vec{\nabla}p_{\bot}+\sigma (\vec{B}\cdot \vec{\nabla})\vec{B}=\vec{J}\times \vec{B}\eeq
 From now on we are going to use the dimensionless $\sigma $ function, which is defined as
 \beq \label{sigmad} \sigma _d (\psi)=\mu _0 \sigma (\psi)\eeq
 Thus, with the use of the identity 

\begin{center}
 $\vec{\nabla}(\vec{F}\cdot \vec{G})=(\vec{G}\cdot \vec{\nabla})\vec{F}+(\vec{F}\cdot \vec{\nabla})\vec{G}+\vec{F}\times (\vec{\nabla}\times \vec{G})+
 \vec{G}\times (\vec{\nabla}\times \vec{F})$,
\end{center} 
 the momentum equation takes the useful form
 \beq \label{momentum} \rho \vec{\nabla}\left(\frac{v^2}{2}\right)-\rho \vec{v}\times (\vec{\nabla}\times \vec{v})+
 \vec{\nabla}p_{\bot}+\sigma _d \left(\frac{B^2}{2\mu _0}\right)+(\sigma _d -1)\vec{J}\times \vec{B}=0\eeq
At last, we define the quantity
\beq \label{effective} \overline{p}=\frac{p_{\|}+p_{\bot}}{2}\eeq
which may interpreted as an effective isotropic pressure, and which should not be confused with the average plasma pressure
\beq \label{average} <p>=\frac{p_{\|}+2p_{\bot}}{3}=\overline{p}-\sigma _d \frac{B^2}{6\mu _0}\eeq
On the basis of Eqs. (\ref{sigma}), (\ref{sigmad}), and (\ref{effective}), the following instructive relations arise for the two scalar pressures:
\beq \label{pper} p_{\bot}=\overline{p}-\sigma _d \frac{B^2}{2\mu _0}\eeq
and
\beq \label{pparallel} p_{\|}=\overline{p}+\sigma _d \frac{B^2}{2\mu _0}\eeq
The above analysis was based on the CGL model \cite{CGL}, which is a collisionless one. It is recalled that this is a good approximation for fusion plasmas since due to Spitzer's law, the frequency of collisions at a plasma is proportional to $T^{-3/2}$, where T is the temperature, and fusion temperatures are of the order of 15 million degrees of Celsius.

\section{Controlled Nuclear Fusion}

\hspace{2em}Fusion is the process that heats the sun and other stars, which can be viewed as `natural fusion reactors'. In the sun 600 million tons of hydrogen is fused into helium each second, and it is the energy from this process that sustains life on our planet. In stars, the fusion process involves other elements too, but, in fact, all matter present in the universe was at one time formed by fusion of the lightest element, hydrogen.

\begin{flushleft}
 {\bfseries {The process}}
\end{flushleft}
 
The nucleus of the atoms is held together by very strong nuclear forces, while the electrons are attracted to the protons by electrical forces which are much weaker than the nuclear forces. In the fusion process, two light atoms fuse together to make a heavier one. The protons in a nucleus have a positive charge, so when two nuclei come close together, the electrical forces try to push them away from each other.  To make the nuclei coming close enough each other so that the nuclear forces prevail over the electrical repulsive forces they need to collide at a very high speed. This means that the plasma need have a very high temperature. Once the nuclei fuse, the process releases a large amount of energy.
\subsection{Controlled Thermonuclear Fusion}

\hspace{2em}As a consequence of the development of civilization, the worldwide demand for energy has been increasing rapidly. The estimated world reserves of fossil fuels-petroleum, coal and gas- probably will be depleted within a century. Thus nuclear energy, is recognized as the most important near-term replacement of fossil fuels. Fission reactors are already being exploited and have an important contribution to the energy reserves. Even so, if a time span of several hundred years or more is considered, this will still not be a sufficient reserve. Furthermore-and this is perhaps the most important constraint-there seems to be no safe way, at present or in the foreseeable future, to dispose of the vast quantities of radioactive waste associated with a fission program. A possible alternative and successor to the fast breeder fission reactor is a fusion reactor.
\par
On earth, the hydrogen-hydrogen reaction would be impractical because it would require too much energy, or too high temperatures, to start. Instead, light nuclides such as deuterium, tritium, helium-3, and lithium are used. Deuterium exists abundantly in nature; for example, it comprises 0.015 atom percent of the hydrogen sea water, while tritium will be made from lithium, which is contained in the earth's crust and the world seas in huge amounts.
Nuclear reactions of interest for fusion reactors are as follows:

\beq \label{re1} D+D\rightarrow T(1.01 MeV)+p(3.03 MeV)\eeq
\beq \label{re2} D+D\rightarrow He^3(0.82 MeV)+n(2.45 MeV)\eeq
\beq \label{re3} D+T\rightarrow He^4(3.52 MeV)+n(14.06 MeV)\eeq
\beq \label{re4} D+ He^3\rightarrow He^4(3.67 MeV)+p(14.67 MeV)\eeq
\beq \label{re5} Li^6+n\rightarrow T+He^4+4.8 MeV\eeq
The D-T reaction has the largest cross section of those shown. Taking into account the difficulties of plasma confinement, the D-T reaction should be the one undertaken in the first  stages of development because of its  higher cross-section at practically feasible temperatures. This reaction needs a temperature of 100 to 150 million degrees Centigrade, while all other reactions need higher temperatures. 

\begin{flushleft}
{\bfseries {Lawson's Condition}}
\end{flushleft}

As already mentioned, a plasma for a fusion reactor should have very high temperature so that the ions have velocities large enough to overcome their mutual Coulomb repulsion in order that they collide and fusion takes place. The energy released must be larger than that necessary to maintain the hot plasma.
The thermal energy per unit volume is $3nT$ - where $n$ is the plasma density and $T$ the plasma temperature - and decreases by heat conduction and also by particle losses. If $P_L$ is the  power loss per unit volume, the loss time $\tau _E $ is defined by the equation,
\beq \label{powerloss} P_L\equiv \frac{3nT}{\tau _E}\eeq
We may also call this time $\tau _E $ the energy confinement time, as we can consider that the plasma is confined during the period $\tau _E $. The largest $\tau _E$ is the lower is the power loss $P_L$.
\par
Lawson derived his condition for the energy balance in a fusion reactor taking the efficiency of the thermal-to-electric energy conversion and heating efficiency into consideration. To be energetically favourable the fusion reaction rate has to be higher than the energy losses from plasma. The fusion energy is three orders of magnitude higher than the mean thermal energy. Thus, the characteristic confinement time of plasma energy $\tau _E$ has to be not less than $10^{-3}$ of complete burn time. So for the $D-T$ self-sustained fusion reaction for which the optimal temperature is $T \approx 10keV $ we have the necessary condition
\beq \label{Lawson} n\tau _E \geq  10^{20}m^{-3}s\eeq 
In order to satisfy Lawson's condition, many problems remain to be solved concerning plasma confinement and heating.

\subsection{Magnetic Confinement - Tokamak Devices}

\hspace{2em}At present the major research effort in the area of controlled nuclear fusion is focused on the confinement of hot plasmas by means of strong magnetic fields. The magnetic devices most actively studied are the toroidal configurations not having an open end (see Fig. 1.1). In the simple toroidal field, ions and electrons drift in opposite directions due to the gradient of the magnetic field. This gradient-$\vec{B}$ drift causes charge separation that induces the electric field $\vec{E}$ directed parallel to the major axis of the torus. The subsequent $\vec{E}\times \vec{B}$ drift tends to carry the plasma ring outward. In order to reduce the $\vec{E}\times \vec{B}$ drift, it is necessary to connect the upper and lower parts of the plasma by lines of magnetic force thus leading to a short circuit of the separated charges along these field lines. Consequently, a poloidal component of the magnetic field is essential to the equilibrium of toroidal plasmas, and toroidal devices may be classified according to the methods used to generate the poloidal field.
In the tokamak devices an inductively driven current in the toroidal direction creates the poloidal magnetic field. 

\subsubsection{Tokamaks}

\hspace{2em}Tokamak is the best investigated device of magnetic confinement. Plasma is produced in the form of an axisymmetrical torus, of circular cross section in the simplest case.  In connection with the tokamak geometry we employ the usual right-handed cylindrical coordinate system $(R,z,\phi)$. Axisymmetry means that $\frac{\partial W}{\partial \phi}=0$ for every physical quantity $W$. The geometry of a tokamak is shown in Fig. 1.1.

\begin{figure}
  \centering
    \includegraphics[width=4in]{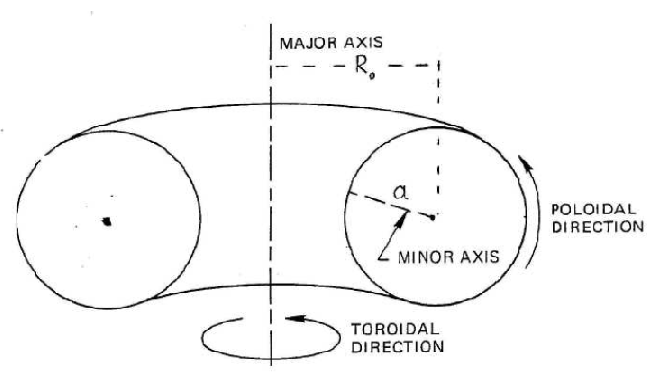}
     \caption{{\small\emph{The cross section of an axisymmetrical torus indicating the toroidal and poloidal directions. Its basic geometrical characteristics are set by the major radius $R_0$ and the minor radius $\alpha $} .}}
\end{figure}

The principal magnetic field in a tokamak is the toroidal field. This field alone does not allow confinement of the plasma. In order to have an equilibrium in which plasma pressure is balanced by magnetic forces it is necessary also to have a poloidal magnetic field. In a tokamak this field is produced mainly by currents in the plasma itself. The pressure gradient force which can be balanced by the magnetic force increases with the strength of the magnetic field. The toroidal magnetic field is limited by technological factors and is on the order of magnitude of 1 Tesla. The resulting magnetic field configuration is illustrated in Fig. 1.2.

\begin{figure}
  \centering
    \includegraphics[width=3in]{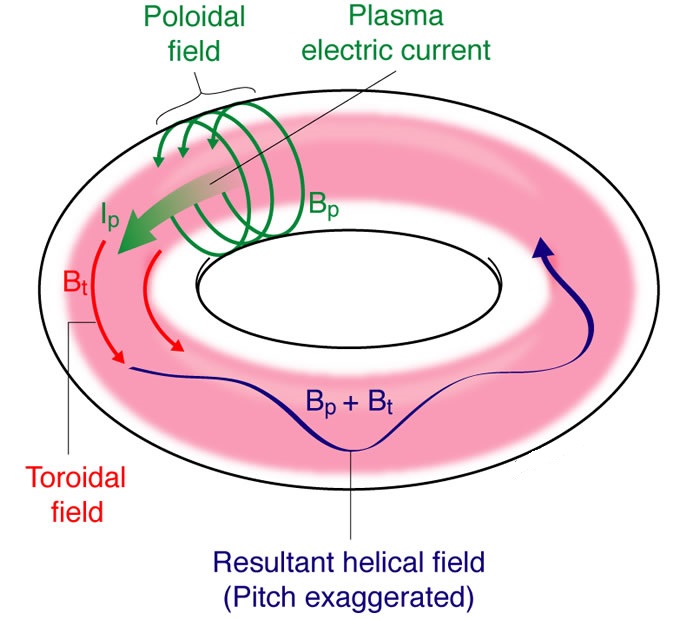}
     \caption{{\small \emph{Toroidal magnetic field $\vec{B} _t $ and poloidal magnetic field $\vec{B} _p $ due to toroidal current $I_p $.}}}
\end{figure}

The plasma current is produced by transformer action. A current is passed through primary coils around the torus. This gives a flux change through the torus and produces a toroidal electric field which drives the plasma current. The plasma shape and position are controlled by additional toroidal currents in suitably placed coils. Tokamak plasmas have particle densities of around $10^{20}m^{-3}$, which is 4-5 orders of magnitude lower than that in the atmosphere. The plasma is therefore enclosed in a vacuum vessel in which very low background pressures must be maintained. Because impurities in the plasma give rise to radiation losses, the restriction of their entry into the plasma plays a fundamental role in the successful operation of tokamaks. This requires a separation of the plasma from the vacuum vessel. Two techniques are currently used. The first is to define an outer boundary of the plasma with a material limiter and the second is to keep the particles away from the region of the vacuum vessel by means of a magnetic divertor. An overall schematic representation of a tokamak device is shown in Fig. 1.3.

\begin{figure}
  \centering
    \includegraphics[width=5in]{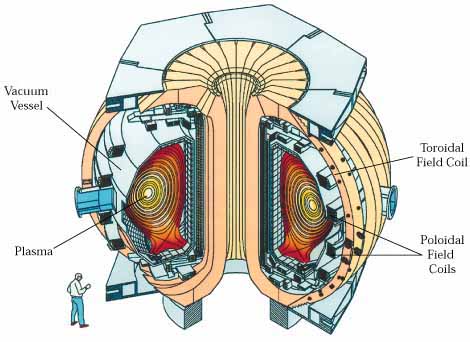}
     \caption{{\small \emph{Schematic representation of a tokamak device.}}}
\end{figure}

\subsubsection{ITER}

\hspace{2em}ITER, which means `the way' in Latin, is an international project involving the European Union (with Switzerland), Japan, the Russian Federation, China, India, the USA, and South Korea, the construction of which has began in Cadarache, southern France. The main goal of ITER is to demonstrate the feasibility of fusion as an energy source. Compared to the largest existing device, JET, the plasma volume of ITER is almost ten times larger, which makes it easier to keep the plasma confined for a longer time. ITER is planned to produce ten times more output power than the power needed to heat the plasma. An important scientific objective of ITER is to study plasmas which are heated by the fusion reactions themselves, instead of by external heating (ignition condition).
ITER will incorporate technology to potentially be used in future power stations: superconducting magnets, high heat-flux components, remote maintenance systems and tritium handling equipment. 
\par
Some of the most important geometrical characteristics and physical parameters of the ITER tokamak are the following:
\begin{center}
Major Radius : $R_0=6.2\ m$
\end{center}
\begin{center}
Minor Radius : $\alpha =2.0\ m$
\end{center}
\begin{center}
Elongation : $\kappa =1.7$
\end{center}
\begin{center}
Triangularity : $t=0.33$
\end{center}
\begin{center}
Vacuum magnetic field on the geometrical center $R_0$: $B_0=5.3\ T$
\end{center}
\begin{center}
Maximum Pressure (on magnetic axis) : $\sim 10^6\ Pa$
\end{center}
The above parameters and other confinement figures of merit such as the plasma beta and the safety factor, will be explained in Chapter 2.
A provisional illustration of ITER is illustrated in Fig. (1.4).
\begin{figure}
  \centering
    \includegraphics[width=4in]{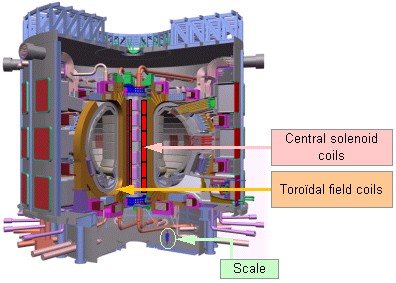}
     \caption{{\small \emph{ITER configuration in comparison with human dimensions.}}}
\end{figure}
\subsubsection{Spherical Tokamaks - NSTX and NSTX-Upgrade}

\hspace{2em}A spherical tokamak is a type of fusion power device based on the tokamak principle. It is notable for its very narrow profile, or``aspect ratio". A traditional tokamak has a toroidal confinement area that gives it an overall shape similar to a donut, having a large hole in the middle.  In the spherical tokamak the size of the hole is almost reduced to zero, resulting in a plasma shape that is almost spherical, often compared with a cored apple. The spherical tokamak is sometimes referred to as a spherical torus (ST). A schematic representation of the ST compared to the conventional tokamak is given in Fig. (1.5).
\begin{figure}
	\centering
	\includegraphics[width=4in]{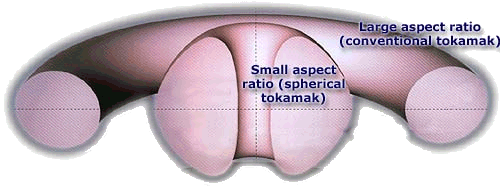}
	\caption{{\small \emph{Shape comparison between spherical and conventional tokamaks.}}}
\end{figure} 
The spherical tokamak is an offshoot of the conventional tokamak design. Proponents claim that it has a number of substantial practical advantages over these devices.
\par
The first is practical. Using the ST layout places the toroidal magnets much closer to the plasma, on average. This greatly reduces the amount of energy needed to power the magnets in order to reach any particular level of magnetic field within the plasma. Smaller magnets cost less, reducing the cost of the reactor.
The other advantages have to do with the stability of the plasma.
\par
On the other hand, the ST also have some distinct disadvantages compared to ``conventional" advanced tokamaks with higher aspect ratios.
The first issue is that the overall pressure of the plasma in an ST is lower than in conventional designs, in spite of higher beta. This is due to the limits of the magnetic field on the inside of the plasma, $B_{max}$. This limit is theoretically the same in the ST and conventional designs, but as the ST has a much lower aspect ratio, the effective field changes more drastically over the plasma volume.
\par
In addition, the ST is so small, at least in the center, that there is little or no room for superconducting magnets. This is not a deal-breaker for the design, as the fields from conventional copper wound magnets is enough for the ST design. However, this means that power dissipation in the central column will be considerable. 
\par
Finally, the highly asymmetrical plasma cross sections and tightly wound magnetic fields require very high toroidal currents to maintain. Normally this would require large amounts of secondary heating systems, like neutral beam injection. These are energetically expensive, so the ST design relies on high bootstrap currents for economical operation. Luckily, high elongation and triangularity are the features that give rise to these currents, so it is possible that the ST will actually be more economical in this regard. This is an area of active research. A recent review on spherical tokamaks is provided in \cite{Ono}.
\par
In the present work we were interested in the National Spherical Torus Experiment (NSTX) and it's upgrade (NSTX-U) designed by the Princeton Plasma Physics Laboratory (PPPL) in collaboration with the Oak Ridge National Laboratory, Columbia University, and the University of Washington at Seattle \cite{Princeton}.
\par
Some of the most important geometrical characteristics and physical parameters of these configurations are presented in Table (1.1):
\begin{table}[ht]
\centering
\begin{tabular}{c c c}
\hline\hline
 & NSTX &NSTX-U  \\ [1ex] 
\hline 
$R_0$ & 0.85m & 0.93m  \\
$\alpha$ & 0.67m & 0.57m \\
$B_0$ & 0.43T & 1.0T  \\
$\kappa$ & 2.2 & 2.5  \\
$t$ & 0.5 & 0.3  \\ 
$P_{max}$ & $\sim 10^4Pa$ & $\sim 10^4Pa$  \\ [1ex]
\hline 
\end{tabular}
\caption{\emph{Geometrical characteristics and physical parameter values for spherical tokamaks NSTX and NSTX-U}}
\end{table}
It is found \cite{Hazeltine} that for a sufficiently isolated plasma, the anisotropy function $\sigma _d$, which is related with the collisions in the plasma, is proportional to the parameter
\begin{center}
 $\delta \sim $ (ion gyroradius)/(plasma scale-size)
 \end{center}
 so it is clear that anisotropy becomes larger in higher $\delta$-values.
 Consequently, we expect the anisotropy in pressure to be larger on spherical tokamaks than on the conventional ones, since their size is smaller.
\section{Objectives and Outline of the Thesis}

\hspace{2em}The MHD equilibria of axisymmetric plasmas,  which can be  starting points of 
stability and transport studies, is governed by the well known  Grad-Shafranov 
(GS) equation. The most widely employed analytic solutions of this equation is the Solovev solution \cite{so} and the Hernegger-Maschke solution \cite{hema}, the former corresponding  to toroidal current density non vanishing on the plasma boundary and the  latter to toroidal current density vanishing thereon.  
In the presence of flow the equilibrium satisfies a generalised  Grad-Shafranov 
(GGS) equation together with a Bernoulli equation involving the pressure 
(see for example \cite{moso,ha,T-Th}). 
For compressible flow the GGS equation can be either elliptic or hyperbolic 
depending on the value of a  Mach function  associated with the poloidal velocity. 
Note that the toroidal velocity is inherently incompressible because of 
axisymmetry. In the presence of compressibility the GGS equation is coupled with the Bernoulli 
equation through the density which is not uniform on magnetic surfaces.
For incompressible flow the density becomes a surface quantity and the  GGS 
equation  becomes elliptic and  decouples from the 
Bernoulli equation (see section 2.1). Consequently one has to solve an easier and 
well posed elliptic boundary value problem. In particular for fixed boundaries,  
convergence to the solution is guaranteed under mild requirements of 
monotonicity for the free functions involved in the GGS equation 
\cite{couhi}. For plasmas with anisotropic pressure the equilibrium equations involve a function associated with this anisotropy [Eq. (\ref{sigma})]. To get a closed set of reduced equilibrium  equations an assumption on the functional dependence of this function is required  (cf. \cite{Cotsaftis,Clement,Kuznetsov} and \cite{zwi}-\cite{fu} for static equilibria and  \cite{clst}, \cite{iabo}-\cite{ivma} for stationary ones).\par

The main motivation of the present work was the perspective of the generalisation of the Grad-Shafranov equation for an equilibrium with pressure anisotropy as well as mass flow, because in most situations the plasma pressure is thought to be isotropic. Since in nowadays and future tokamak experiments the external momentum sources employed for heating and current drive usually induce both plasma flow and pressure anisotropy, understanding their combined effects is of practical importance. Another motive is that once such an equilibrium is constructed, then one can examine the impact of anisotropy on the equilibrium characteristics and compare with that of the flow. Also this work may be seen as a future incentive for the generalisation of the sufficient stability condition \cite{stability} in order for an equilibrium with pressure anisotropy to be linearly stable.
\par
In the present chapter we made an introduction to plasma physics and controlled thermonuclear fusion. The main work of the thesis has been organized into two basic parts; the first one includes chapter 2, and the second chapters 3 and 4. The main conclusions are reported in chapter 5 together with a brief proposal for potential extensions of the study.
\par
In the second chapter we derive a new GGS equation by including both anisotropic pressure and incompressible flow of arbitrary direction. This equation consists of six arbitrary surface quantities and recovers  known equations as particular cases, as well as the usual GS equation for a static isotropic plasma. Together we obtain a Bernoulli equation for the quantity $\overline{p}$ [Eq. (\ref{effective})]. Recall that this may be interpreted as an effective isotropic pressure. For the derivation we assume that the function of pressure anisotropy is uniform on magnetic surfaces. In fact, as it will be shown,  for static equilibria as well as for stationary equilibria either with toroidal flow or incompressible flow parallel to the magnetic field, this property of the anisotropy function  follows if
the current density shares the same surfaces with the magnetic field. In addition, with a generalised transformation this equation will be transformed in u-space, where u is the flux function that does not affect but relabels the magnetic surfaces in place of $\psi$.  For convenience the generalised GS equation will be put in completely dimensionless form. At last, a brief presentation of basic equilibrium parameters and figures of merit such as the plasma beta and the safety factor, to be calculated at the second part of the thesis, is made.
Then for appropriate choices of the free functions involved we obtain an extended Solovev solution describing configurations with a non-predefined boundary, and  an extended Hernegger-Maschke solution with a fixed boundary possessing an X-point imposed by  Dirichlet type boundary conditions.  On the basis of these solutions we construct ITER-like, as well as NSTX and NSTX-Upgrade-like equilibria for arbitrary flow, both diamagnetic and paramagnetic.  
\par
In chapter 3 we construct Solovev-like equilibria, in which the free functions are chosen to be linear with respect to $u$, considering the free-boundary problem (without boundary conditions) and solving the normalized -with respect to the magnetic axis- Grad-Shafranov equation, both for diamagnetic and paramagnetic plasmas, for parallel and non-parallel flows, and for ITER and NSTX configurations.
\par
Thereafter, in chapter 4 we solve the normalized -with respect to the geometric center of the configurations- GGS equation for quadratic choice of the free functions involved, to construct Hernegger-Maschke-like equilibria for flow parallel to the magnetic field and diamagnetic plasmas, both for ITER and NSTX-U configurations. Pertinent up-down asymmetric configurations having a lower X-point will be derived by imposing an appropriate set of boundary conditions on the general solution. Because of the fixed boundary the geometric center remains fixed too, thus justifying the normalization adopted.
\par
In both of the above mentioned chapters on the basis of the equilibrium solutions constructed we will examine the impact of both pressure anisotropy and plasma flow on the basic equilibrium quantities and confinement figures of merit, such as the pressure, the magnetic field, the current density, the parameter beta and the safety factor. The main conclusions are that the pressure anisotropy and the flow act on equilibrium in an additive way, with the anisotropy having a stronger impact than that of the flow. Also the effects of flow and anisotropy are in general more noticeable  in spherical tokamaks than in conventional ones. 
\par
Finally, in chapter 5 we will present the overall conclusions of the current work, and will briefly outline potential projects for future work.
\newpage\null\newpage

\chapter{Equilibrium - Generalised Grad-Shafranov Equation}

\section{ Derivation of Grad-Shafranov Equation in the Precence of Pressure Anisotropy and Plasma Flow}

\hspace{2em}The ideal MHD equilibrium states of plasma flows and anisotropic pressure are governed by the set of equations (\ref{continuity}), (\ref{Ampere})-(\ref{Ohm}), and (\ref{momentum}), presented in section 1.3.
The Grad-Shafranov equation is a two-dimensional, quasilinear, elliptic partial differential equation obtained from the reduction of these equations for the case of axisymmetry. The geometry of interest is illustrated in Fig. (2.1).
\begin{figure}
  \centering
    \includegraphics[width=4in]{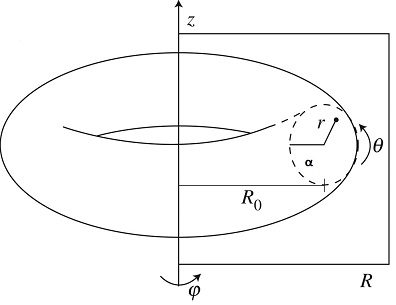}
     \caption{{\small \emph{Geometry for axisymmetric toroidal equilibrium.}}}
\end{figure}
Here, $R$, $\phi$, $z$ denote the usual right-handed cylindrical coordinate system and the assumption of toroidal axisymmetry implies that $\partial W/ \partial \phi =0$, for every quantity $W$.
\par
The derivation of the generalised Grad-Shafranov equation has been organized as follows: 1) we express the divergence-free fields, the magnetic field, the current density and the momentum density field in terms of scalar functions, 2) we identify some integrals of the system in the form of surface quantities, and 3) by using these integrals we derive a generalised Grad-Shafranov equation together with a Bernoulli equation for the effective pressure. The derivation is presented below.
\par
 The axisymmetric assumption implies that Eq. (\ref{BGauss}) can be written as
\beq \label{magnetic1} \frac{1}{R}\frac{\partial (RB_R)}{\partial R}+\frac{\partial B_z}{\partial z}=0\eeq
It is useful to introduce a stream function $\psi$ for the poloidal magnetic field
\beq \label{magnetic2} B_R=\frac{1}{R}\frac{\partial \psi}{\partial z} , B_z=-\frac{1}{R}\frac{\partial \psi}{\partial R}\eeq
where $\psi=-RA_{\phi}$ and $A_{\phi}$ is the toroidal component of vector potential.
\par
Owing to axisymmetry, the divergence-free fields, i.e., the magnetic field, the current density $\vec{J}$ and the momentum density $\rho \vec{v}$ can be expressed in terms of the stream functions $\psi (R,z)$, $I(R,z)$, $F(R,z)$ and $\Theta (R,z)$ as
\beq \label{magnetic field} \vec{B}=I\vec{\nabla}\phi +\vec{\nabla}\phi \times \vec{\nabla}\psi\eeq
\beq \label{current density} \vec{J}=\frac{1}{\mu _0}(\Delta ^{*}\psi \vec{\nabla}\phi -\vec{\nabla}\phi \times \vec{\nabla}I)\eeq
and
\beq \label{velocity field} \rho \vec{v}=\Theta \vec{\nabla}\phi +\vec{\nabla}\phi \times \vec{\nabla}F\eeq
Here constant $\psi$ surfaces are the magnetic surfaces; $F$ is related to the poloidal flux of the momentum density field, $\rho \vec{v}$;   the quantity $I=RB_\phi$ is related to the net poloidal current flowing in   the plasma and the toroidal field coils;   $\Theta =\rho R v_{\phi} $;   $\Delta^{*}$ is the elliptic operator defined by $\Delta^{*}\equiv R^2\vec{\nabla}\cdot (\vec{\nabla}/R^2)$; and  $\vec{\nabla}\phi \equiv \hat{e_{\phi}}/R$.
The resulting forms of $\vec{J}$, Eq. (\ref{current density}), and $\rho \vec{v}$, Eq. (\ref{velocity field}), are due to the Ampere's law, $\vec{J}=\frac{1}{\mu _0}\vec{\nabla}\times \vec{B}$, and the mass conservation equation, $\vec{\nabla}\cdot (\rho \vec{v})=0$.
\begin{figure}
  \centering
    \includegraphics[width=3.5in]{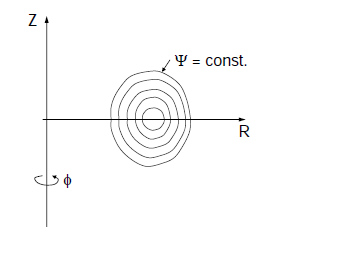}
     \caption{{\small \emph{Magnetic Surfaces.}}}
\end{figure}

\begin{flushleft}
{\bfseries {The meaning of $\psi $}}
\end{flushleft}

For a magnetically confined plasma the contours   $\psi (R,z) =$const. are closed and nested as shown in Fig. (2.2). Therefore in space the magnetic field lies on nested toroidal surfaces called magnetic surfaces. These surfaces are well defined because of axisymmetry. 
We can calculate the magnetic flux through a disk lying in the $z=0$ plane, as shown in Fig. (2.3).
\begin{figure}
  \centering
    \includegraphics[width=3.5in]{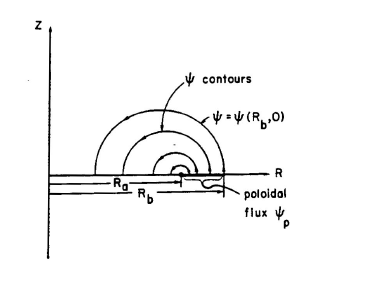}
     \caption{{\small \emph{Washer-shaped surface through which the poloidal flux $\psi_{pol}$ passes. }}}
\end{figure}
By direct computation using the form of $\vec{B}$ Eq. (\ref{magnetic field}) and $d\vec{A}=RdRd\phi \hat{e_z}$, the poloidal flux is the integral over the disk shown
\begin{center}
$\psi _{pol}=\int \vec{B_{pol}}\cdot d\vec{A}=-2\pi \int \frac{\partial \psi}{\partial R}dR=-2\pi \int d\psi$
\end{center}
\begin{center}
$\Longrightarrow \psi _{pol}=-2\pi \psi$ 
\end{center}
Thus, $\psi$ is the negative of the poloidal magnetic flux per radian.
The magnetic field lie inside this $\psi =$constant surfaces, since it follows form Eq. (\ref{magnetic field}) that
\beq \label{magnetic surfaces} \vec{B}\cdot \vec{\nabla}\psi =0\eeq
and therefore these surfaces are called magnetic surfaces.
\par
In addition, the current lies inside well defined closed nested surfaces $I=$constant, called current surfaces, since it follows from Eq. (\ref{current density}) that
\beq \label{current surfaces} \vec{J}\cdot \vec{\nabla}I =0\eeq
From Eq. (\ref{velocity field}) in a similar way it turns out that the function $F$ is related to the poloidal flux of the momentum density field $\rho \vec{v}$, while the quantity $I$ is related to the net poloidal current flowing in the plasma and the toroidal field coils. Specifically, the current flows through a disk-shaped surface lying in the $z=0$ plane extending out to an arbitrary $I$ contour defined by $I =I (R_b,0)$, see Fig. (2.4). One finds 
\begin{figure}
  \centering
    \includegraphics[width=3.5in]{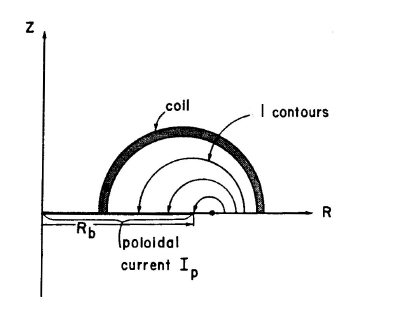}
     \caption{{\small \emph{Disk-shaped surface through which the total poloidal current $I_{pol}$ flows. }}}
\end{figure}
\begin{center}
$I_{pol}=\int \vec{J}_{pol}\cdot d\vec{A}=\frac{2\pi}{\mu _0}\int \frac{\partial I}{\partial R}dR\Rightarrow I_{pol}=\frac{2\pi}{\mu _0}I$
\end{center} 
We have to note that there exist three sets of surfaces in the plasma, namely, the isobaric surfaces (surfaces of constant pressure), the magnetic surfaces and the current surfaces, which in general do not coincide.
\par
The next step of the derivation consists in projecting the Ohm's law, Eq. (\ref{Ohm}), and the momentum equation (\ref{momentum}), into three directions: $\vec{\nabla} \phi$, $\vec{B}$, and $\vec{\nabla} \psi$; that is along the toroidal direction, parallel to the magnetic field and  perpendicular to a magnetic surface respectively.
The projection of Ohm's law to $\vec{\nabla}\phi$ and $\vec{B}$, with the use of Eq. (\ref{Faraday}) yields
\beq \label{un3} \vec{\nabla}\phi \cdot (\vec{\nabla}F \times \vec{\nabla}\psi ) =0\eeq
and
\beq \label{un4} \vec{B}\cdot \vec{\nabla}\Phi =0\eeq
From the above equations we identify two integrals of the system, which are surface quantities, and these are:
\beq \label{integral1} \Phi =\Phi (\psi)\eeq
and
\beq \label{integral2} F=F(\psi)\eeq
It is recalled that surface quantities are quantities remaining uniform on magnetic surfaces.
\par
An additional surface quantity is found from the component of Eq. (\ref{Ohm}) perpendicular to a magnetic surface:
\beq \label{integral3} \Phi ^{'}=\frac{1}{\rho R^2}(IF^{'}-\Theta )\eeq
where the prime denotes differentiation with respect to $\psi$.
Using equations (\ref{velocity field}) and (\ref{integral3}), one obtains a more useful relation for the velocity field: 
\beq \label{velocity1} \vec{v}=\frac{F^{'}}{\rho}\vec{B}-R^2\Phi ^{'}\vec{\nabla}\psi\eeq
which decomposes $\vec{v}$ into a component parallel to $\vec{B}$ and a non parallel one associated with the electric field in consistence with the Ohm's law (\ref{Ohm}).
\par
Next, by projecting the momentum conservation equation (\ref{momentum}) along $\vec{\nabla}\phi $ we get
\beq \label{un5} \vec{\nabla}\cdot \left\lbrace \left[(1-\sigma _d -M_p ^2)I+\mu _0R^2F^{'}\Phi^{'}\right]\vec{B}_{pol}\right\rbrace =0\eeq
Here we have introduced the poloidal Mach function as:
\beq \label{Mach} M_p ^2\equiv \frac{v_{pol}^2}{v_{Apol}^2}=\frac{v_{pol}^2}{B_{pol}^2/\mu_0 \rho}= \mu_0 \frac{(F^{'})^2}{\rho}\eeq
where $v_{Apol}=\frac{B_{pol}}{\sqrt{\mu_0\rho}}$ is the Alfv\' en velocity associated with the poloidal magnetic field.
\par
With the use of the identity $\vec{\nabla}\cdot (f\vec{G})=f(\vec{\nabla}\cdot \vec{G})+\vec{G}\cdot \vec{\nabla}f$ and of $\vec{\nabla}\cdot \vec{B}_{pol}=0$, we obtain from Eq. (\ref{un5})
\beq \label{un6} \vec{B}_{pol}\cdot \vec{\nabla}\left[(1-\sigma _d -M_p ^2)I+\mu _0R^2F^{'}\Phi^{'}\right]=0\eeq
and so we have found a fourth surface quantity of the system, which is:
\beq \label{integral4} X(\psi)\equiv (1-\sigma _d -M_p ^2)I+\mu _0R^2F^{'}\Phi^{'}\eeq
From Eq. (\ref{integral4}) it follows that, unlike the case in static equilibria, $I$ is not a surface quantity, and neither is $\Theta $, since on the basis of (\ref{integral4}) they take the following forms
\beq \label{Ipsi} I(\psi ,R)=\frac{X-\mu _0R^2 F^{'}\Phi ^{'}}{1-\sigma _d-M_p^2}\eeq
and
\beq \label{Thetapsi} \Theta (\psi ,R)=\frac{XF^{'}-(1-\sigma _d)\rho R^2 \Phi ^{'}}{1-\sigma _d-M_p^2}\eeq
Subsequently, by projecting the momentum conservation equation (\ref{momentum}) along $\vec{B}$, we obtain the following relation
\beq \label{Bernoulli1} \vec{B}\cdot \left\lbrace \vec{\nabla}\left[\frac{v^2}{2}+\frac{\Theta \Phi ^{'}}{\rho}\right]+\frac{1}{\rho}\vec{\nabla}\overline{p}\right\rbrace =0\eeq
where the effective pressure is defined as $\overline{p}=\frac{p_{\parallel}+p_{\bot}}{2}$.
\par
The last step is to project the momentum equation (\ref{momentum}) perpendicular to a magnetic surface, but first we put (\ref{momentum}) in a slightly different form.
From Eq. (\ref{velocity field}) we get
\beq \label{un7} -\rho \vec{v}\times (\vec{\nabla}\times \vec{v})=-\frac{1}{2}\frac{\rho}{R^2}\vec{\nabla}\left(\frac{\Theta}{\rho}\right)^2+\left[\vec{B}_{pol}\cdot \vec{\nabla}\left(\frac{\Theta F^{'}}{\rho}\right)\right]\vec{\nabla}\phi -\left\lbrace\vec{\nabla}\cdot \left[\frac{(F^{'})^2}{\rho}\frac{\vec{\nabla}\psi}{R^2}\right]-\frac{F^{'}F^{''}}{\rho}\frac{|\vec{\nabla}\psi|^2}{R^2}\right\rbrace \vec{\nabla}\psi \eeq
From Eqs. (\ref{magnetic field}) and (\ref{current density}) it follows
\beq \label{un8} -(1-\sigma _d)\vec{J}\times \vec{B}=\frac{1}{\mu _0}\vec{\nabla}\cdot \left[(1-\sigma _d)\frac{\vec{\nabla}\psi}{R^2}\right]\vec{\nabla}\psi -\frac{(1-\sigma _d)^{'}}{\mu _0}\frac{|\vec{\nabla}\psi|^2}{R^2}\vec{\nabla}\psi +\frac{(1-\sigma _d)}{2\mu _0 R^2}\vec{\nabla}I^2-\frac{(1-\sigma _d)}{\mu _0}\left[\vec{B}_{pol}\cdot \vec{\nabla}I\right]\vec{\nabla}\phi \eeq
Also, by the use of the identity $\vec{\nabla}(fg)=f\vec{\nabla}g+g\vec{\nabla}f$ we obtain
\beq \label{un9} \sigma _d \vec{\nabla}\left(\frac{B^2}{2\mu _0}\right)=\vec{\nabla}\left[\sigma _d \frac{B^2}{2\mu _0}\right]-\sigma _d ^{'}\frac{B^2}{2\mu _0}\vec{\nabla}\psi \eeq
With equations (\ref{un7})-(\ref{un9}), the momentum equation (\ref{momentum}) can be written in a convenient for the next step form:
\begin{eqnarray} \label{un10}
 \rho \vec{\nabla}\left(\frac{v^2}{2}\right)
 &-&\frac{\rho}{2R^2}\vec{\nabla}\left(\frac{\Theta}{\rho}\right)^2
 +\frac{(1-\sigma _d)}{2\mu _0 R^2}\vec{\nabla}I^2+\left\lbrace \vec{B}_{pol}\cdot \left[\vec{\nabla}\left(\frac{\Theta F^{'}}{\rho}\right) -\frac{(1-\sigma _d)}{\mu _0}\vec{\nabla}I\right] \right\rbrace \vec{\nabla}\phi \nonumber \\
&+&\vec{\nabla}\overline{p}+\vec{\nabla}\cdot \left[\left(\frac{1-\sigma _d}{\mu _0}-\frac{(F^{'})^2}{\rho}\right)\frac{\vec{\nabla}\psi}{R^2}\right]\vec{\nabla}\psi -\left(\frac{(1-\sigma _d)^{'}}{\mu _0}-\frac{F^{'}F^{''}}{\rho}\right)\frac{|\vec{\nabla}\psi |^2}{R^2}\vec{\nabla}\psi \nonumber \\
&-&\sigma _d ^{'}\frac{B^2}{2\mu _0}\vec{\nabla}\psi =0
\end{eqnarray}
Thus, by projecting the momentum conservation equation (\ref{un10}) normal to a magnetic surface, one obtains the following equation:

\begin{eqnarray} \label{GS1}
&&\left\lbrace \vec{\nabla} \cdot \left[(1-\sigma _d -M_p ^2)\frac{\vec{\nabla}\psi}{R^2}\right]+
\left[\mu _0 \frac{F^{'}F^{''}}{\rho}-(1-\sigma _d )^{'}\right]\frac{|\vec{\nabla}\psi |^2}{R^2}
-\mu _0 \sigma _d^{'}\frac{B^2}{2\mu _0}\right\rbrace |\vec{\nabla}\psi |^2 \nonumber \\
&+&\left\lbrace  \mu _0 \rho \vec{\nabla}\left(\frac{v^2}{2}\right)-\frac{\mu _0 \rho}{2R^2}\vec{\nabla}\left(\frac{\Theta}{\rho}\right)^2+\frac{(1-\sigma _d )}{2R^2}\vec{\nabla}I^2 +\mu _0 \vec{\nabla}\overline{p}\right\rbrace \cdot \vec{\nabla}\psi =0
\end{eqnarray}
Therefore, irrespective of compressibility the equilibrium is governed by the  equations (\ref{Bernoulli1}) and (\ref{GS1}) coupled through the density, $\rho$, and the pressure anisotropy function, $\sigma_d$. 
Equation (\ref{GS1})  has a singularity when $\sigma_d + M_p ^2 =1$, and so we must assume that 
$\sigma _d + M_p ^2 \neq 1$.

\begin{flushleft}
{\bfseries {Incompressible Flows}}
\end{flushleft}

In order to reduce the equilibrium equations further, we employ the incompressibility condition
\beq \label{incompressibility} \vec{\nabla}\cdot \vec{v}=0\eeq
Then Eq. (\ref{continuity}) implies that the density is a surface quantity, 
\beq \label{rhopsi} \rho =\rho (\psi)\eeq
and so is the Mach function
\beq \label{Machpsi} M_p ^2=M_p ^2(\psi)\eeq
Recall that in order to obtain a closed set of equations we also assume that $\sigma _d$ is uniform on magnetic surfaces. For static equilibria this follows from Eq. (\ref{integral4}), which becomes $X(\psi)=-I\sigma_d$,  if in the presence of anisotropy the current density  remains on the magnetic surfaces ($I=I(\psi)$).  Since $M_p=M_p(\psi)$, the  same implication for $\sigma_d$ holds for parallel incompressible flow as well as for purely toroidal flow.
\par
In addition, from Eqs. (\ref{integral3}) and (\ref{integral4}) it follows that axisymmetric equilibria with purely poloidal flow $(\Theta =0)$ cannot exist because of the following contradiction: from Eq. (\ref{integral4}) it follows that  $I=\frac{X(\psi)}{1-\sigma _d (\psi)}$ is a surface function, but also, $I=\frac{\rho (\psi)\Phi ^{'}(\psi)}{F^{'}(\psi)}R^2$ from Eq. (\ref{integral3}), implying that $I$ has an explicit dependence on $R$; so it cannot be a surface function.$  $
On the other hand, there can exist an equilibrium with purely toroidal flow, either ``compressible" with uniform temperature $T(\psi )$, but density that varies on the magnetic surfaces, or an incompressible one with uniform density $\rho (\psi )$, but varying temperature. ``Compressible" here means that the density varies on magnetic surfaces; otherwise for purely toroidal flow the incompressibility condition $\vec{\nabla}\cdot \vec{v}=0$ is identically satisfied. For isotropic plasmas both kinds of these  equilibria were examined in \cite{pothta}. In a future project we may extent these studies for plasmas with anisotropic pressure.
\par
We also note that when the electric field term associated with non parallel flows vanishes, Eq. (\ref{Ipsi}) implies that $I$ is a surface function, $I=I(\psi)$. Then it follows from (\ref{current surfaces}) that $\vec{J}\cdot\vec{\nabla}\psi=0$. Thus, for static or field-aligned incompressible flows the current surfaces coincide with the magnetic surfaces.
\par
On the basis of Eq. (\ref{rhopsi}), Eq. (\ref{Bernoulli1}) can be put in the form:
\beq \label{Bernoulli2} \vec{B}\cdot \vec{\nabla}\left[\rho \frac{v^2}{2}+\frac{XF^{'}\Phi ^{'}}{1-\sigma _d -M_p ^2}-
\frac{(1-\sigma _d)\rho R^2 (\Phi ^{'})^2}{1-\sigma _d -M_p ^2}+\overline{p}\right]=0\eeq
which implies that the scalar, which its gradient appears in (\ref{Bernoulli2}), is a surface quantity to be called  $f(\psi)$. 
So the effective pressure is written in the following form:
\beq \label{Bernoulli3} \overline{p}=\underbrace{f(\psi)-\frac{X(\psi)F^{'}(\psi)\Phi ^{'}(\psi)}{1-\sigma _d (\psi) -M_p ^2 (\psi)}}_{\text{function of $\psi \equiv \overline{p}_s (\psi) $}}
-\underbrace{\rho \left[\frac{v^2}{2}-\frac{(1-\sigma _d) R^2 (\Phi ^{'})^2}{1-\sigma _d -M_p ^2}\right]}_{\text{flow term}}\eeq
Thus, we obtain a Bernoulli equation for $\overline{p}$ which is:
\beq \label{Bernoullipsi} \overline{p}=\overline{p}_s (\psi)-\rho \left[\frac{v^2}{2}-\frac{(1-\sigma _d) R^2 (\Phi ^{'})^2}{1-\sigma _d -M_p ^2}\right]\eeq
Therefore, in the presence of flow the magnetic surfaces in general do not coincide with the  surfaces on which $\overline{p}$ is uniform. In this respect, the term containing $\overline{p}_s (\psi)$ is the static part of the effective pressure which does not vanish when $\vec{v}=0$.
\par
Finally, by inserting  Eq. (\ref{Bernoullipsi}) into Eq. (\ref{GS1}) after some algebraic manipulations,  the latter reduces to the following elliptic differential equation,

\begin{eqnarray} \label{GGS psi}
&&(1-\sigma _d -M_p ^2)\Delta ^{*}\psi +\frac{1}{2} (1-\sigma _d -M_p ^2)^{'}|\vec{\nabla}\psi |^2
+\frac{1}{2}\left(\frac{X^2}{1-\sigma _d -M_p ^2}\right)^{'} \nonumber \\
&+&\mu _0 R^2 \overline{p}_s ^{'}+
\mu _0 \frac{R^4}{2}\left[\frac{(1-\sigma _d)\rho (\Phi ^{'})^2}{1-\sigma _d -M_p ^2}\right]^{'}=0
\end{eqnarray}
This is the GGS equation that governs the equilibrium for an axisymmetric plasma with pressure anisotropy and incompressible flow. For flow parallel to the magnetic field the $R^4$-term vanishes. For vanishing flow Eq. (\ref{GGS psi}) reduces to the one derived in \cite{Clement}, when the pressure is isotropic it reduces to the one obtained in \cite{T-Th}, and when both anisotropy and flow are absent it reduces to the well known GS equation. Equation (\ref{GGS psi}) contains six arbitrary surface quantities, namely: $X(\psi)$, $\Phi (\psi)$, $\overline{p}_s (\psi)$, $\rho (\psi)$, $M_p ^2 (\psi)$ and $\sigma _d (\psi)$, which can be assigned as functions of $\psi$ to obtain analytically solvable  linear  forms of the equation or from other physical considerations.

\begin{flushleft}
{\bfseries{Isodynamicity}}
\end{flushleft}

There is  a special class of static equilibria called isodynamic for which the magnetic field magnitude is a surface quantity ($|\vec{B}|=|\vec{B}(\psi)|$) \cite{Palumbo}. This feature can have beneficial effects on confinement because the grad-$B$ drift vanishes and consequently plasma transport perpendicular  to the magnetic surfaces is reduced. Also, it was proved that the only possible isodynamic equilibrium is axisymmetric \cite{PaBo}. In the presence of flow,  assuming  that the plasma obeys to the ideal gas law, $P=\hat{R}\rho T$, and isothermal magnetic surfaces $T=T(\psi)$, then it follows that the scalar pressure becomes also a surface function, $P=P(\psi )$. Thus, from Bernoulli equation it follows that the magnitude of the magnetic field can be written in the form

\beq \label{iso1} |\vec{B}|^2=\Xi (\psi )+R^2 \left(\frac{\rho (\psi) \Phi ^{'}(\psi)}{F^{'}(\psi)}\right)^2\eeq
implying that in the case of field-aligned flows, $\Phi ^{'}=0$, the equilibrium becomes isodynamic.  Isodynamic -like equilibria with a variety of side-conditions including $P=P(\psi)$ were studied in \cite{THR-TAS}. 
\par
Here we are interested in what happens in the anisotropic pressure case. For fusion plasmas the thermal conduction along $\vec{B}$ is fast compared to the heat transport perpendicular to a magnetic surface, so a good assumption is that the parallel temperature is a surface function, $T_{\parallel}=T_{\parallel}(\psi )$. Then, it follows that the parallel pressure becomes also a surface function, $p_{\parallel}=p_{\parallel}(\psi )$. We are going to examine the properties of $B$ and $p_{\bot}$.
\par
On the basis of Eq. (\ref{Bernoulli1}), and due to the assumptions $T_{\parallel}=T_{\parallel}(\psi )$, $p_{\parallel}=p_{\parallel}(\psi )$, we get
\beq \label{iso2} \vec{B}\cdot \vec{\nabla}\left[\rho \frac{v^2}{2}+\rho \frac{\Theta \Phi ^{'}}{\rho}+\frac{p_{\bot}}{2}\right]=0\Rightarrow \rho \left[\frac{v^2}{2}+\frac{\Theta \Phi ^{'}}{\rho}\right]+\frac{p_{\bot}}{2}\equiv G(\psi)\eeq
Then, it follows form equations (\ref{velocity1}) and (\ref{iso2}) that the magnitude of the magnetic field is related with the perpendicular pressure as
\beq \label{iso3} |\vec{B}|^2=\frac{2G(\psi)}{M_p ^2(\psi)}-\left(p_{\bot}-\rho R^2 (\Phi ^{'})^2\right)\frac{1}{M_p ^2(\psi)}\eeq
A first observation that can be made, is that $|\vec{B}|^2$ becomes a surface function when the perpendicular pressure satisfies the relation $p_{\bot}=\rho R^2 (\Phi ^{'})^2$. 
This implies that 
\beq \label{iso4} \sigma _d=\sigma _d(\psi ,R)=\mu _0 \frac{p_{\parallel} (\psi)}{|\vec{B}|^2(\psi)}-R^2\mu _0\frac{\rho(\psi)(\Phi ^{'})^2(\psi)}{|\vec{B}|^2(\psi)}\eeq
which is in contradiction with the hypothesis that function $\sigma _d$ is a surface quantity.
Consequently, the only possibility for isodynamic magnetic surfaces to exist is that for field aligned flows, $\Phi ^{'}=0$, because then Eq. (\ref{iso3}) reduces to 
\beq \label{iso5} |\vec{B}|^2=\frac{2G(\psi)}{M_p ^2(\psi)}-\frac{p_{\bot}}{M_p ^2(\psi)}\eeq
Eqs. (\ref{sigma}) and (\ref{iso5})
imply that both $|\vec{B}|^2=|\vec{B}|^2(\psi)$ and 
$p_{\bot}=p_{\bot}(\psi)$. \par
Thus, the conclusions for the isotropic case \cite{THR-TAS} are generalised for 
anisotropic pressure, i.e. all three $B$, $p_{\parallel}$ and $p_{\bot}$ become surface quantities. We note here that the more physically pertinent case that $B$ and $p_{\bot}$ remain arbitrary functions would require either compressibility or eliminating the assumption $\sigma _d=\sigma _d(\psi)$. However, in this case tractability is lost and the problem requires numerical treatment.

\section{Generalised Transformation}

\hspace{2em}Using the transformation  
\beq \label{transformation} u(\psi)=\int_{0}^{\psi} \sqrt{1-\sigma _d (g)-M_p ^2 (g)}dg,\qquad \sigma _d +M_p ^2<1\eeq
Eq. (\ref{GGS psi}) reduces to 
\beq \label{GGSu} \Delta ^{*}u+\frac{1}{2}\frac{d}{du}\left(\frac{X^2}{1-\sigma _d -M_p ^2}\right)+\mu _0 R^2\frac{d\overline{p}_s}{du}+\mu _0 \frac{R^4}{2}\frac{d}{du}\left[(1-\sigma _d )\rho \left(\frac{d\Phi}{du}\right)^2\right]=0\eeq
Also, Eq. (\ref{Bernoullipsi}) is put in the form
\beq \label{Bernoulliu} \overline{p}=\overline{p}_s (u)-\rho \left[\frac{v^2}{2}-(1-\sigma _d) R^2 \left(\frac{d\Phi}{du}\right)^2\right]\eeq
 Transformation (\ref{transformation}) does not affect the magnetic surfaces, it just relabels them by the flux function $u$, and is a generalisation of that introduced in \cite{Simintzis} for isotropic equilibria with incompressible flow ($\sigma_d=0$) and that introduced in \cite{Clement} for static anisotropic equilibria ($M_p^2=0$). 
 Note that no quadratic term as $|\vec{\nabla}u|^2$ appears anymore in (\ref{GGSu}).
 \par
 Once a solution of this equation is found, the equilibrium can be completely constructed  with calculations in the $u$-space by using (\ref{transformation}) and the inverse transformation 
 $\label{inv_transformation}\psi(u)=\int_{0}^{u} (1-\sigma _d (g)-M_p^2 (g))^{-1/2}dg $.
 Thus, the equilibrium quantities presented on section (2.1) take the following form:
\beq \label{Bu} \vec{B}=I\vec{\nabla}\phi +(1-\sigma _d -M_p ^2)^{-1/2}\vec{\nabla}\phi \times \vec{\nabla}u\eeq
\beq \label{Xu} X=(1-\sigma _d -M_p ^2)\left[I+\mu _0 R^2\left(\frac{dF}{du}\right)\left(\frac{d\Phi}{du}\right)\right]\eeq
\beq \label{Iu} I=\frac{X}{1-\sigma _d -M_p ^2}-\mu _0 R^2\left(\frac{dF}{du}\right)\left(\frac{d\Phi}{du}\right)\eeq
\beq \label{dFu} \frac{dF}{du}=(1-\sigma _d -M_p ^2)^{-1/2}\sqrt{\frac{\rho}{\mu _0}}M_p\eeq
\beq \label{vu} \vec{v}=\frac{M_p}{\sqrt{\rho \mu _0}}\vec{B}-R^2 (1-\sigma _d -M_p ^2)^{1/2}\left(\frac{d\Phi}{du}\right)\vec{\nabla}\phi \eeq
\beq \label{Eu} \vec{E}=-\frac{d\Phi}{du}\vec{\nabla}u\eeq
\beq \label{Ju} \vec{J}=\frac{1}{\mu _0}\left[(1-\sigma _d -M_p ^2)^{-1/2}\Delta ^{*}u-\frac{1}{2}(1-\sigma _d -M_p ^2)^{-3/2}\frac{d}{du}(1-\sigma _d -M_p ^2)|\vec{\nabla}u|^2\right]\vec{\nabla}\phi -\frac{1}{\mu _0}\vec{\nabla}\phi \times \vec{\nabla}I\eeq

\section{Normalized Equations}

\hspace{2em}Before continuing to the presentation of the analytical solutions, we find it convenient to make a normalization to all equilibrium quantities. To this end we introduce dimensionless quantities with the use of constant reference ones to be defined  later.  
We adopt the following normalization:
\beq \label{nor1} \xi =\frac{R}{R_i}\qquad ,\qquad \zeta =\frac{z}{R_i}\eeq
\beq \label{nor2} \widetilde{\overline{p}}=\frac{\overline{p}}{B_i ^2 /\mu _0}\qquad ,\qquad \widetilde{\rho}=\frac{\rho}{\rho _i}\eeq
\beq \label{nor3} \widetilde{u}=\frac{u}{B_i R_i ^2}\qquad ,\qquad \widetilde{I}=\frac{I}{B_i R_i}\eeq
\beq \label{nor4} \widetilde{\vec{E}}=\frac{\vec{E}}{v_{A_i}B_i}\qquad ,\qquad \widetilde{\vec{B}}=\frac{\vec{B}}{B_i}\eeq
\beq \label{nor5} \widetilde{\vec{J}}=\frac{\vec{J}}{B_i /\mu _0 R_i}\qquad ,\qquad \widetilde{\vec{v}}=\frac{\vec{v}}{v_{A_i}}\eeq
\beq \label{nor6} \widetilde{\Phi}=\frac{\Phi}{v_{A_i}B_i R_i}\qquad ,\qquad \widetilde{F}=\frac{F}{\rho _i v_{A_i}R_i ^2}\eeq
We note that $X$ is normalized in the same way as $I$. Also, $p_{\parallel}$, $p_{\bot}$, $\overline{p}_s$, are normalized in the same way as $\overline{p}$, and the functions $\sigma _d $, and $M_p ^2 $ need not be normalized since they are already dimensionless.
The index $i$ is either $i=a,0$, where $a$ denotes the magnetic axis, and 0 the geometric center of a configuration. This is because in chapter 3 and in Solovev solution we use a normalization with respect to the magnetic axis, since we solve a free-boundary problem. Thus, the magnetic axis can be held in a fixed position, $\xi _a =1$. Also, in chapter 4 for the Hernegger-Maschke solution we normalize with respect to the geometric center since we fix the boundary, and it is inconvenient to predefine the position of the magnetic axis in this situation.
Thus, the normalization constants  are defined as follows: $R_i$ is the radial coordinate of the configuration's magnetic axis/geometric center, and $B_i$, $\rho _i$, $v_{A_i}=\frac{B_i}{\sqrt{\mu _0 \rho _i}}$ are the magnitude of the magnetic field, the  plasma density, and the Afv\' en velocity thereon.
 \par
 On the basis of Eqs. (\ref{nor1})-(\ref{nor6}), equations (\ref{GGSu})-(\ref{Ju}) take the following normalized forms:
\beq \label{GGSu nor} \widetilde{\Delta ^{*}}\widetilde{u}+\frac{1}{2}\frac{d}{d\widetilde{u}}\left(\frac{\widetilde{X}^2}{1-\sigma _d -M_p ^2}\right)+\xi ^2\frac{d\widetilde{\overline{p}_s}}{d\widetilde{u}}+\frac{\xi ^4}{2}\frac{d}{d\widetilde{u}}\left[(1-\sigma _d )\widetilde{\rho} \left(\frac{d\widetilde{\Phi}}{d\widetilde{u}}\right)^2\right]=0\eeq
\beq \label{Bernoulliu nor} \widetilde{\overline{p}}=\widetilde{\overline{p}_s} (\widetilde{u})-\widetilde{\rho} \left[\frac{\widetilde{v}^2}{2}-(1-\sigma _d) \xi ^2 \left(\frac{d\widetilde{\Phi}}{d\widetilde{u}}\right)^2\right]\eeq
\beq \label{Bu nor} \widetilde{\vec{B}}=\widetilde{I}\widetilde{\vec{\nabla}}\phi +(1-\sigma _d -M_p ^2)^{-1/2}\widetilde{\vec{\nabla}}\phi \times \widetilde{\vec{\nabla}}\widetilde{u}\eeq
\beq \label{Xu nor} \widetilde{X}=(1-\sigma _d -M_p ^2)\left[\widetilde{I}+\xi ^2\left(\frac{d\widetilde{F}}{d\widetilde{u}}\right)\left(\frac{d\widetilde{\Phi}}{d\widetilde{u}}\right)\right]\eeq
 \beq \label{dFu nor} \frac{d\widetilde{F}}{d\widetilde{u}}=(1-\sigma _d -M_p ^2)^{-1/2}\sqrt{\widetilde{\rho}}M_p\eeq
\beq \label{vu nor} \widetilde{\vec{v}}=\frac{M_p}{\sqrt{\widetilde{\rho}}}\widetilde{\vec{B}}-\xi ^2 (1-\sigma _d -M_p ^2)^{1/2}\left(\frac{d\widetilde{\Phi}}{d\widetilde{u}}\right)\widetilde{\vec{\nabla}}\phi \eeq
\beq \label{Ju nor} \widetilde{\vec{J}}=\left[(1-\sigma _d -M_p ^2)^{-1/2}\widetilde{\Delta ^{*}}\widetilde{u}-\frac{1}{2}(1-\sigma _d -M_p ^2)^{-3/2}\frac{d}{d\widetilde{u}}(1-\sigma _d -M_p ^2)|\widetilde{\vec{\nabla}}\widetilde{u}|^2\right]\widetilde{\vec{\nabla}}\phi -\widetilde{\vec{\nabla}}\phi \times \widetilde{\vec{\nabla}}\widetilde{I}\eeq
\beq \label{Eu nor} \widetilde{\vec{E}}=-\frac{d\widetilde{\Phi}}{d\widetilde{u}}\widetilde{\vec{\nabla}}\widetilde{u}\eeq
where $\widetilde{\Delta ^{*}}=\frac{\partial ^2}{\partial \xi ^2}+\frac{\partial ^2}{\partial \zeta ^2}-\frac{1}{\xi}\frac{\partial}{\partial \xi}$.

\section{Equilibrium Parameters and Figures of Merit}

\hspace{2em}Once given an MHD equilibrium, it is possible to define a number of global quantities depending only upon the flux surface label. These surface quantities represent important physical parameters which can be related to experiment and be used as figures of merit for confinement. Some of the most important are presented as follows:

\begin{flushleft}
{\bfseries {Plasma Beta}}
\end{flushleft}

The quantity $\beta $ is a global plasma parameter whose value is critical for a fusion reactor. It measures the efficiency of plasma confinement by the magnetic field. Interestingly, there is actually no unique definition of plasma $\beta $ that is agreed upon by the fusion community. Various definitions are distinguished by different geometric factors whose choice is motivated by a given configuration's aspect ratio and cross sectional shape.
The plasma beta is defined as the ratio of plasma pressure to magnetic pressure. In the case of anisotropic pressure we represent the plasma pressure by the effective pressure, so that the plasma beta can be defined as
\beq \label{beta} \beta \equiv \frac{\overline{p}}{B^2/2\mu _0}\eeq
A more accurate definition is the average beta, defined as the ratio of the averaged plasma energy to the averaged magnetic energy:
\beq \label{betaav} \beta _{av}\equiv \frac{2\mu _0 <p>}{<B_{tor} ^2+B_{pol} ^2>}\eeq
where the average pressure is defined as 
\beq \label{avpress} <p>=\frac{1}{3}Tr(\stackrel{\textstyle\leftrightarrow}{\rm {\mathbb P}})=\frac{p_{\parallel}+2p_{\bot}}{3}\eeq
and $B_{tor}$, $B_{pol}$ are averaged over the total plasma volume.
\par
It is often useful to define separate toroidal and poloidal $\beta$'s measuring plasma confinement efficiency with respect to each component of the magnetic field. In the present thesis we are interested only in the calculation of the local toroidal beta, defined as
\beq \label{betator} \beta _t =\frac{\overline{p}}{B_0 ^2/2 \mu _0}\eeq
and particularly the value of $\beta _t$ on the magnetic axis of each configuration.
\par
Above, we have made the usual, convenient choice for the toroidal magnetic pressure for any cross section, $B_{tor} ^2\rightarrow B_0 ^2$, where $B_0$ is the vacuum toroidal field at the geometric center of the chamber confining the plasma.
In general, high values of $\beta $ are desirable for fusion reactor economics and technology. However, there is a maximum allowable value of $\beta $ set by MHD equilibrium requirements and by MHD instabilities driven by the pressure gradients. Recent experiments on the National Spherical Torus Experiment (NSTX) have made significant progress in reaching high toroidal beta $\beta _t \leq 35\% $ \cite{Menard}, while on ITER the beta parameter is expected to take low values, $\beta _t \sim 2\% $ \cite{ITERbeta}.

\begin{flushleft}
{\bfseries {Safety Factor}}
\end{flushleft}

For good confinement in a closed system, it is required that the field lines form a set of nested toroidal surfaces. The rotational-transform angle, $\vartheta$, defines the poloidal angle through which a field line rotates in the course of one complete transit in the toroidal direction. In work on tokamaks and toroidal pinches, one encounters the inverse quantity called the safety factor
\beq q \equiv \frac{2\pi}{\vartheta}\eeq
The safety factor $q$ plays an important role in determining stability, and also it is involved in transport theory. It is a topological property of the surface and cannot change in time unless field lines are cut and reconnected.
\par
An alternative expression for $q$ can be obtained in terms of the magnetic fluxes. Specifically, consider an infinitesimal annulus between two flux surfaces; the safety factor can be expressed as the rate of change of toroidal flux with respect to poloidal flux
\beq \label{q2} q\equiv \frac{d\psi _{tor}}{d\psi _{pol}}\eeq
For an axisymmetric configuration Eq. (\ref{q2}) yields
\beq \label{q3} q=\frac{d\int \vec{B_{tor}}\cdot d\vec{s}}{d\psi _{pol}}=\frac{\oint \frac{Idld\psi _{pol}}{R|\vec{\nabla}\psi _{pol}|}}{d\psi _{pol}}=\frac{1}{2\pi}\oint \frac{Idl}{R|\vec{\nabla}\psi |}\eeq
since $\psi _{pol}=-2\pi \psi $, and $\oint dl$ is the line integral along the intersection curve of the toroidal flux surface with the poloidal plane. The length element $dl$ can be written in Shafranov coordinates $(r,\theta )$ \cite{Fr}, defined in Fig. (2.5), as
\beq \label{dl1} (dl)^2=(dr)^2+(rd\theta )^2 \Rightarrow dl=d\theta \sqrt{r^2+\left(\frac{dr}{d\theta}\right)^2}\eeq

\begin{figure}
  \centering
    \includegraphics[width=2.5in]{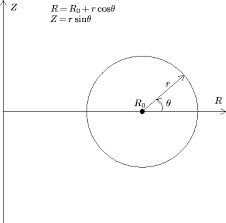}
     \caption{{\small \emph{Shafranov coordinates $(r,\theta)$ related with cylindrical coordinates $(R,z)$ on the poloidal cross section.}}}
\end{figure}
On a streamline $\psi (r,\theta) =$constant, implying that $d\psi =\frac{\partial \psi}{\partial r}dr+\frac{\partial \psi}{\partial \theta}d\theta=0$, so then, Eq. (\ref{dl1}) yields
\beq \label{dl2} dl=d\theta\sqrt{r^2+\left(\frac{\psi _{\theta}}{\psi _{r}}\right)^2}\eeq
where, $\psi _{\theta}=\frac{\partial \psi}{\partial \theta}$, and $\psi _{r}=\frac{\partial \psi}{\partial r}$.
Substituting Eq. (\ref{dl2}) into Eq. (\ref{q3}), we get the following expression for the safety factor
\beq \label{qpsi} q=\frac{1}{2\pi}\int_{0}^{2\pi}\frac{I\sqrt{r^2+\left(\frac{\psi _{\theta}}{\psi _{r}}\right)^2}}{R|\vec{\nabla}\psi |}d\theta \eeq
On the basis of the transformation (\ref{transformation}), the safety factor $q$ is expressed in u-space as:
\beq \label{qu} q=\frac{1}{2\pi}\int_{0}^{2\pi}\frac{I(u,R)\sqrt{r^2+\left(\frac{u _{\theta}}{u _{r}}\right)^2}}{(1-\sigma _d -M_p ^2)^{-1/2}R|\vec{\nabla}u |}d\theta \eeq
Hence, with respect to the adopted normalization, Eq. (\ref{qu}) takes the following form:
\beq \label{qu nor} q=\frac{1}{2\pi}\int_{0}^{2\pi}\frac{\widetilde{I}(\widetilde{u},\xi)\sqrt{\widetilde{r}^2+\left(\frac{\widetilde{u} _{\theta}}{\widetilde{u} _{r}}\right)^2}}{(1-\sigma _d -M_p ^2)^{-1/2}\xi|\widetilde{\vec{\nabla}}\widetilde{u} |}d\theta \eeq
In order to calculate numerically the safety factor profile we constructed a simple Do loop programme in Wolfram Mathematica suite, presented in Appendix A.
\par
There also exists a simpler individual relation for the local value of the safety factor, on the magnetic axis of a given configuration, which reads
\beq \label{qa psi} q_a =\frac{I}{R}\left\lbrace \frac{\partial ^2\psi}{\partial R^2}\frac{\partial ^2\psi}{\partial z^2}\right\rbrace _{R=R_a,z=z_a} ^{-1/2}\eeq
which in u-space becomes
\beq \label{qa u} q_a =(1-\sigma _d-M_p ^2)^{1/2}\frac{I}{R}\left\lbrace \frac{\partial ^2u}{\partial R^2}\frac{\partial ^2u}{\partial z^2}\right\rbrace _{R=R_a,z=z_a} ^{-1/2}\eeq
On the basis of the normalization adopted Eq. (\ref{qa u}) takes the form
\beq \label{qa u nor} q_a =(1-\sigma _d-M_p ^2)^{1/2}\frac{\widetilde{I}}{\xi}\left\lbrace \frac{\partial ^2\widetilde{u}}{\partial \xi ^2}\frac{\partial ^2\widetilde{u}}{\partial \zeta ^2}\right\rbrace _{\xi =\xi _a,\zeta =\zeta _a} ^{-1/2}\eeq
The proof of relation (\ref{qa psi}) is presented in Appendix B. \par
From equations (\ref{Iu}) and (\ref{qa u}), one observes that when the flows are parallel to the magnetic field, $\frac{d\Phi }{du}=0 $, then the value of $q_a$ has no dependence on the anisotropy. Indeed $q_a$ becomes independent on $\sigma _{d_a}$, where $\sigma _{d_a}$ is the local value of the anisotropy function on the magnetic axis, $\sigma _{d_a}=\sigma _d|_{R=R_a,z=z_a}$.
\par
In general, higher values of the safety factor are desirable for an equilibrium to be stable. In order for a tokamak equilibrium to be stable there exists a necessary condition known as the Kruskal-Shafranov criterion, which implies that $q$ must be higher that the unit:
\beq \label{Kruskal1} q>1\eeq
In conventional tokamaks the profile of the safety factor is usually monotonically increasing from the magnetic axis to the plasma boundary, so the Kruskal-Shafranov limit is satisfied if
\beq \label{Kruskal2} q_a >1\eeq
since $q$ takes its lower value on the magnetic axis.
In view of (\ref{Kruskal2}) later, on chapter 4, we will practically demand $q_a=1.1$, in order to construct a desirable Hernegger-Maschke -like equilibrium.

\section{Conclusions}

\hspace{2em}In the present chapter we derived a new generalised Grad-Shafranov (GGS) equation that governs plasma equilibrium in the presence of pressure anisotropy and incompressible mass flow [Eq. (\ref{GGS psi})]. This is an elliptic partial differential equation containing six arbitrary surface quantities: $X(\psi)\,,\Phi(\psi)\,,\overline{p}_s(\psi)\,,\rho(\psi)\,,M_p^2(\psi)$, together with the anisotropy function $\sigma_d(\psi)$, assumed to be a surface quantity.
\par
 The derivation of the GGS equation based on axisymmetric ideal MHD equations with anisotropic pressure and convective plasma flow, was conducted in the following three steps. First, express the divergence-free fields in terms of scalar functions, second, project the Ohm's law and the momentum equation into three directions -parallel to the magnetic field, along the toroidal direction and normal to a magnetic surface- to identify four integrals of the system in form of surface quantities, and third, with the aid of these integrals obtain the GGS equation together with a generalised Bernoulli equation for the effective pressure [Eq. (\ref{Bernoullipsi})].
The GGS equation recovers known GS-like ones governing static anisotropic equilibria \cite{Clement} and isotropic equilibria with plasma flow \cite{T-Th}. Also for static isotropic equilibria the equation is reduced to the usual GS equation.
\par
In addition, by employing a generalised transformation [Eq. (\ref{transformation})], the GGS equation is put in a simpler form [Eq. (\ref{GGSu})]. This transformation does not affect the magnetic surfaces but relabels them, and consists extension of the transformations introduced in \cite{Clement} for static anisotropic equilibria and \cite{Simintzis} for isotropic equilibria with incompressible flow. The form of the equation containing the sum $M_p^2+\sigma_d$ indicates that pressure anisotropy and flow act additively with the only exception the $R^4$-term associated with non parallel flows.
\par
After assigning the free functions, the GGS equation derived can be solved under appropriate boundary conditions, e.g. Dirichlet ones. Analytic solutions of linearised forms of this equation will be constructed and studied in the following two chapters.
\newpage

\chapter{Solovev-like Solution}

\hspace{2em}The equilibrium of an axisymmetric toroidal plasma with anisotropic pressure and incompressible flow of arbitrary direction is governed by the generalised Grad-Shafranov (GGS) [Eq. (\ref{GGSu nor})], which is a non-linear, partial differential equation of second order. In order to solve analytically such an equation, we first have to linearise it for several reasonable choices of the surface functions $\widetilde{X}(\widetilde{u})$, $\widetilde{\overline{p}_s}(\widetilde{u})$, $\widetilde{\rho}(\widetilde{u})$, $\widetilde{\Phi}(\widetilde{u})$, $M_p(\widetilde{u})$, and $\sigma_d(\widetilde{u})$.
In this chapter Solovev-like equilibria with a free toroidal boundary will be constructed, assigning the free functions of the GGS equation to be linear in $\widetilde{u}$. Since the boundary is not fixed we adopt a normalization with respect to the magnetic axis of a given configuration, as presented on section 2.3. The original Solovev solution was introduced in \cite{so} for the case of static and isotropic equilibria, and that solution as well as extensions of it in the presence of plasma flow have been extensively studied in the literature \cite{Sri}-\cite{Pfi}. After such an equilibrium is constructed we are going to examine the impact of pressure anisotropy on its characteristics and compare them with that of the flow. Both diamagnetic and paramagnetic ITER and NSTX configurations with field-aligned and/or non-parallel to the magnetic field flows will be examined.

\section{Construction of the Solution}

\hspace{2em}According to the Solovev ansatz the free functions in the GGS equation are chosen as
\beq \label{ans1} \widetilde{\overline{p}}_s=\widetilde{\overline{p}}_0+\widetilde{\overline{p}}_1\frac{\widetilde{u}}{\widetilde{u}_b},\qquad \widetilde{u}\geq 0\eeq
\beq \label{ans2} \frac{\widetilde{X}^2}{1-\sigma _d-M_p ^2}=\frac{2\epsilon \widetilde{\overline{p}}_{s_a}}{(1+\delta ^2)}\frac{\widetilde{u}}{\widetilde{u}_b}+1\eeq
\beq \label{ans3} \widetilde{\rho} (1-\sigma _d)\left(\frac{d\widetilde{\Phi}}{d\widetilde{u}}\right)^2=-\frac{2\lambda \widetilde{\overline{p}}_{s_a}}{(1+\delta ^2)}\frac{\widetilde{u}}{\widetilde{u}_b}+\widetilde{\Phi} _0\eeq
Here,  $a$ denotes the magnetic axis and $b$ the plasma boundary; $\delta$ determines the elongation of the magnetic surfaces near the magnetic axis; for $\epsilon >0\ (<0)$ the plasma is diamagnetic (paramagnetic); and $\lambda$ is a parameter related with the non-parallel component of the flow.  In addition, we impose that the solution $\widetilde{u}$ vanishes on the magnetic axis, $\widetilde{u}_a=0$.
\par
Since the plasma is extended up to the boundary, pressure and density should vanish there, $\widetilde{\overline{p}}_{s_b}=\widetilde{\rho} _b=0$, so that one finds that $\widetilde{\overline{p}}_1=-\widetilde{\overline{p}}_0=-\widetilde{\overline{p}}_{s_a}$ and $\widetilde{\Phi} _0=\frac{2\lambda \widetilde{\overline{p}}_{s_a}}{(1+\delta ^2)}$.
Thus, relations (\ref{ans1})-(\ref{ans3}) take the form
\beq \label{sol1} \widetilde{\overline{p}}_s=\widetilde{\overline{p}}_{s_a}\left(1-\frac{\widetilde{u}}{\widetilde{u}_b}\right),\qquad \widetilde{u}\geq 0\eeq
\beq \label{sol2} \frac{\widetilde{X}^2}{1-\sigma _d-M_p ^2}=\frac{2\epsilon \widetilde{\overline{p}}_{s_a}}{(1+\delta ^2)}\frac{\widetilde{u}}{\widetilde{u}_b}+1\eeq
\beq \label{sol3} \widetilde{\rho} (1-\sigma _d)\left(\frac{d\widetilde{\Phi}}{d\widetilde{u}}\right)^2=\frac{2\lambda \widetilde{\overline{p}}_{s_a}}{(1+\delta ^2)}\left(1-\frac{\widetilde{u}}{\widetilde{u}_b}\right)\eeq
From Eq.(\ref{sol3}) it follows that $\lambda$ is restricted to non-negative values, $\lambda \geq 0$.
The unit constant in Eq. (\ref{sol2}) results due to the adopted normalization.
With this linearising ansatz the GGS equation (\ref{GGSu nor}) reduces to
\beq \label{linearsol} \widetilde{\Delta ^*}\widetilde{u}+\frac{\widetilde{\overline{p}}_{s_a}}{\widetilde{u}_b}\left[\frac{\epsilon}{(1+\delta ^2)}-\xi ^2-\xi ^4 \frac{\lambda}{(1+\delta ^2)}\right]=0\eeq
which admits the generalised Solovev solution valid for arbitrary $\widetilde{\rho}$, $\sigma_d$ and $M_{p}^2$:
\beq \label{solovevsolution} \widetilde{u}(\xi ,\zeta)=\frac{\widetilde{\overline{p}}_{s_a}}{2(1+\delta ^2)\widetilde{u}_b}\left[\zeta ^2(\xi ^2-\epsilon)+\frac{\delta ^2+\lambda}{4}(\xi ^2 -1)^2+\frac{\lambda}{12}(\xi ^2 -1)^3\right]\eeq
To completely determine the equilibrium 
we choose the plasma density, the Mach function and the anisotropy function profiles to be peaked on the magnetic axis and vanishing on the plasma boundary as 
\beq \label{rho sol} \widetilde{\rho} (\widetilde{u})=\widetilde{\rho} _a\left(1-\frac{\widetilde{u}}{\widetilde{u}_b}\right)^g \eeq
\beq \label{Mach sol} M_p^2(\widetilde{u})=M_{p_a}^2\left(1-\frac{\widetilde{u}}{\widetilde{u}_b}\right)^\mu \eeq
\beq \label{sigma sol} \sigma _d(\widetilde{u})=\sigma _{d_a}\left(1-\frac{\widetilde{u}}{\widetilde{u}_b}\right)^n \eeq
with $\widetilde{\rho}_a$ and $\widetilde{u}_b$ constant quantities.
It is noted  here that the above chosen  density function, peaked on the magnetic axis and vanishing on the boundary is  typical for tokamaks. Also, the Mach function adopted having a similar shape is reasonable at least in connection with experiments with  on axis focused  external momentum sources.
\par
The functions $\widetilde{\rho}$, $M_p^2$ and $\sigma_{d}$ chosen depend on two free parameters; their maximum on axis and an exponent associated with the shape of the profile; the exponent of the function  $M_p^2$, connected with flow shear, is held fixed at $\mu =2$, while we also choose $g=1/2$ and $\widetilde{\rho} _a =1$ due to the normalization with respect to the magnetic axis. The value of $M_{p_a}$ depends on the kind of tokamak  (conventional or spherical). On account of experimental evidence \cite{Brau}-\cite{Eriksson} the toroidal rotation velocity in tokamaks is approximately $10^4-10^6\,ms^{-1}$ which for large conventional ones implies  $M_{p_a}^2 \sim 10^{-4}$, while the flow is stronger  for spherical tokamaks ($M_{p_a}^2 \sim 10^{-2}$) \cite{Menard}. As we will see in sections (3.2) and (3.3) below, similar to the flow, also pressure anisotropy takes higher values on spherical tokamaks than on conventional ones. An argument why the flow and pressure anisotropy are stronger in spherical tokamaks rather than in the usual ones, is that in the former the magnetic field is strongly inhomogeneous, as their aspect ratio is too small. 
\par
One can also see by inspection of the ansatz (\ref{sol3}), that the parameter $\lambda$ is of the order of $\lambda\sim \left(\frac{R_a}{\alpha}\right)^2\frac{v_\phi^2}{B_a^2/\rho}=\left(\frac{R_a}{\alpha}\right)^2M_t^2$, where $M_t$ is the toroidal Mach function the values of which are on the same order of magnitude as the poloidal one. This is because according to experimental evidence in tokamaks the poloidal velocity is of one order of magnitude less than the toroidal one, while the poloidal magnetic field is of one order higher than the toroidal one. Thus, since for ITER it is $\left(\frac{R_a}{\alpha}\right)^2\sim10$, while for NSTX $\left(\frac{R_a}{\alpha}\right)^2\sim2.5$, $\lambda$ will be of the order of $10^{-2}$, for both configurations. In order to have a clear estimate of the impact of the electric field on equilibrium here we will permit $\lambda$ to be one order of magnitude higher $(\lambda=0.5)$. Also, it may be noted that the electric fields associated with non parallel flow play an important role in the transitions to the improved confinement regimes of tokamaks.

\begin{flushleft}
{\bfseries{Geometrical Characteristics}}
\end{flushleft}

The generalised Solovev solution (\ref{solovevsolution}) does not include enough free parameters  to impose desirable  boundary conditions, but has the property that a separatrix is spontaneously formed. Thus, as already mentioned, we can predefine the position of the magnetic axis, $(\xi _a =1,\zeta_a=0)$, chosen as normalization point and the plasma extends from the magnetic axis up to a closed magnetic surface which we will choose to coincide with the separatrix.
\par
In this subsection we discuss the parameters that characterize the shape of the projection of a magnetic surface on the poloidal plane. Most tokamak configurations today have a D-shaped poloidal cross section. The ``midplane'' is defined as the plane that passes through the geometric center and is perpendicular to the symmetric axis ($z$ axis). For an up-down symmetric (about the midplane $\zeta=0$) magnetic surface, its shape can be  characterized by four parameters, namely, the $\xi$ coordinates of the innermost and outermost points on the midplane, $\xi_{in}$ and $\xi_{out}$, and the $(\xi,\zeta)$ coordinates of the highest (upper) point of the plasma boundary, $(\xi_{up},\zeta_{up})$ (see Fig. 3.1).
\par
In terms of these four parameters we can define the normalized major radius
\beq \label{major} \xi _0=\frac{\xi _{in}+\xi _{out}}{2}\eeq
which is the radial coordinate of the geometric center, the minor radius
\beq \label{minor} \widetilde{\alpha} =\frac{\xi _{out}-\xi _{in}}{2}\eeq
the triangularity of a magnetic surface
\beq \label{triangularity} t=\frac{\xi _0-\xi _{up}}{\widetilde{\alpha }}\eeq
defined as the horizontal distance between the geometric center and the highest point of the magnetic surface normalized  with respect to minor radius, and the elongation of a magnetic surface
\beq \label{elongation} \kappa =\frac{\zeta _{up}}{\widetilde{\alpha }}\eeq
The poloidal magnetic field coil system may be used to shape the plasma cross section. A circular one is the simplest but not the optimal one from the plasma equilibrium and stability point of view. To improve stability with respect to the interchange modes a D-shape is advantageous because of the smaller curvature on the high field side. The D-shaped plasma combines elongation along the vertical axis with triangularity. Usually, we specify the values of $R_0$, $\alpha$, $t$, and $\kappa$, instead of ($\xi _{in}$, $\xi_ {out}$, $\xi _{up}$, $\zeta _{up}$) to characterize the shape of the outermost magnetic surface. On the basis of solution (\ref{solovevsolution}) the latter quantities  can be expressed in terms of $\epsilon$, $\delta$, $\lambda$. Subsequently, in order to make an estimate of realistic values for the free parameters $\epsilon$, $\delta$  and the radial coordinate of the magnetic axis $R_a$ in connection with the tokamaks under consideration,
we employ the relations (\ref{major})-(\ref{elongation}) to find  ($\epsilon$, $\delta$ and $R_a$)  in terms of  the known parameters  ($R_0$, $\alpha$, $t$ and $\kappa$), already presented in subsection (1.4.2) for the ITER, NSTX and NSTX-U devices. Besides using $\widetilde{\alpha}$, and $\xi _0$, we can define another useful parameter $\varepsilon \equiv \widetilde{\alpha} /\xi _0$, which is called the inverse aspect ratio. Note that the major radius $\xi _0$ is different from $\xi _a$ (the radial coordinate of the magnetic axis). Usually $\xi _a >\xi _0$, and the so called Shafranov shift is the radial displacement of the magnetic axis from the geometric center due to the toroidicity
 \beq \label{shift} \Delta \xi =\xi _a -\xi _0\eeq
 and depends on the plasma pressure.
 \begin{figure}
  \centering
    \includegraphics[width=3.5in]{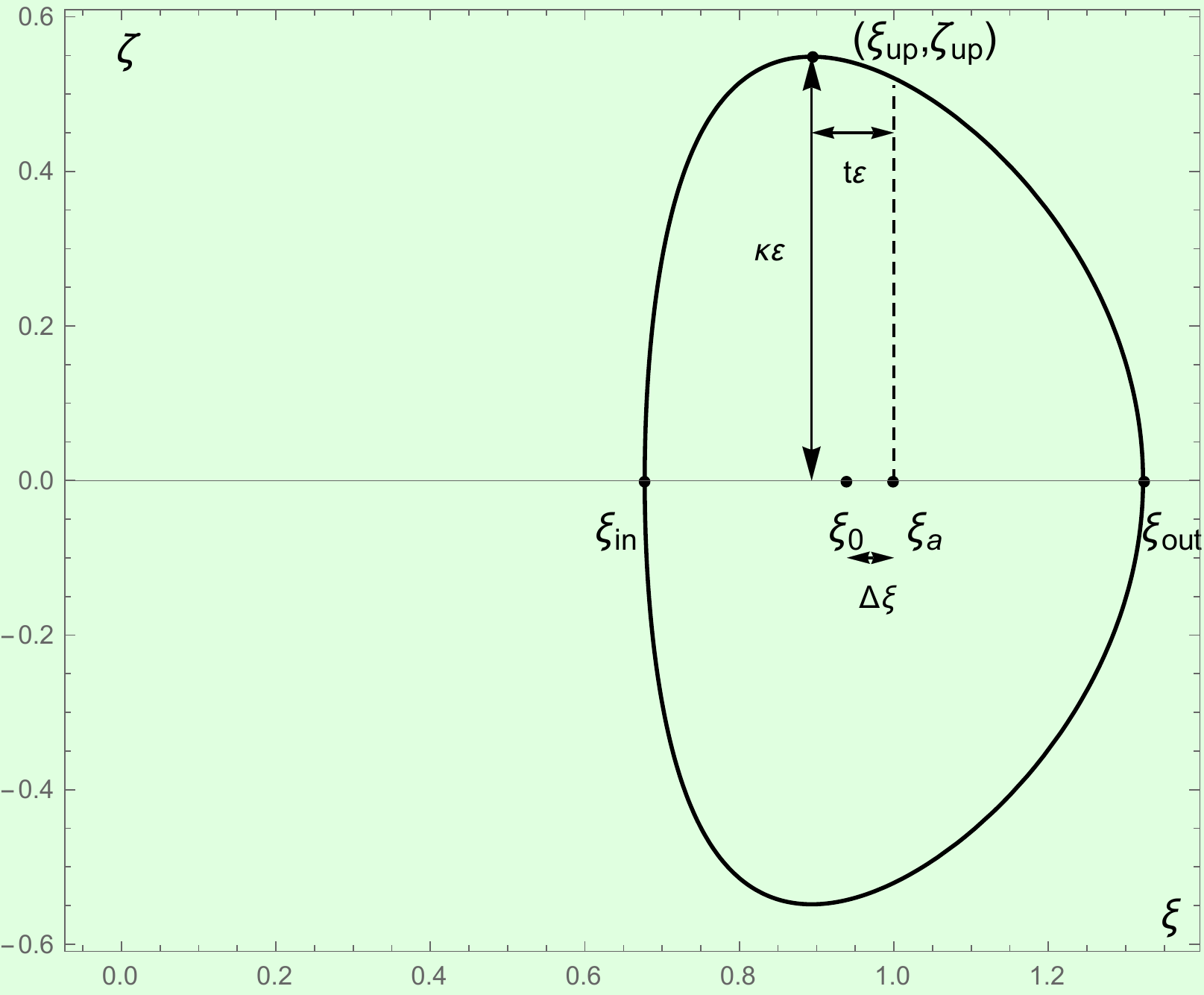}
     \caption{{\small \emph{Characteristic points of the cross section of a D-shaped magnetic surface with the poloidal plane, and shaping parameters.}} }
\end{figure}
Inside the last closed surface having the X-point, named the separatrix, the magnetic surfaces represent nested tori and beyond the separatrix they are open ones and the magnetic lines may continue up to the chamber walls. In the present work, we assume that the plasma extends up to the separatrix. Thus, the last magnetic surface, called plasma boundary is assumed to coincide with the separatrix.

\begin{flushleft}
{\bfseries{Stagnation Points}}
\end{flushleft}

Depending on the value of $\lambda$ the equilibrium configurations can have either one or two stagnation points located:
\newline
\newline
1) at the fixed position $(\xi =1,\zeta =0)$, which is the magnetic axis, and
\newline
\newline
2) the flow-dependent position $(\xi =\sqrt{-1-\frac{2\delta ^2}{\lambda}},\zeta=0)$.
\newline
\newline
For our ansatz, and because the plasma density is chosen to vanish on the boundary, $\lambda \geq 0$, it is apparent that the second stagnation point remains always at $\infty$.
For $\lambda<0$ and free boundary, the equilibrium has an X-point additional to the magnetic axis \cite{Simintzis,Arapoglou}.
\par
Expanding around a given stagnation point one finds that its kind is determined by the sign of the determinant
\beq \label{determinant} D=\left(\frac{\partial ^2\widetilde{u}}{\partial \xi \partial \zeta}\right)^2-\left(\frac{\partial ^2\widetilde{u}}{\partial \xi ^2}\right)\left(\frac{\partial ^2\widetilde{u}}{\partial \zeta ^2}\right)\eeq
If $D<0(>0)$ at a point $(\xi ,\zeta )$, the function $\widetilde{u}$ has an X-point or a magnetic axis there.
\par
One also finds that solution (\ref{solovevsolution}) present the following stationary points (on $ \xi \geq 0$ axis), at which the poloidal magnetic field equals to zero
\newline
\newline
1) $(0,0)$ which is the configuration's center (axis of symmetry),
\newline
\newline
2) $(\sqrt{\epsilon}, \frac{1}{2}\sqrt{2\delta ^2(1-\epsilon)+\lambda(1-\epsilon ^2)})$, and
\newline
\newline
3)$(\sqrt{\epsilon},- \frac{1}{2}\sqrt{2\delta ^2(1-\epsilon)+\lambda(1-\epsilon ^2)})$
\newline
\newline
The second and the third stationary points are symmetrical with respect to $\xi$-axis and depend on the values of $\epsilon$, $\delta$, and $\lambda$. All the above are points of the separatrix and as we will see, in a diamagnetic configuration the separatrix innermost point is at the position $(\sqrt{\epsilon},0)$, while in a paramagnetic one it extends up to the axis of symmetry, $(0,0)$, presenting a corner there.

\begin{flushleft}
{\bfseries{Diamagnetic - Paramagnetic Characterization of a Plasma}}
\end{flushleft}

When the plasma parameter beta in a tokamak is high, for given magnetic field the kinetic pressure is also high and so usually are the pressure gradients inside the plasma. Thus, the plasma is diamagnetic due the homonymous current $ \sim \vec{\nabla} \overline{p} \times \vec{B}/|\vec{B}|^2$. This current flows in such a way as to reduce the imposed field. On the other hand, at low $\beta$ the diamagnetic current is negligible, and the large current in tokamaks is induced by the toroidal loop voltage. It flows parallel to the magnetic field and the poloidal component of the parallel current causes the toroidal magnetic field to increase from its vacuum value; thus, the plasma becomes paramagnetic. Experimental observations of plasma paramagnetism have been reported in \cite{Holly}.
\par
The average kinetic pressure of the plasma is principally balanced by the magnetic pressure associated with the poloidal field. The toroidal field contributes to the pressure balance only to the extent that the toroidal field within the plasma is affected by the poloidal current, $\vec{J}_{pol}$, so that its volume average value differs form the average over the plasma surface. That is, a plasma is paramagnetic when $<B_{tor}^2>_s-<B_{tor}^2>_v<0$, and diamagnetic when $<B_{tor}^2>_s-<B_{tor}^2>_v>0$, where $<>_s$ and $<>_v$ represent averages over the plasma surface and volume, respectively. Since the toroidal field $\widetilde{B}_\phi$ equals $\widetilde{I}(\widetilde{u})/\xi$ (i.e. in the static and isotropic case $\widetilde{X}(\widetilde{u})= \widetilde{I}(\widetilde{u})$), the above conditions can be rewritten as:
\begin{center}
$\widetilde{I}\frac{d\widetilde{I}}{d\widetilde{u}}<0$, paramagnetic
\end{center}
\begin{center}
$\widetilde{I}\frac{d\widetilde{I}}{d\widetilde{u}}>0$, diamagnetic
\end{center}
For the Solovev ansantz (\ref{sol2}) in the static case, it is $\widetilde{I}\frac{d\widetilde{I}}{d\widetilde{u}}\sim \epsilon$. Thus, for $\epsilon >0$ the plasma is diamagnetic, and for $\epsilon<0$ it is paramagnetic.

\section{Diamagnetic Configurations}

\hspace{2em}When $\epsilon >0$ the inner point of the separatrix on the midplane is located at $\xi =\sqrt{\epsilon}$. So, if $\widetilde{u}_s$ is the flux function on the separatrix, then
\beq \label{us diam} \widetilde{u}_s =\widetilde{u}(\sqrt{\epsilon},0)\Rightarrow \widetilde{u}_s =\frac{\widetilde{\overline{p}}_{s_a}(-1+\epsilon)^2[3\delta ^2+(2+\epsilon )\lambda]}{24\widetilde{u}_b(1+\delta ^2)}\eeq
We assume that the plasma extends up to the separatrix of a configuration, thus, 
\beq \label{ub diam} \widetilde{u}_b=\widetilde{u}_s\Rightarrow \widetilde{u}_b=\left[\frac{\widetilde{\overline{p}}_{s_a}(-1+\epsilon)^2[3\delta ^2+(2+\epsilon )\lambda]}{24(1+\delta ^2)}\right]^{1/2}\eeq
We recall that in a Solovev-like solution the value of $\widetilde{u}$ on the magnetic axis is zero by choice. On the boundary it is given by (\ref{ub diam}) and is dependent on the values of $\epsilon$, $\delta$, and $\lambda$.
\par
In order to find the radial coordinates of the innermost and outermost points of the plasma boundary, which are located on the midplane, we have to solve the equation $\widetilde{u}_b=\widetilde{u}(\xi ,0)$, from which we find that
\beq \label{inner diam} \xi _{in}=\sqrt{\epsilon}\eeq
and
\beq \label{outter diam} \xi _{out}=\left[-\frac{\epsilon}{2}-\frac{3\delta ^2}{2\lambda}+\frac{\sqrt{\epsilon}\sqrt{3\delta ^4+8\delta ^2\lambda -2\delta ^2 \epsilon \lambda +4\lambda ^2-\epsilon ^2\lambda ^2}}{2\lambda}\right]^{1/2}\eeq
By using Eqs. (\ref{major}), (\ref{minor}) and (\ref{elongation}) we find analytical relations for the major radius, the minor radius and the Shafranov shift as functions of $\epsilon$, $\delta$, and $\lambda$
\beq \label{d1} \xi _0=\frac{1}{4}\left\lbrace 2\sqrt{\epsilon}+\left[\frac{-6\delta ^2-2\epsilon \lambda +2\sqrt{3}\sqrt{3\delta ^4-2\delta ^2(-4+\epsilon)\lambda -(-4+\epsilon ^2)\lambda ^2}}{\lambda}\right]^{1/2}\right\rbrace \eeq ,
\beq \label{d2} \widetilde{\alpha} =\frac{1}{2}\left\lbrace -\sqrt{\epsilon}+\left[-\frac{\epsilon}{2}-\frac{3\delta ^2}{2\lambda}+\frac{\sqrt{3}\sqrt{3\delta ^4+8\delta ^2\lambda -2\delta ^2\epsilon \lambda +4\lambda ^2-\epsilon ^2\lambda ^2}}{2\lambda}\right]^{1/2}\right\rbrace \eeq 
and
\beq \label{d3} \Delta \xi =1-\frac{1}{4}\left\lbrace 2\sqrt{\epsilon}+\left[\frac{-6\delta ^2-2\epsilon \lambda +2\sqrt{3}\sqrt{3\delta ^4-2\delta ^2(-4+\epsilon)\lambda -(-4+\epsilon ^2)\lambda ^2}}{\lambda}\right]^{1/2}\right\rbrace \eeq
Until now, we have found two out of the four characteristic parameters of the boundary shaping, $\xi _{in}$ and $\xi _{out}$, so the next step is to find respective analytical relations for the coordinates of the upper point of the boundary $(\xi _{up},\zeta _{up})$.
At first, by solving $\widetilde{u}=\widetilde{u}_b$ for $\zeta$ we find the parametric equation 
\beq \label{d4} \zeta _b(\xi)=\frac{\left[3\delta ^2-\frac{3\delta ^2}{(1-\epsilon )^2}+2\lambda -\frac{2\lambda}{(1-\epsilon )^2}+\epsilon \lambda +\frac{6\delta ^2\xi ^2}{(1-\epsilon )^2}-\frac{3\delta ^2\xi ^4}{(1-\epsilon )^2}-\frac{\lambda \xi ^6}{(1-\epsilon )^2}\right]^{1/2}}{\left[-\frac{12\epsilon}{(1-\epsilon )^2}+\frac{12\xi ^2}{(1-\epsilon )^2}\right]^{1/2}}\eeq
Thus, in order to find the radial coordinate of the upper point (in which there exists an extremum) we have to solve the equation $\frac{d\zeta _b(\xi)}{d\xi}=0$.
\par
 To solve this equation is not trivial even by using Mathematica. So we decide to solve it in the static limit, for $\lambda \rightarrow 0$, in which equations (\ref{outter diam})-(\ref{d4}) reduce to the following ones:
\beq \label{ds1} \xi _{out(s)}=\sqrt{2-\epsilon}\eeq
\beq \label{ds2} \xi _{0(s)}=\frac{1}{2}(\sqrt{2-\epsilon}+\sqrt{\epsilon})\eeq
\beq \label{ds3} \widetilde{\alpha} _{(s)}=\frac{1}{2}(\sqrt{2-\epsilon}-\sqrt{\epsilon})\eeq
\beq \label{ds4} \Delta \xi _{(s)}=\frac{1}{2}(2-\sqrt{2-\epsilon}-\sqrt{\epsilon})\eeq
and
\beq \label{ds5} \zeta _{b(s)}(\xi)=\sqrt{\frac{3\delta ^2(1-\epsilon )^2-3\delta ^2+6\delta ^2\xi ^2-3\delta ^2\xi ^4}{12(\epsilon +\xi ^2)}}\eeq
where $(s)$ denotes the static limit.
By solving the equation $\frac{d\zeta_{b(s)}(\xi)}{d\xi}=0$ we find the radial coordinate of the upper point for the case $\lambda =0$ (i.e. for parallel flows); this is
\beq \label{ds6} \xi _{up(s)}=\sqrt{\epsilon}\eeq
Thus, from equation (\ref{triangularity}) it follows that the triangularity of a diamagnetic configuration is $t=1$. Consequently, from Eq. (\ref{ds5}) we can find the $\zeta$-coordinate of the upper point for $\lambda=0$, and this is
\beq \label{ds7} \zeta _{up(s)}=\sqrt{\frac{-\delta ^2(-1+\epsilon)}{2}}\eeq
We observe that in the static limit, $\xi _{in}$, $\xi _{out}$, $\xi _0$, $\xi _{up}$, and $\widetilde{\alpha}$, do not depend anymore on the value of $\delta$ parameter.
In addition, elongation of the magnetic surfaces as a function of $\epsilon$ and $\delta$ in the static limit is given by
\beq \label{elo static diam} \kappa_{(s)}=\frac{\sqrt{-2\delta ^2(-1+\epsilon )}}{\sqrt{2-\epsilon}-\sqrt{\epsilon}}\eeq
As already mentioned, we usually specify the shape of a configuration using known parameters such as the elongation, the triangularity, the major and the minor radius. For this reason we want to find analytical relations for the parameters $\epsilon$ and $\delta$ as function of $\kappa$, $t$, $R_0$, and $\alpha$. Due to the normalization with respect to the magnetic axis adopted we have
\beq \label{minor static diam} \widetilde{\alpha} _{(s)}=\frac{\alpha _{(s)}}{R_a}\Rightarrow \alpha _{(s)}=\frac{R_a}{2}(\sqrt{2-\epsilon}-\sqrt{\epsilon})\eeq
and
\beq \label{major static diam} \xi _{0(s)}=\frac{R_{0(s)}}{R_a}\Rightarrow R_{0(s)}=\frac{R_a}{2}(\sqrt{2-\epsilon}+\sqrt{\epsilon})\eeq
Thus, by solving the system of equations [$\kappa=\kappa_{(s)}$, $\alpha=\alpha _{(s)}$, $R_0=R_{0(s)}$], we find the following relations
\beq \label{epsilon diam} \epsilon =\frac{(R_0-\alpha)^2}{R_0^2+\alpha ^2}\eeq
\beq \label{delta diam} \delta =\kappa \sqrt{\frac{\alpha}{R_0}}\eeq
and
\beq \label{Ra diam} R_a=\sqrt{R_0^2+\alpha ^2}\eeq
From these relations we can now determine the values of $\epsilon$ and $\delta$ parameters for any given configuration for which its major radius, minor radius, and elongation are prescribed. Afterwards, we can also find the values of the four characteristic parameters of the boundary and of the Shafranov shift by using Eqs. (\ref{inner diam}), (\ref{ds1})-(\ref{ds4}), and (\ref{ds6})-(\ref{ds7}). The above relations will also be  employed to assign  values of the free parameters $\epsilon$, $\delta$ and $R_a$  for  non parallel flows ($\lambda \neq 0)$ because in this case the above estimation procedure becomes complicated.
Since the magnetic field at the geometric center of each device is known, we can also estimate its value on the magnetic axis, and then the value of $\widetilde{\overline{p}}_{s_a}$ from the relations
\beq \label{Ba} B_a=B_0\frac{R_0}{R_a}\eeq
and
\beq \label{pmax} \widetilde{\overline{p}}_{s_a}=\frac{\overline{p}_{s_a}}{B_a^2/\mu _0}\eeq
The maximum equilibrium pressure, on the magnetic axis, for ITER is about $\overline{p}_a\approx 10^6\ Pa$, while for the NSTX spherical tokamak it is about $\overline{p}_a\approx 10^4\ Pa$. Since the flow term in the Bernoulli equation (\ref{Bernoulliu nor}) is small, we usually adopt the approximation $\overline{p}_a\approx\overline{p}_{s_a}$.
\par
On the basis of the above analysis we are ready to calculate the values of all the parameters mentioned above for the ITER and NSTX tokamaks for which we are going to examine the influence of pressure anisotropy. These values are presented on Table (3.1) for the case $\lambda=0$. 
\begin{table}[ht]
\centering
\begin{tabular}{c c c}
\hline\hline
 & ITER &NSTX  \\ [1ex] 
\hline 
$\epsilon$ & 0.415646 & 0.0276592  \\
$\delta$ & 0.965535 & 1.95322 \\
$R_a\ (m)$ & 6.5146 & 1.08231  \\
$B_a\ (T)$ & 5.04406 & 0.337703  \\
$\widetilde{\overline{p}}_{s_a}$ & 0.049353 & 0.110104  \\ 
$\widetilde{u}_b$ & 0.0318804 & 0.101537  \\ 
$\xi _0$ & 0.951709 & 0.785356 \\
$\widetilde{\alpha}_0$ & 0.307 & 0.619 \\
$\xi _{in}$ & 0.644706 & 0.166311 \\
$\xi _{out}$ & 1.25871 & 1.4044 \\
$\xi _{up}$ & 0.644706 & 0.166311 \\
$\zeta _{up}$ & 0.521905 & 1.3619 \\[1ex]
\hline 
\end{tabular}
\caption{{\small \emph{Characteristic values of the shaping parameters for ITER and NSTX spherical tokamaks for $\lambda=0$. }}}
\end{table}
Thus, we can fully determine the solution $\widetilde{u}$ from Eq. (\ref{solovevsolution}), 
as well as the position of the characteristic points of the boundary and obtain the ITER-like and NSTX-like diamagnetic configurations, whose poloidal cross-section with  a set of magnetic surfaces are shown in Figs. (3.2)-(3.3). We note that by expansions around the magnetic axis it turns out that the magnetic surfaces in the vicinity of the magnetic axis have elliptical cross-sections (see also \cite{Lao}-\cite{Bizarro}).
\par
 Also, the separatrix presents two up-down symmetrical corners, with its inner part to be defined by the vertical line $\xi=\sqrt{\epsilon}$; for the NSTX spherical tokamak this is located very close to the axis of symmetry in accordance with the small hole of spherical tokamaks. For $\epsilon=0$ it will coincide with the axis of symmetry and the configuration will become compact. When $\lambda>0$ the configurations are similar to the ones presented for parallel flows, since the plasma is extended always up to the separatrix, but as $\lambda$ takes higher values the magnetic surfaces are more elongated parallel to $\zeta$ axis as compared with the static-and parallel flow-equilibrium ones.

\begin{figure}
  \centering
    \includegraphics[width=3.5in]{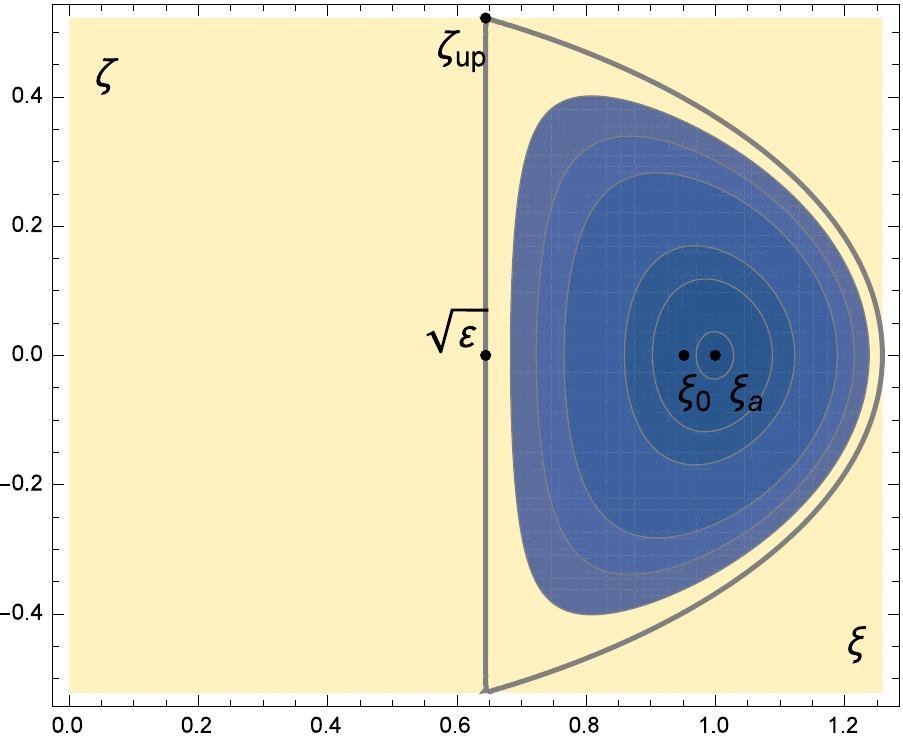}
     \caption{{\small \emph{Diamagnetic ITER-like equilibrium configuration determined by Eq. (3.8) for the parameter values of Table (3.1).}} }
\end{figure}
\begin{figure}
  \centering
    \includegraphics[width=2.5in]{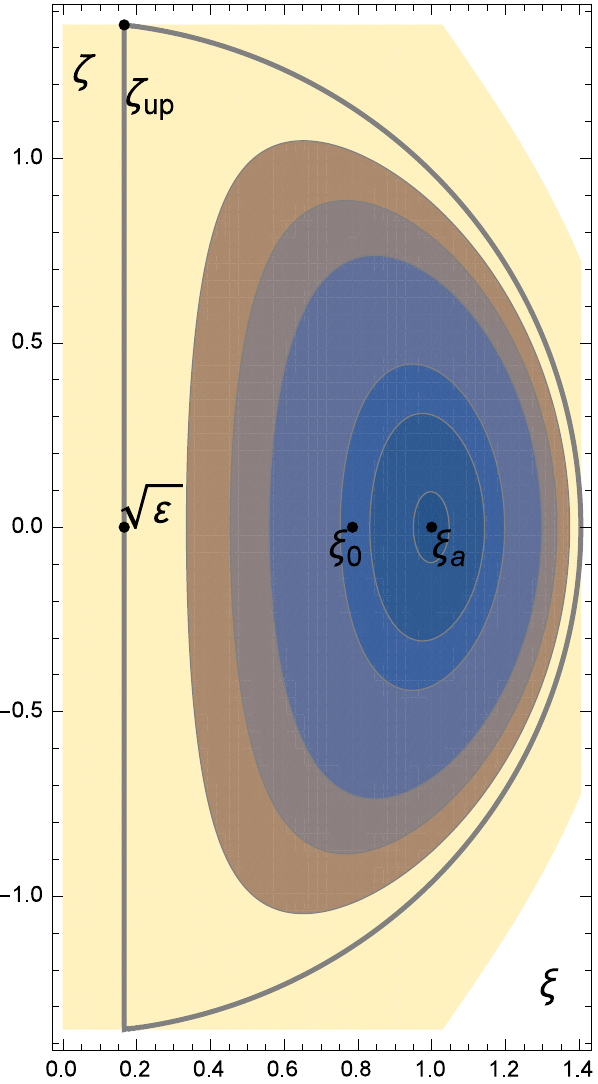}
     \caption{{\small \emph{Diamagnetic NSTX-like equilibrium configuration determined by Eq. (3.8) for the parameter values presented on Table (3.1).}} }
\end{figure}

\subsection{Effects of Pressure Anisotropy and Flow}

\hspace{2em}In this subsection we are going to examine the impact of pressure anisotropy on the equilibrium quantities and confinement figures of merit, via the variation of the free parameters $\sigma_{d_a}$ and $n$, and compare it with that of the flow. From the requirement of positiveness for all pressures within the whole plasma region, we find that the pressure anisotropy parameter $\sigma_{d_a}$ takes higher values on spherical tokamaks than in the conventional ones, as shown on Table (3.2). Also it must be $n\geq 2$, so we will let $2\leq n \leq 10$.
\begin{table}
	\centering
	\begin{tabular}{|l||l|l||l|l|}
		\hline
		&\multicolumn{2}{l|}{Parallel flow $(\lambda=0)$}&\multicolumn{2}{l|}{Non-parallel flow $(\lambda=0.5)$}\\
		\cline{2-5}
		&ITER&NSTX&ITER&NSTX\\
		\hline\hline
		$\sigma_{d_a}^{max}$&0.08&0.11&0.10&0.13\\
		\hline
	\end{tabular}
	\caption{{\small \emph{Approximate maximum permissible values of the free parameter $\sigma_{d_a}$ for the extended Solovev solution in connection with the non negativeness  of  pressure. }}}
\end{table}
\subsubsection{Magnetic Field}

\hspace{2em}When the plasma is diamagnetic the toroidal magnetic field inside the plasma decreases from its vacuum value as  
\beq \label{Bfi} \widetilde{B}_\phi =\frac{\widetilde{I}}{\xi}\eeq
Consequently the profile of the function $\widetilde{I}$ is expected to be hollow.
 As shown in Fig. (3.4), 
\begin{figure}
  \centering
    \includegraphics[width=3.5in]{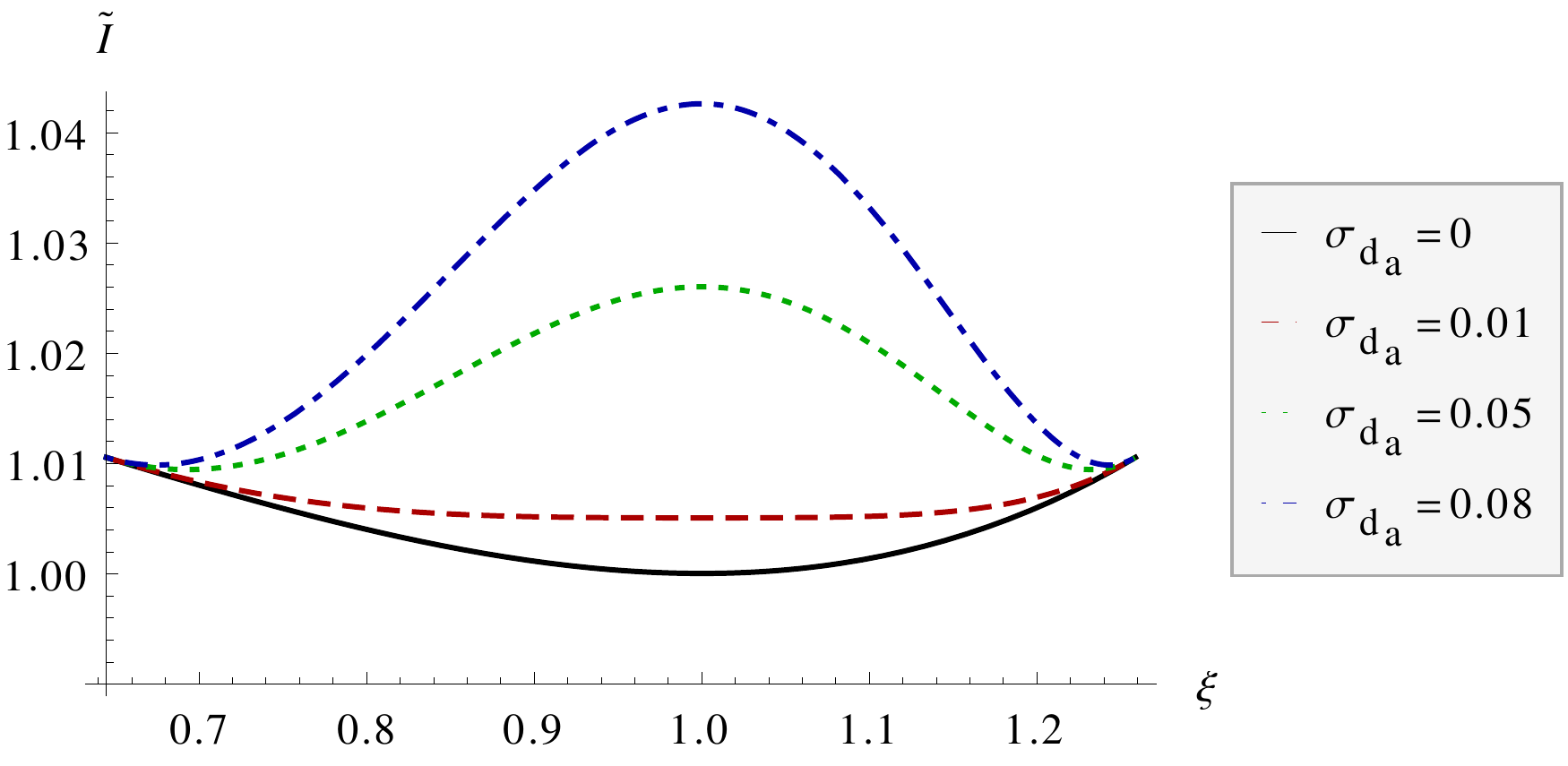}
     \caption{{\small \emph{Variation of $\widetilde{I}$ with $\sigma_{d_a}$ in ITER-like diamagnetic configuration, for field aligned flows ($\lambda=0$), $M_{p_a}^2=10^{-4}$, $n=2$, on the midplane $\zeta=0$. }   }} 
\end{figure} 
as $\sigma_{d_a}$ becomes larger the field increases, and for sufficient high $\sigma_{d_a}$ it becomes peaked on the magnetic axis. This means that increasing pressure anisotropy acts paramagnetically in terms of its maximum value on axis, $\sigma_{d_a}$. The same result can also be inferred directly from $\widetilde{B}_\phi$ profile, but not so clearly since it falls like $\frac{1}{\xi}$, see Figure (3.5). 
 \begin{figure}
  \centering
    \includegraphics[width=3.5in]{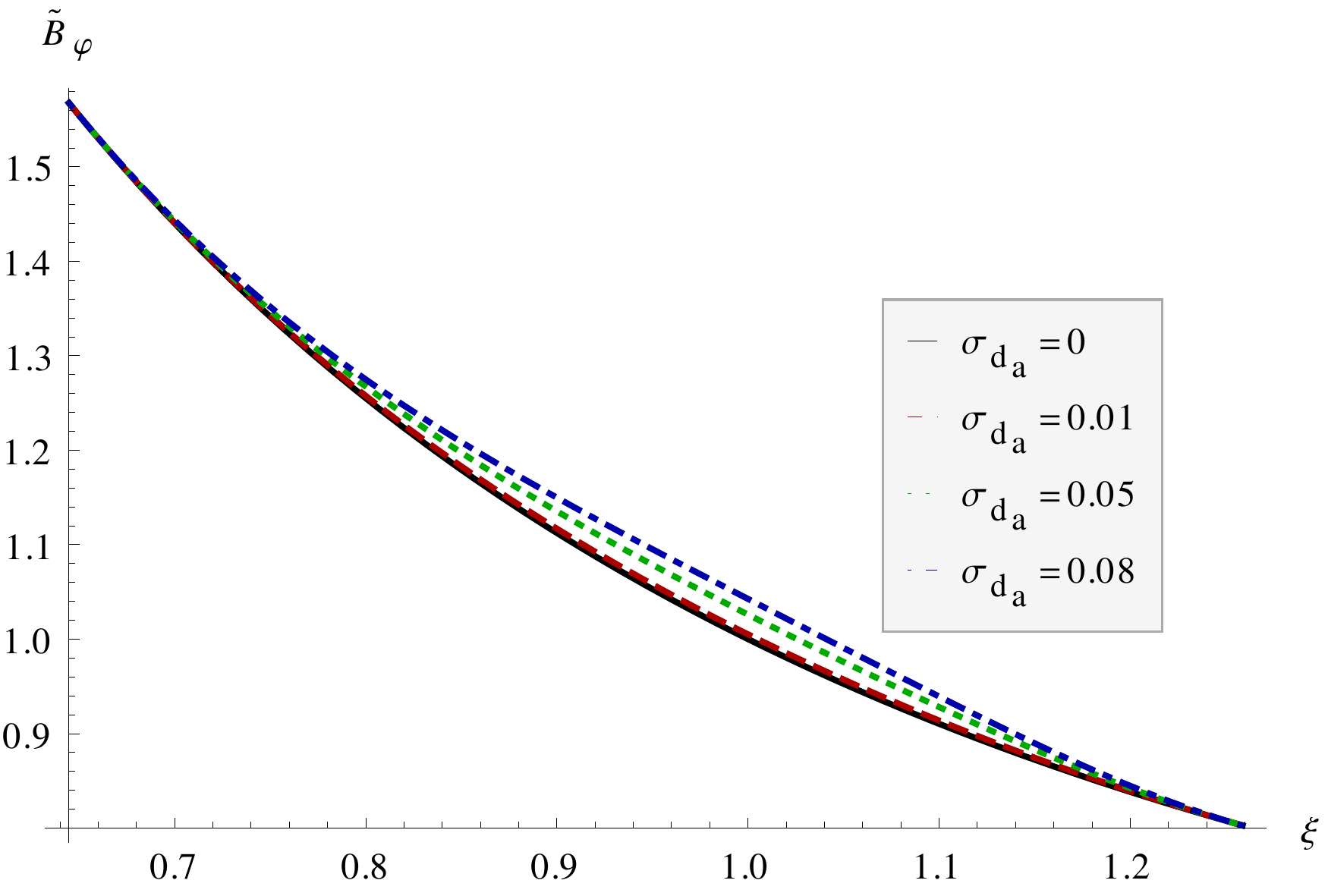}
     \caption{{\small \emph{Variation of $\widetilde{B}_\phi$ with $\sigma_{d_a}$ in ITER-like diamagnetic configuration, for field aligned flows ($\lambda=0$), $M_{p_a}^2=10^{-4}$, $n=2$, on the midplane $\zeta=0$. }}}    
\end{figure}
Additionally, plasma flow through $M_{p_a}^2$ also acts paramagnetically, but its effects are weaker than that of pressure anisotropy, as shown in Fig. (3.6) for the ITER configuration, since $\sigma_{d_a}$ is two orders of magnitude higher than $M_{p_a}^2$. This result could be expected, because from Eq. (\ref{Iu}) function $\widetilde{I}$ is written as
 \beq \label{Iu nor2} \widetilde{I}=\frac{\widetilde{X}}{1-\sigma _d-M_p^2}-\xi ^2\left(\frac{d\widetilde{F}}{d\widetilde{u}}\right)\left(\frac{d\widetilde{\Phi}}{d\widetilde{u}}\right)\eeq
 The non-parallel flow term has a negligible effect. Thus, pressure anisotropy and plasma flow have a cumulative paramagnetic effect on equilibrium.
\begin{figure}
  \centering
    \includegraphics[width=3.5in]{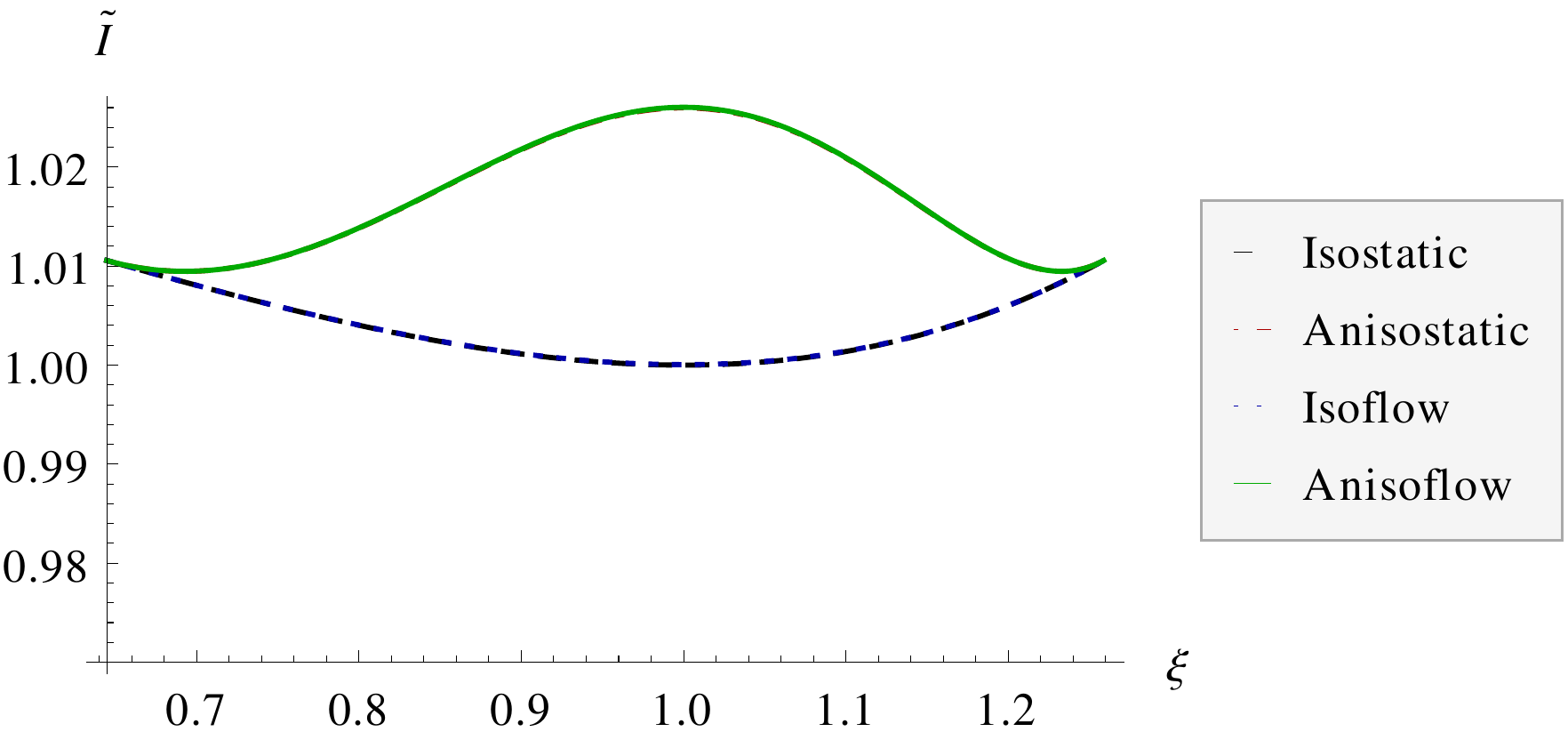}
     \caption{{\small \emph{Pressure anisotropy vs flow impact on $\widetilde{I}$ function for ITER-like diamagnetic equilibria. The notations Iso and Aniso refer to whether plasma pressure is isotropic $(\sigma_{d_a}=0)$/anisotropic $(\sigma_{d_a}\neq 0)$, while the notations static and flow refer to the absence $(M_{p_a}^2=0)$ or presence $(M_{p_a}^2\neq 0)$ of the flow.} }}
\end{figure}
The comparison between the impacts of pressure anisotropy and plasma flow on function $\widetilde{I}$ is more clear for the NSTX tokamak, see Fig. (3.7). 
\begin{figure}
  \centering
    \includegraphics[width=3.5in]{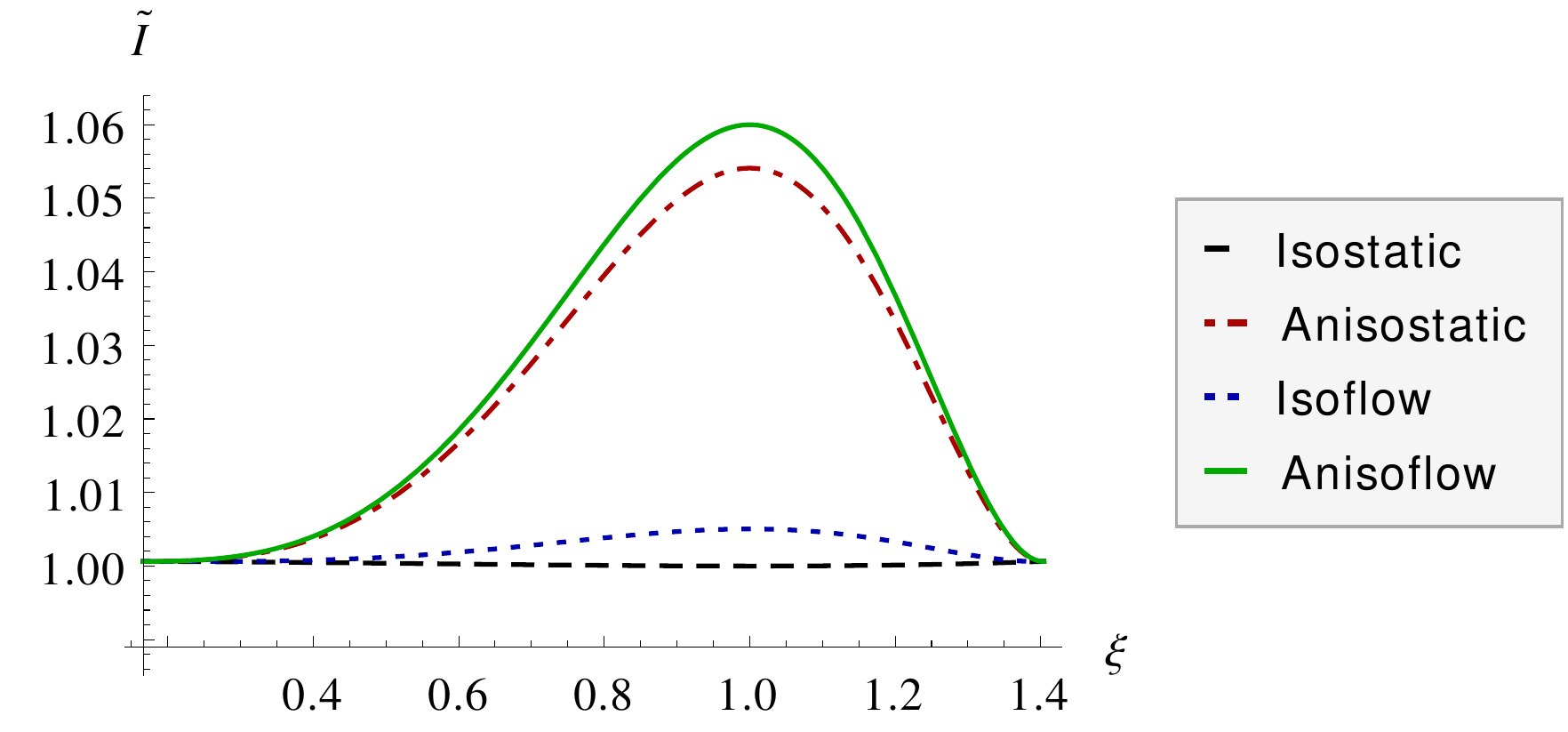}
     \caption{{\small \emph{The additive paramagnetic action of anisotropy and flow for NSTX-like diamagnetic equilibria, on the midplane $\zeta=0$. Note that anisotropy (red-dashed-dotted curve) has a stronger impact than the flow (Blue Dotted curve) on equilibrium. The maximum paramagnetic action is found when both anisotropy and flow are present (green-straight curve). } }}
\end{figure}
 On the other side, pressure anisotropy may also act diamagnetically through the shaping parameter $n$ when $\sigma_{d_a}$ is fixed, see Fig. (3.8).
\begin{figure}
  \centering
    \includegraphics[width=3.5in]{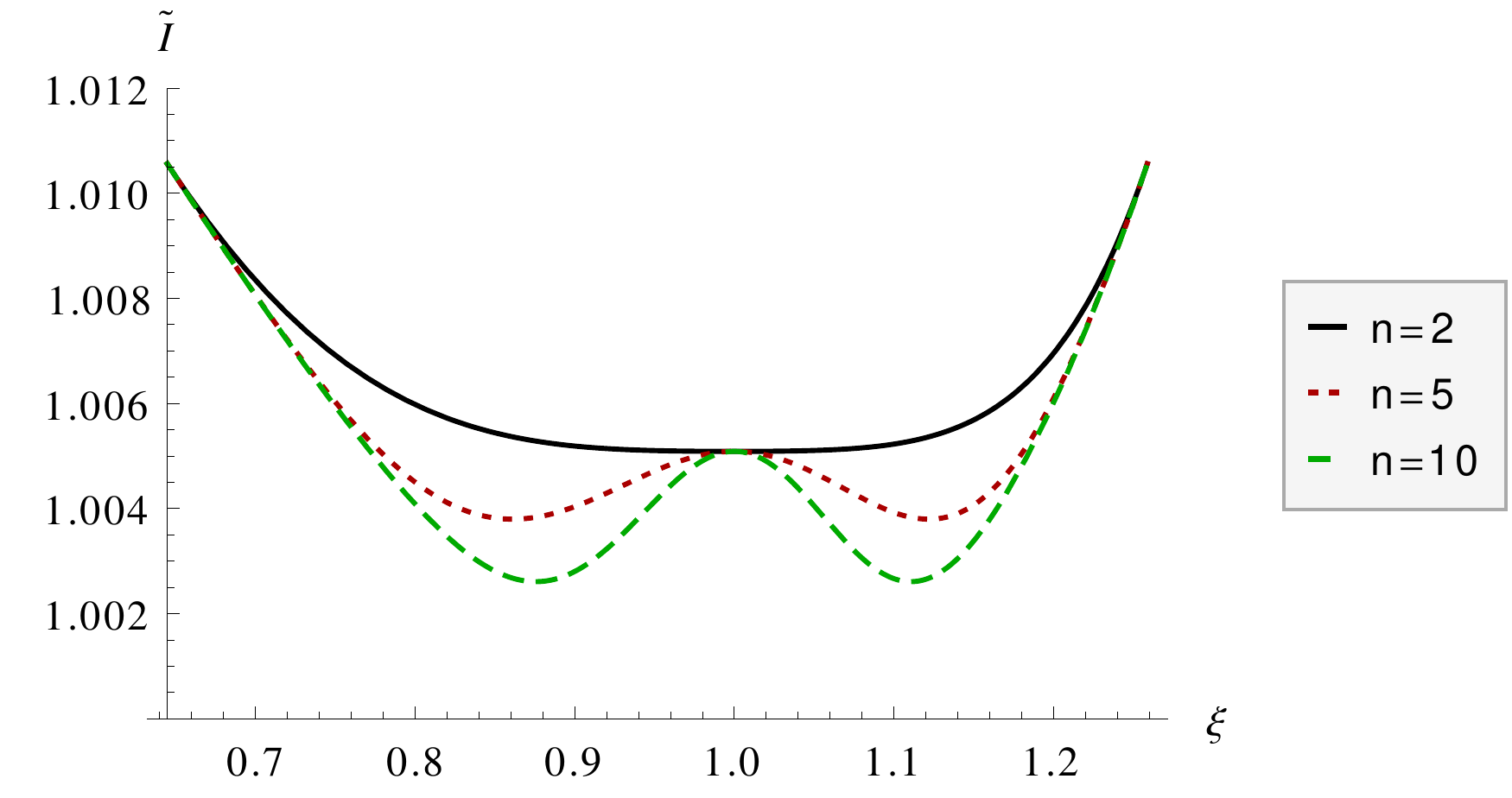}
     \caption{{\small \emph{Raising the free parameter $n$ of the anisotropy function, decreases the toroidal magnetic field in the off-axis region, leading to a diamagnetic action. }}}    
\end{figure} 
Regarding the poloidal magnetic field, 
\begin{center}
	$\widetilde{B}_\xi =\frac{1}{\xi}(1-\sigma_d-M_p^2)^{-1/2}\frac{\partial \widetilde{u}}{\partial \zeta}$
\end{center}
\beq \label{Bpol} \widetilde{B}_\zeta =-\frac{1}{\xi}(1-\sigma_d-M_p^2)^{-1/2}\frac{\partial \widetilde{u}}{\partial \xi}\eeq
the pressure anisotropy through both parameters $\sigma_{d_a}$ and $n$ has a noticeable effect neither on $\widetilde{B}_\xi$ nor on $\widetilde{B}_\zeta$.
Also, it may be noted that the $\xi$-component of $\widetilde{\vec{B}}$ is zero on the midplane $\zeta=0$ (Eq. (\ref{Bpol})).

\subsubsection{Current Density}

\hspace{2em}From Eq. (\ref{Ju nor}) we see that the toroidal current density is 
\beq \label{Jfi} \widetilde{J}_\phi =\frac{1}{\xi}\left[(1-\sigma _d -M_p ^2)^{-1/2}\widetilde{\Delta ^{*}}\widetilde{u}-\frac{1}{2}(1-\sigma _d -M_p ^2)^{-3/2}\frac{d}{d\widetilde{u}}(1-\sigma _d -M_p ^2)|\widetilde{\vec{\nabla}}\widetilde{u}|^2\right]\eeq
and its poloidal components are
\begin{center}
$\widetilde{J}_\xi =-\frac{1}{\xi}\frac{\partial \widetilde{I}}{\partial \zeta}$
\end{center}
\beq \label{Jpol} \widetilde{J}_\zeta =\frac{1}{\xi}\frac{\partial \widetilde{I}}{\partial \xi}\eeq
In the static and isotropic case, using the linearised GGS equation (\ref{linearsol}), for the values of $\epsilon$, $\delta$, $\widetilde{u}_b$, and $\widetilde{\overline{p}}_{s_a}$ given in Table (3.1), Eq. (\ref{Jfi}) takes the following form on the midplane
\beq \label{Jfi2} \widetilde{J}_\phi =1.548\xi -\frac{0.333}{\xi}\eeq
so, we expect the toroidal current density to monotonically increase from $\xi_{in}$ to $\xi_{out}$. The flow has no important effect on $\widetilde{J}_\phi$, both on ITER and NSTX configurations. On the other hand, when anisotropy is present there is noticeable difference from its isotropic profile. More precisely, there are three regions where it displays different behaviour: for $\xi_{in}<\xi<\xi_1$ and $\xi_2<\xi<\xi_{out}$ it decreases, while for $\xi_1<\xi<\xi_2$ it increases, compared with the isotropic case. The variation of $\widetilde{J}_\phi$ is shown in Fig. (3.9).
\begin{figure}
  \centering
    \includegraphics[width=3.5in]{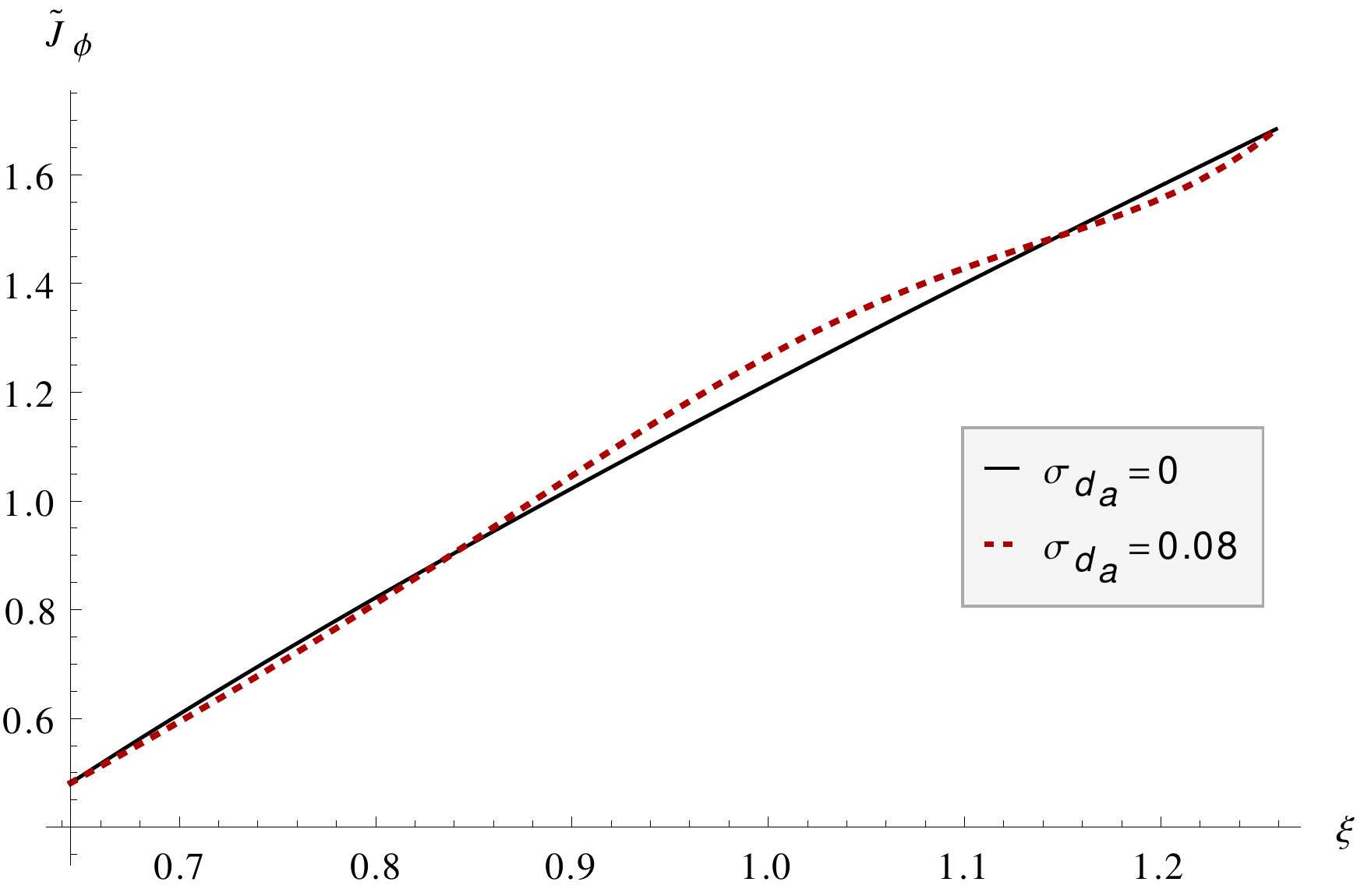}
     \caption{{\small \emph{Variation of $\widetilde{J}_\phi$ with $\sigma_{d_a}$ in ITER-like diamagnetic configuration, for field aligned flows ($\lambda=0$), $M_{p_a}^2=10^{-4}$, $n=2$, on the midplane $\zeta=0$. }   }} 
\end{figure} 
The intersection points for ITER equilibrium with parallel flow are $\xi_1=0.838$ and $\xi_2=1.148$, while for non-parallel flows they move closer to the magnetic axis as $\xi_1=0.850$ and $\xi_2=1.133$. In contrast to $\sigma_{d_a}$, the change of $n$ doesn't affect $\widetilde{J}_\phi$. At last, we have to mention that due to the Solovev-like ansatz, the current density is non zero on the plasma boundary, although plasma density vanishes thereon.  
\par
The isotropic $\widetilde{J}_{\zeta}$ changes from negative to positive values, while it is zero on the plane of the magnetic axis $\xi=1$, in consistence with diamagnetism. The anisotropic $\widetilde{J}_{\zeta}$ exhibits two extrema, both in region $\xi<\xi_a$ and $\xi>\xi_a$, presented in Fig. (3.10), with $|\widetilde{J}_{\zeta}|$ increasing with $\sigma_{d_a}$, so that pressure anisotropy acts paramagnetically.
\begin{figure}
  \centering
    \includegraphics[width=3.5in]{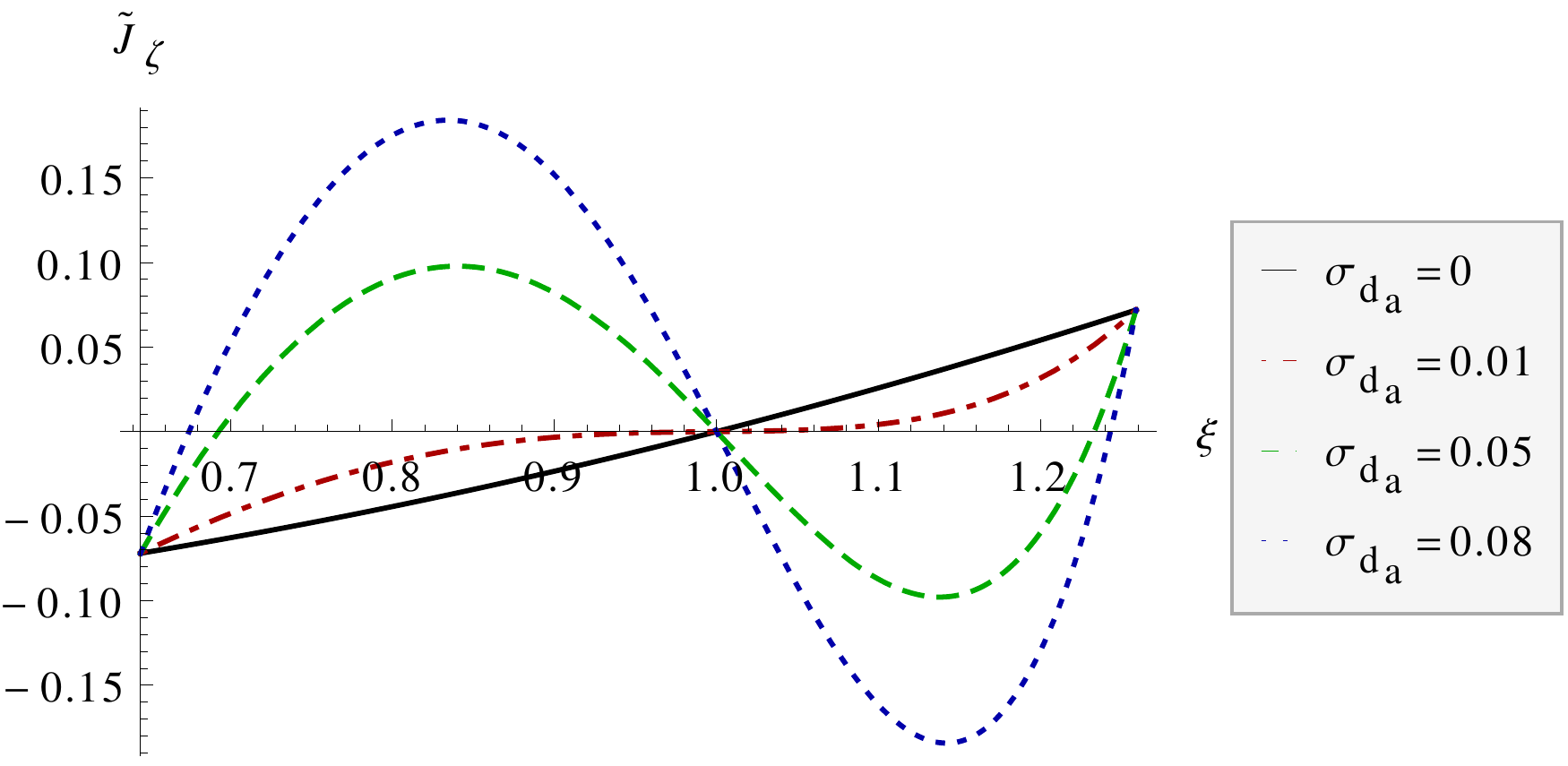}
     \caption{{\small \emph{Variation of $\widetilde{J}_\zeta$ with $\sigma_{d_a}$ in ITER-like diamagnetic configuration, for field aligned flows ($\lambda=0$), $M_{p_a}^2=10^{-4}$, $n=2$, on the midplane $\zeta=0$. }   }} 
\end{figure} 
For fixed $\sigma_{d_a}$ these two extrema get closer to the magnetic axis as $n$ takes higher values, as shown in Fig. (3.11).
\begin{figure}
  \centering
    \includegraphics[width=3.5in]{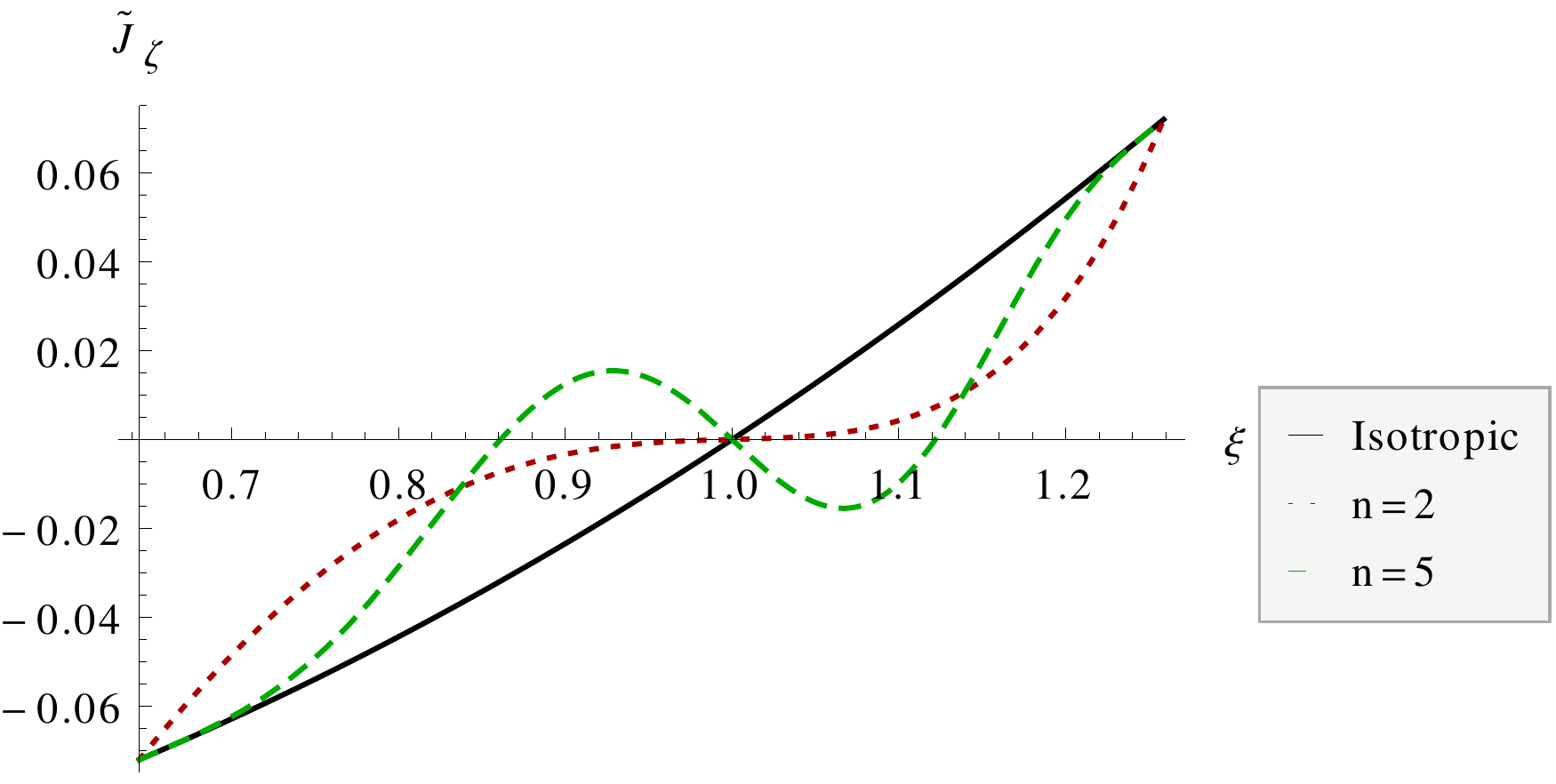}
     \caption{{\small \emph{Variation of $\widetilde{J}_\zeta$ with $n$ in ITER-like diamagnetic configuration, for field aligned flows ($\lambda=0$),  $M_{p_a}^2=10^{-4}$, $\sigma_{d_a}=0.01$, on the midplane $\zeta=0$. } }}   
\end{figure} 
The behavior of $\widetilde{J}_\zeta$ through $\sigma_{d_a}$ and $n$ is the same for the NSTX, with the only difference that the isotropic $\widetilde{J}_\zeta$ is positive in the region $\xi<\xi_a$ and negative in the region $\xi>\xi_a$.

\subsubsection{Velocity}

\hspace{2em}Using Eq. (\ref{vu nor}) we see that the toroidal component of the velocity can be written as
\beq \label{vfi} \widetilde{v}_\phi =\frac{\widetilde{I}}{\xi}\frac{M_p}{\sqrt{\widetilde{\rho}}}-\xi (1-\sigma_d -M_p^2)^{1/2}\left(\frac{d\widetilde{\Phi}}{d\widetilde{u}}\right)\eeq
and the poloidal ones as
\begin{center}
$\widetilde{v}_\xi =\frac{1}{\xi}\frac{M_p}{\sqrt{\widetilde{\rho}}}(1-\sigma_d -M_p^2)^{-1/2}\frac{\partial \widetilde{u}}{\partial \zeta}$
\end{center}
\beq \label{vpol} \widetilde{v}_\zeta =-\frac{1}{\xi}\frac{M_p}{\sqrt{\widetilde{\rho}}}(1-\sigma_d -M_p^2)^{-1/2}\frac{\partial \widetilde{u}}{\partial \xi}\eeq
As we observe, $\widetilde{v}_\phi$ depends on the Mach function, the anisotropy function, the toroidal magnetic field, and the electric field through the second term. On the basis of the chosen ansatz (3.6), Eq. (\ref{vfi}) is put in the following form
\beq \label{vfi2} \widetilde{v}_\phi =\frac{\widetilde{I}}{\xi}\frac{M_p}{\sqrt{\widetilde{\rho}}}-\xi \sqrt{1-\frac{M_p^2}{1-\sigma _d}}\left(\frac{2\lambda \widetilde{\overline{p}}_{s_a}}{\widetilde{\rho}(1+\delta ^2)}\left(1-\frac{\widetilde{u}}{\widetilde{u}_b}\right)\right)^{1/2}\eeq
Thus, for parallel flows the second term in (3.49) vanishes and $\widetilde{v}_\phi$ behaves nearly like $\widetilde{I}$ as far as its dependence on $M_{p_a}^2$, $\sigma_{d_a}$ and $n$ is considered. We can see the increase of the maximum value of the toroidal velocity with $\sigma_{d_a}$, displaced on the left side of the magnetic axis, in Fig. (3.12), for an ITER diamagnetic configuration. 
\par
For the NSTX the impact of anisotropy on $\widetilde{v}_{\phi}$ is qualitatively similar but quantitatively slightly stronger because of the higher values of $\sigma_{d_a}$.
\begin{figure}
  \centering
    \includegraphics[width=3.5in]{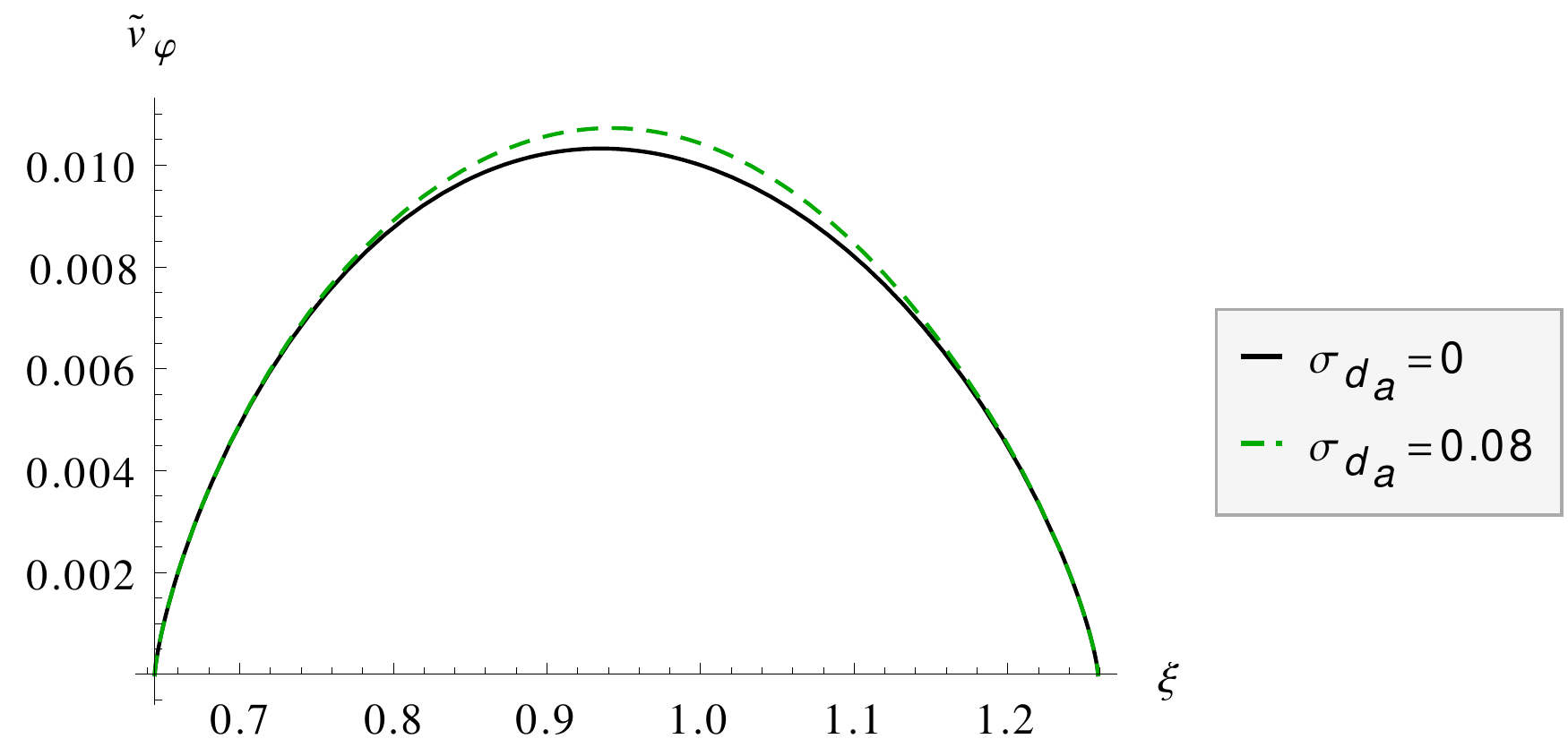}
     \caption{{\small \emph{Variation of $\widetilde{v}_\phi$ with $\sigma_{d_a}$ in ITER-like diamagnetic configuration, for field aligned flows ($\lambda=0$),  $M_{p_a}^2=10^{-4}$, $n=2$, on the midplane $\zeta=0$. } }}   
\end{figure} 
For flows non-parallel to the magnetic field $\widetilde{v}_\phi$ changes sign because of the negative second term. In this situation, as $\sigma_{d_a}$ takes larger values, the second term becomes less negative and the toroidal velocity becomes more positive. However, quantitatively this increase is negligible.
\par
Similar to the case of $\widetilde{B}_p$, the impact of anisotropy on $\widetilde{v}_p$ is negligible. As we observe in Fig. (3.13), the poloidal velocity is one order of magnitude smaller than the toroidal one in accordance with experimental results for tokamaks \cite{Brau,Pantis}. It may be noted here that poloidal rotation damping times are much shorter than the toroidal damping times. 
\begin{figure}
  \centering
    \includegraphics[width=3.5in]{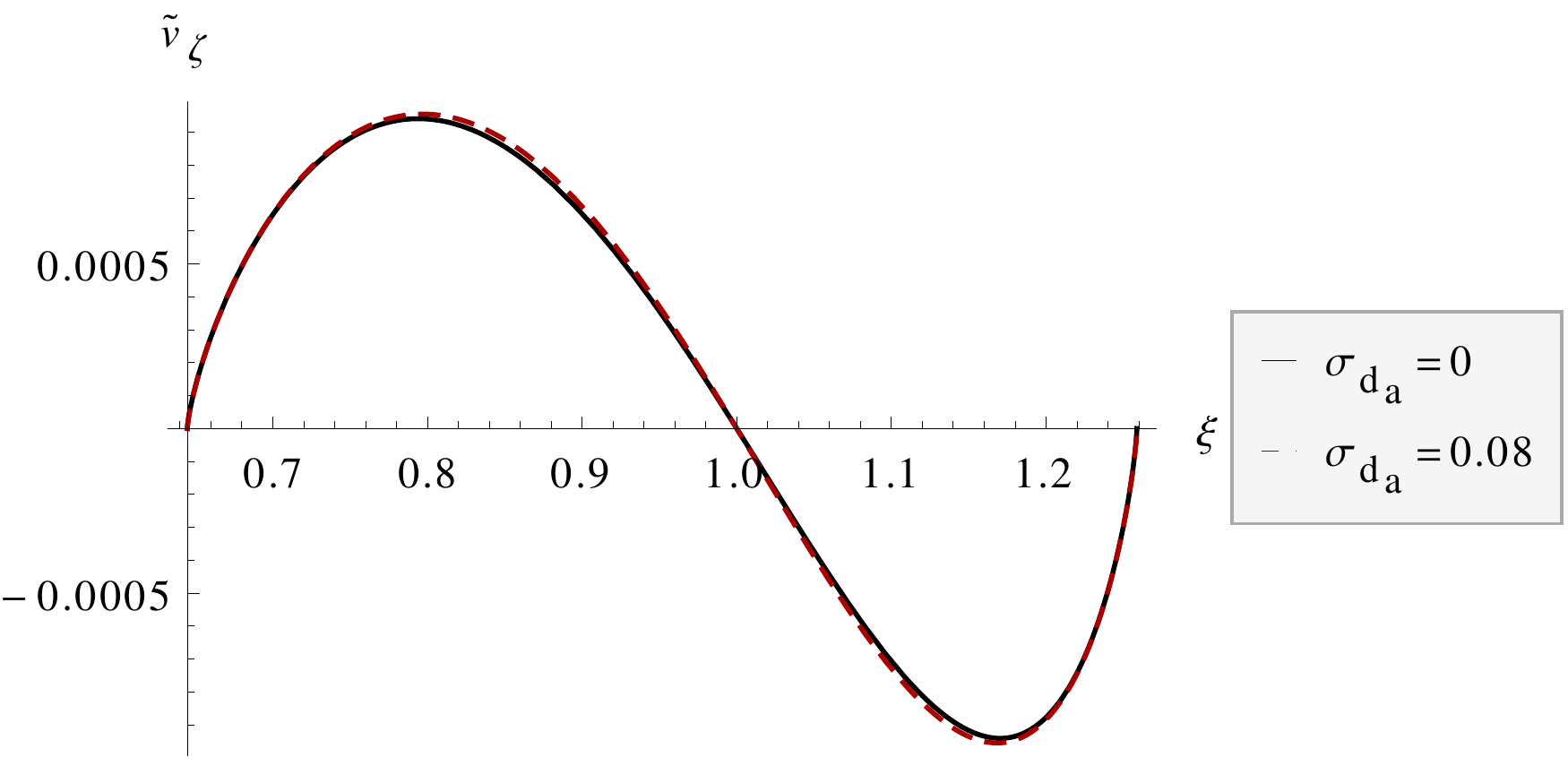}
     \caption{{\small \emph{Variation of $\widetilde{v}_\zeta$ with $\sigma_{d_a}$ in ITER-like diamagnetic configuration, for field aligned flows ($\lambda=0$),  $M_{p_a}^2=10^{-4}$, $n=2$, on the midplane $\zeta=0$. }  }}  
\end{figure}

\subsubsection{Pressures}

\hspace{2em}Considering the Bernoulli equation (\ref{Bernoulliu nor}), the effective pressure 
\beq\label{p1} \widetilde{\overline{p}}=\widetilde{\overline{p}_s} (\widetilde{u})-\widetilde{\rho} \frac{\widetilde{v}^2}{2}+\widetilde{\rho}(1-\sigma _d) \xi ^2 \left(\frac{d\widetilde{\Phi}}{d\widetilde{u}}\right)^2\eeq
consists of three terms; the static one, and two flow terms, the last of which is associated with non-parallel flows.
For vanishing flow, the overall $\widetilde{\overline{p}}$ is identical to $\widetilde{\overline{p}}_s$, which is peaked on the magnetic axis and does not depend on pressure anisotropy, due to the Solovev ansatz, see Fig. (3.14). The static effective pressure $\widetilde{\overline{p}}_s$ depends only on the flow parameter $\lambda$ through the function $\widetilde{u}$, and thus, for non-parallel flows it is decreased from its static or parallel-flow value $(\lambda=0)$.
\begin{figure}
  \centering
    \includegraphics[width=3.5in]{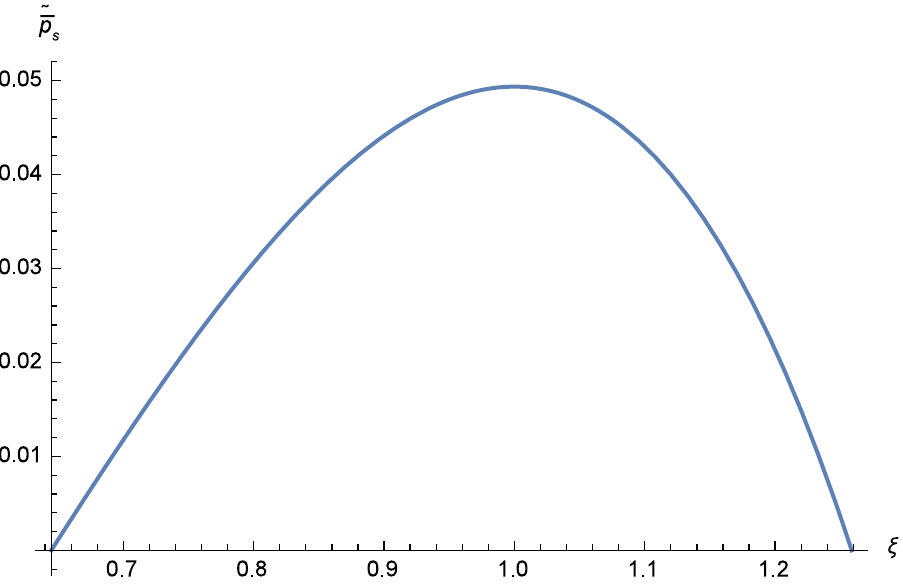}
     \caption{{\small \emph{The static effective pressure on the midplane $\zeta=0$ for the ITER-like diamagnetic configuration.}}}    
\end{figure}
For field-aligned flows, $\widetilde{\overline{p}}$ is a little lower than its static value because of the second negative term in Eq. (\ref{p1}). Both $M_{p_a}^2$ and $\sigma_{d_a}$ have a decreasing influence on $\widetilde{\overline{p}}$, since they additively raise $\widetilde{v}$, though weakly. 
\par
For non-parallel flows there is a positive contribution to the overall pressure because of the third term in (\ref{p1}), so that in this situation  $\widetilde{\overline{p}}$ becomes higher than in the parallel-flow case.
This term does not depend on pressure anisotropy since it has the following form
\beq \label{p2} \widetilde{\rho}(1-\sigma _d) \xi ^2 \left(\frac{d\widetilde{\Phi}}{d\widetilde{u}}\right)^2=\xi ^2 \frac{2\lambda \widetilde{\overline{p}}_{s_a}}{(1+\delta ^2)}\left(1-\frac{\widetilde{u}}{\widetilde{u}_b}\right)\eeq
Thus,  $\widetilde{\overline{p}}$ depends on pressure anisotropy only through the magnitude of velocity. One can see that plasma flow has a stronger influence than anisotropy for the NSTX diamagnetic configuration (Fig. 3.15).
\begin{figure}
  \centering
    \includegraphics[width=3.5in]{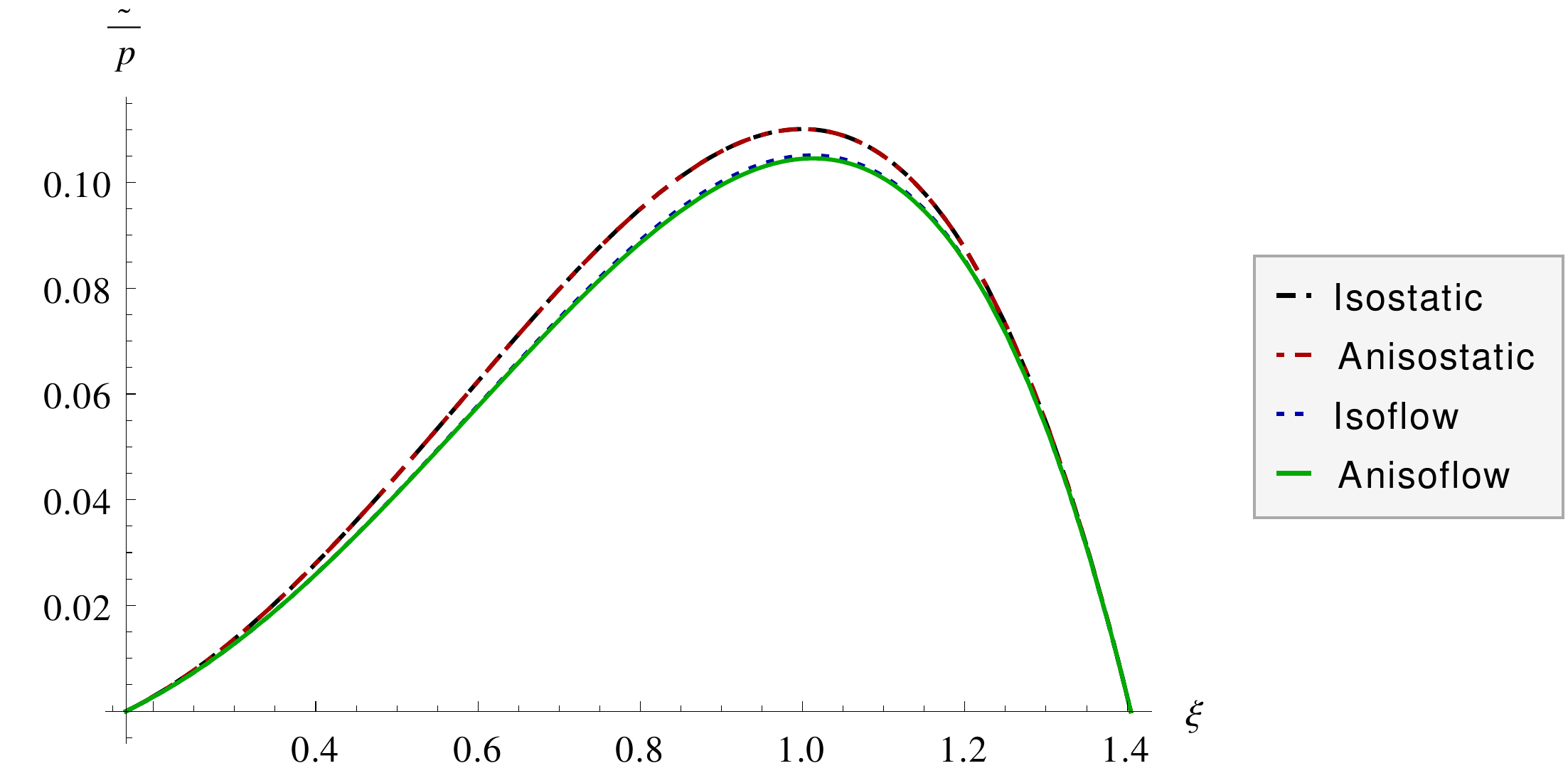}
     \caption{{\small \emph{The impact of pressure anisotropy compared to that of the flow on $\widetilde{\overline{p}}$, on the midplane $\zeta=0$, for the NSTX-like diamagnetic equilibria with parallel flows ($\lambda=0$). When the flow is present the overall effective pressure decreases from its static value, while the presence of pressure anisotropy does not have an important effect on it. The maximum attainable values for the parameters $M_{p_a}^2$ and $\sigma_{d_a}$ were found by requiring non negative pressure.}  }}
\end{figure}
Pressure anisotropy has an appreciable impact on the various pressures, with $\widetilde{\overline{p}}_{\parallel}$ increasing, while $\widetilde{\overline{p}}_{\bot}$ and $<\widetilde{p}>$ decreasing  with $\sigma_d$ as expected by Eqs. (1.46)-(1.48) [see Figs. 3.16 - 3.18]. One observes that as $\sigma_{d_a}$ enhances, $\widetilde{p}_{\bot}$ has a peculiar behaviour -presenting three extrema- because of the stronger anisotropy. For constant $\sigma_{d_a}=$ (i.e. $\sigma_{d_a}=0.01$) the ratio of the scalar pressures parallel and perpendicular to the magnetic field is approximately equal for the two kinds of tokamak: {\tiny$\left(\frac{\widetilde{p}_{\parallel}}{\widetilde{p}_{\bot}}\right)_{ITER}\approx 1.227 $\normalsize},  {\tiny$\left(\frac{\widetilde{p}_{\parallel}}{\widetilde{p}_{\bot}}\right)_{NSTX}\approx 1.099$\normalsize}.
 In addition, the ratio of the maximum values of the average pressures for these two tokamaks is {\tiny$\frac{<\widetilde{p}>_{NSTX}}{<\widetilde{p}>_{ITER}}\approx 2.17$\normalsize}.
 \par
Furthermore, raising $n$ makes $\widetilde{p}_{\parallel}$ and $<\widetilde{p}>$ to decrease, while it makes $\widetilde{p}_{\bot}$ to increase, though the changes are quantitatively small.
The pressure profiles for NSTX anisotropic equilibria with parallel flows are given in Fig. (3.20).
\begin{figure}
  \centering
    \includegraphics[width=3.5in]{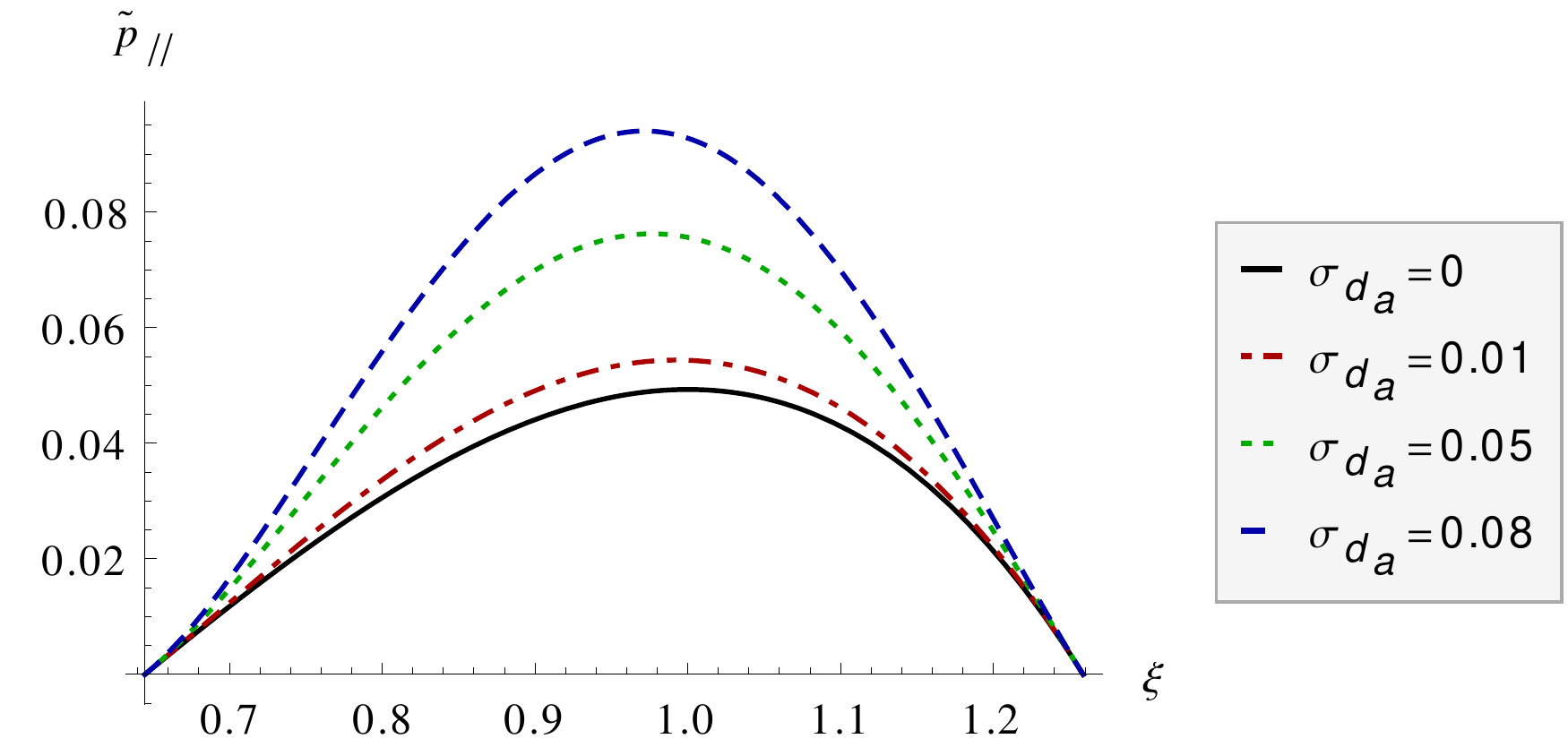}
     \caption{{\small \emph{$\widetilde{p}_{\parallel}$ dependence on $\sigma_{d_a}$, on the mideplane $\zeta=0$, for parallel flows ($\lambda=0$), $M_{p_a}^2=10^{-4}$, $n=2$, for ITER-like diamagnetic equilibria. As $\sigma_{d_a}$ increases it's maximum value gets closer to the innermost point.}  }}
\end{figure}
\begin{figure}
  \centering
    \includegraphics[width=3.5in]{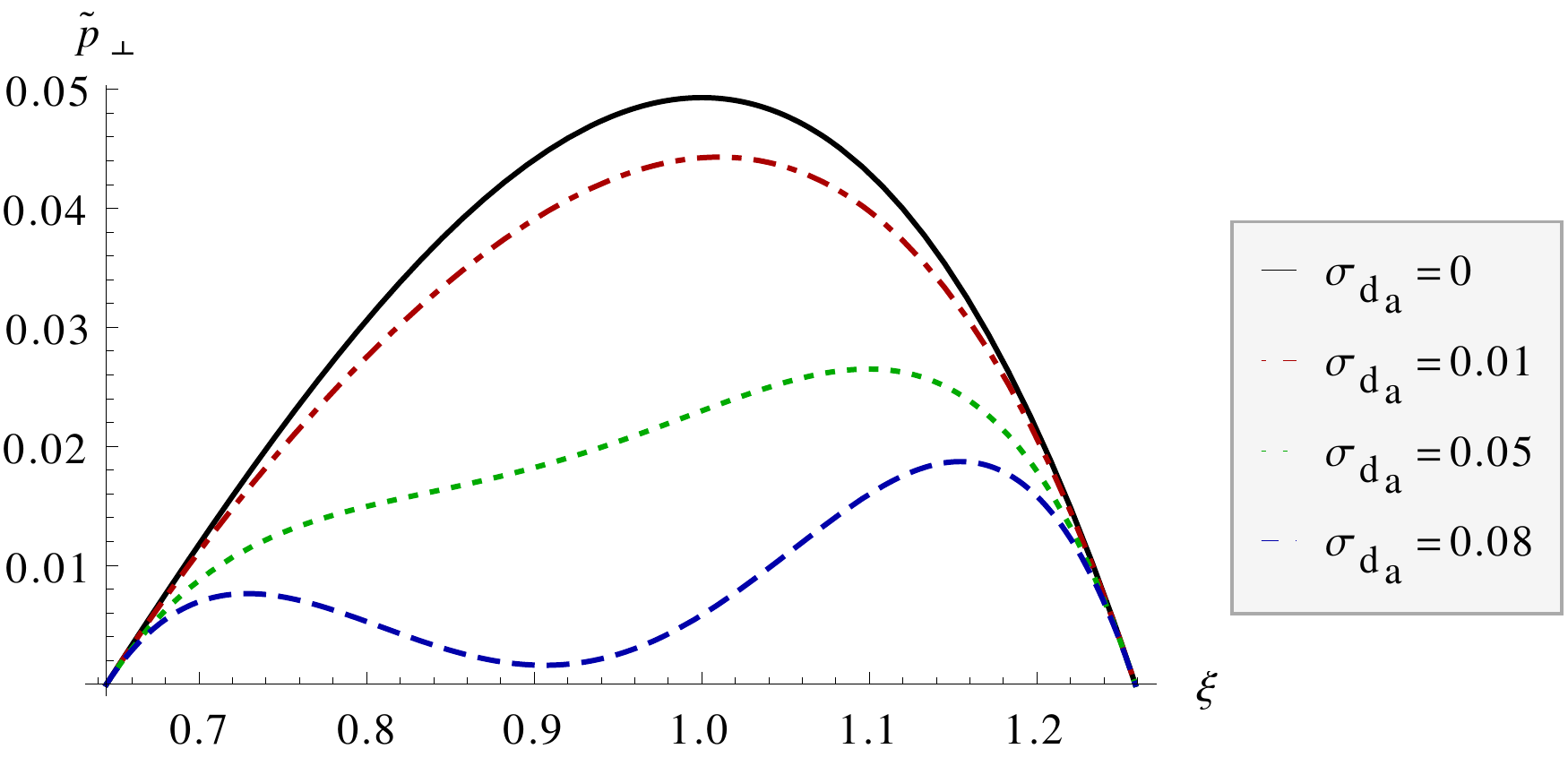}
     \caption{{\small \emph{$\widetilde{p}_{\bot}$ dependence on $\sigma_{d_a}$ anisotropy parameter, on the mideplane $\zeta=0$, for parallel flows ($\lambda=0$), $M_{p_a}^2=10^{-4}$, $n=2$, for ITER-like diamagnetic equilibria.   As $\sigma_{d_a}$ increases it's maximum value gets closer to the outermost point.}}}  
\end{figure}
\begin{figure}
  \centering
    \includegraphics[width=3.5in]{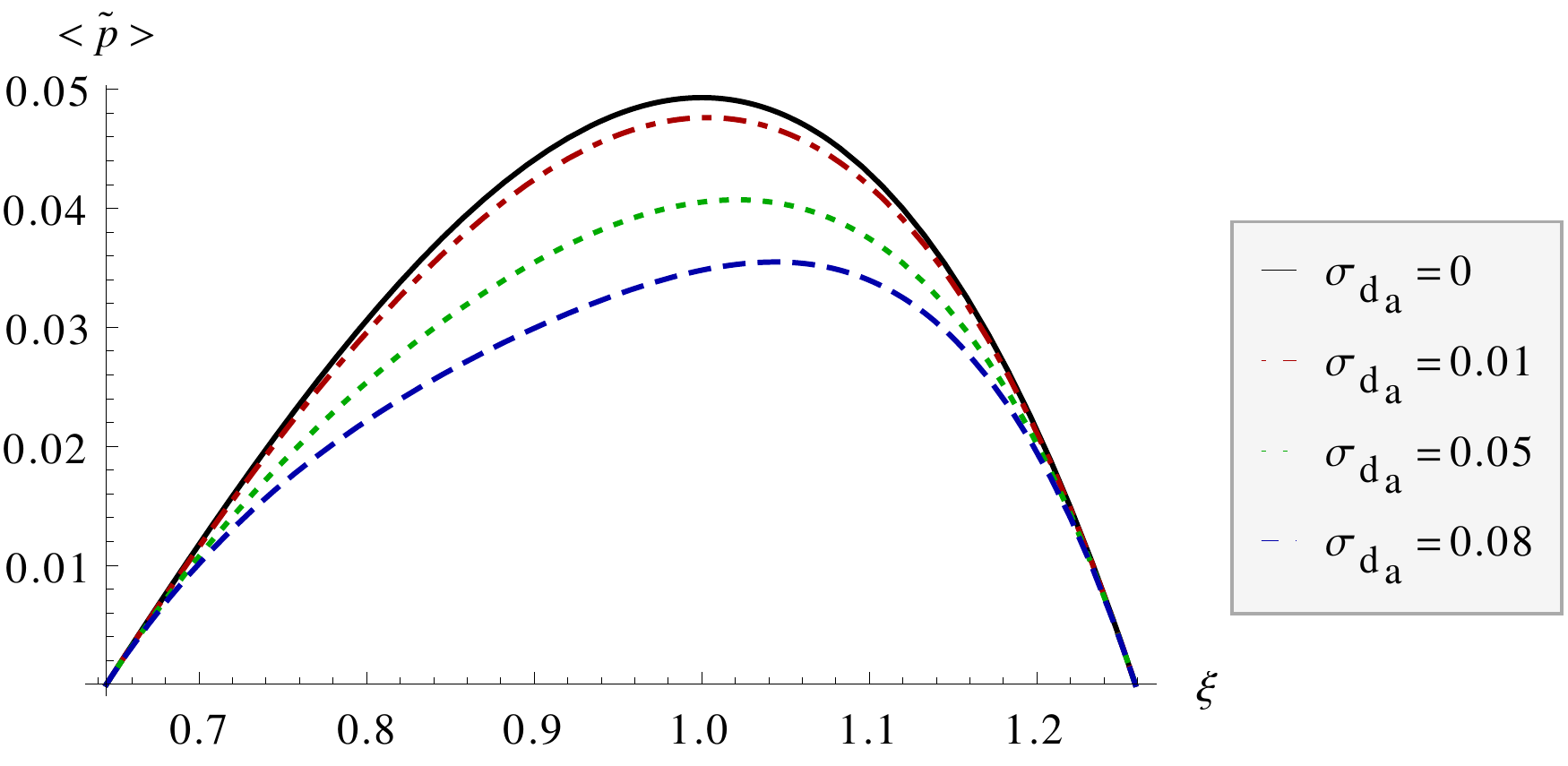}
     \caption{{\small \emph{Impact of the anisotropy through $\sigma_{d_a}$ on $<\widetilde{p}>$ for parallel flows ($\lambda=0$), $M_{p_a}^2=10^{-4}$, $n=2$, for ITER-like diamagnetic equilibria.}}}  
\end{figure}
\begin{figure}
  \centering
    \includegraphics[width=3.5in]{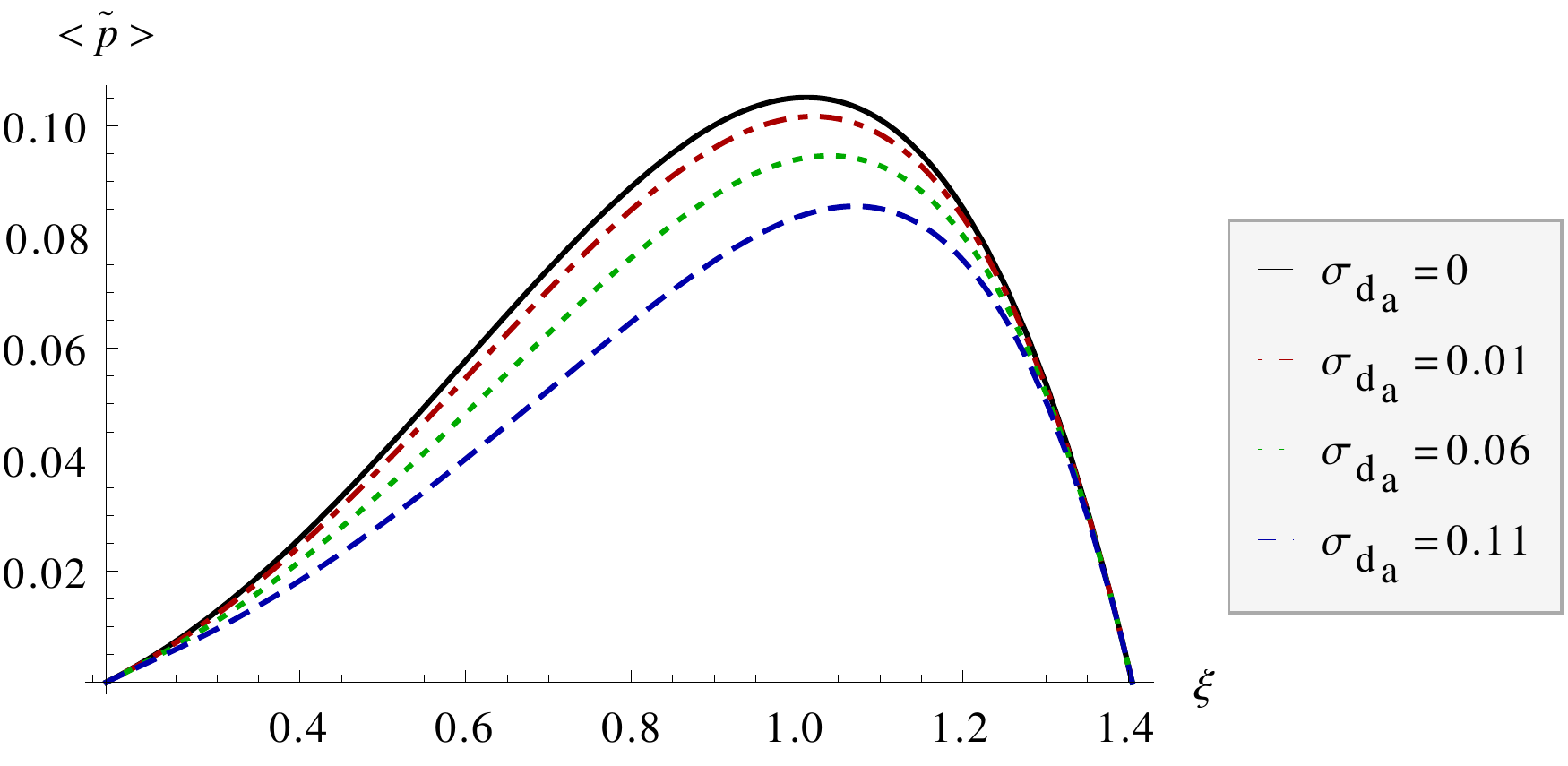}
     \caption{{\small \emph{Impact of the anisotropy through $\sigma_{d_a}$ on $<\widetilde{p}>$ for parallel flows ($\lambda=0$), $M_{p_a}^2=10^{-2}$, $n=2$, for NSTX-like diamagnetic equilibria.} }} 
\end{figure}
\begin{figure}
  \centering
    \includegraphics[width=3.5in]{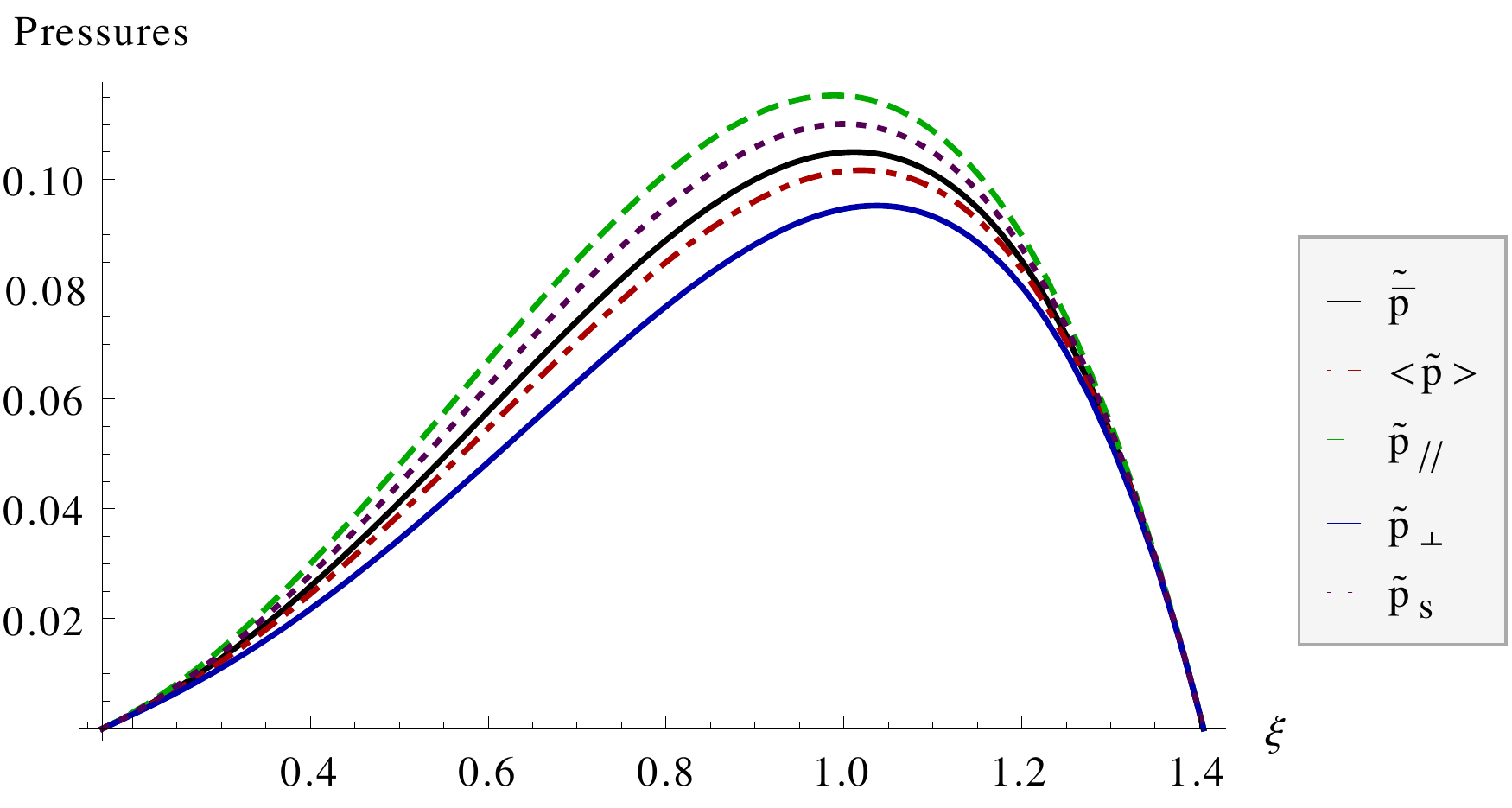}
     \caption{{\small \emph{Normalized pressure profiles on the mideplane $\zeta=0$ for NSTX-like diamagnetic equilibria with parallel flows ($\lambda=0$), $M_{p_a}^2=10^{-2}$, $\sigma_{d_a}=0.02$, $n=2$.}  }}
\end{figure}

\subsubsection{Electric Field}

\hspace{2em}Non parallel flows are associated with an electric field $\vec{E}=-\vec{v}\times\vec{B}$, as it follows from Ohm's law (\ref{Ohm}).
The electric field defined by Eq. (\ref{Eu nor}), on the basis of the ansatz (\ref{sol3}) can be written in the following form
\beq \label{E1} \widetilde{\vec{E}}=-\sqrt{\frac{2\lambda \widetilde{\overline{p}}_{s_a}}{(1+\delta ^2)}\frac{\left(1-\frac{\widetilde{u}}{\widetilde{u}_b}\right)^{1-g}}{1-\sigma _{d_a}\left(1-\frac{\widetilde{u}}{\widetilde{u}_b}\right)^n}}\widetilde{\vec{\nabla}}\widetilde{u}\eeq
where $\widetilde{\vec{\nabla}}\widetilde{u}=\frac{\partial \widetilde{u}}{\partial \xi}\hat{e}_\xi +\frac{\partial \widetilde{u}}{\partial \zeta}\hat{e}_\zeta$. Eq. (\ref{E1}) implies that $\widetilde{\vec{E}}$ increases with $\sigma_{d_a}$, while decreases with $n$. The variation of the component $\widetilde{E}_\xi$ on the midplane $\zeta=0$ shown in Fig. (3.21) is nearly insensitive to anisotropy. The $\zeta$-component of the electric field is zero on this plane. In addition, the electric field gets larger as $\lambda$ takes higher values.
\begin{figure}
  \centering
    \includegraphics[width=3.5in]{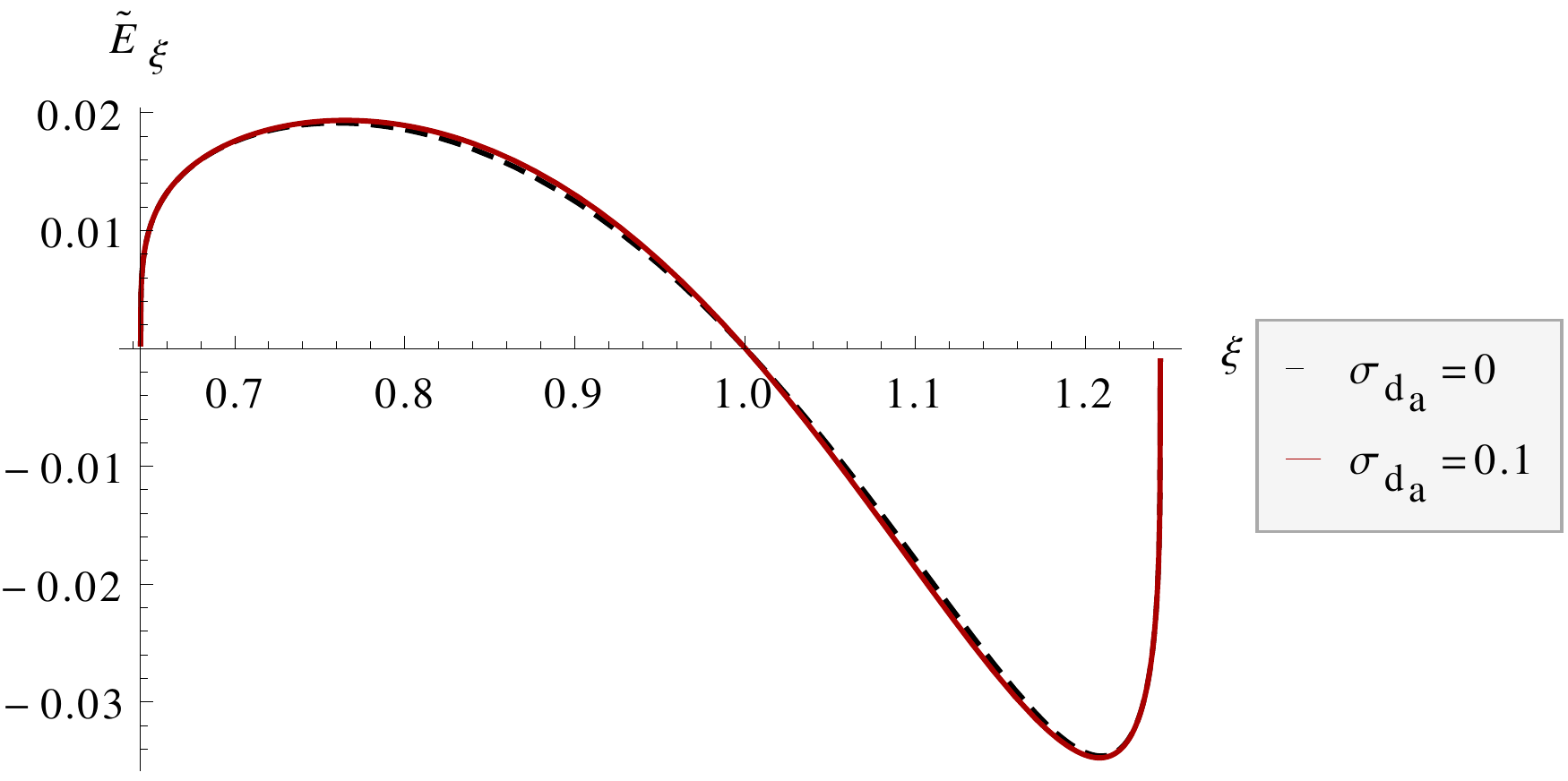}
     \caption{{\small \emph{The $\widetilde{E}_\xi$ profile on the mideplane $\zeta=0$ for ITER-like diamagnetic equilibria with non-parallel flows, $\lambda=0.5$, $M_{p_a}^2=10^{-2}$, $n=2$. The increase of the parameter $\sigma_{d_a}$ does not affect $\widetilde{E}_\xi$.}}}  
\end{figure}

\subsubsection{Toroidal Beta - Safety Factor}

\hspace{2em}The toroidal beta defined by Eq. (\ref{betator}) here is written as
\beq \label{b1} \beta _{t}=\frac{\widetilde{\overline{p}}}{(B_0^2/B_a^2)/2}\eeq
Here we are interested in its local value on the magnetic axis
\beq \label{b2} \beta _{t_a}=\frac{\widetilde{\overline{p}}_a}{(B_0^2/B_a^2)/2}\eeq
which for the ITER diamagnetic parameters from Table (3.1), becomes
\beq \label{b3} \beta _{t_a}=0.089\frac{1-0.55\sigma _{d_a}}{1-\sigma _{d_a}}\eeq
for field aligned flows, $\lambda=0$, and
\beq \label{b4} \beta _{t_a}=1.78\left(0.076+\frac{0.013\left(0.06+0.99\sqrt{\frac{1}{1-\sigma _{d_a}}}(0.01\sigma _{d_a}-1)\right)^2}{\sigma _{d_a}-0.99}\right)\eeq
for non-parallel flows, $\lambda=0.5$.
Thus, the pressure anisotropy through $\sigma_{d_a}$ makes $\beta_{t_a}$ slightly lower for parallel flows and higher for non parallel ones (Fig. 3.22).
\begin{figure}
  \centering
    \includegraphics[width=3.5in]{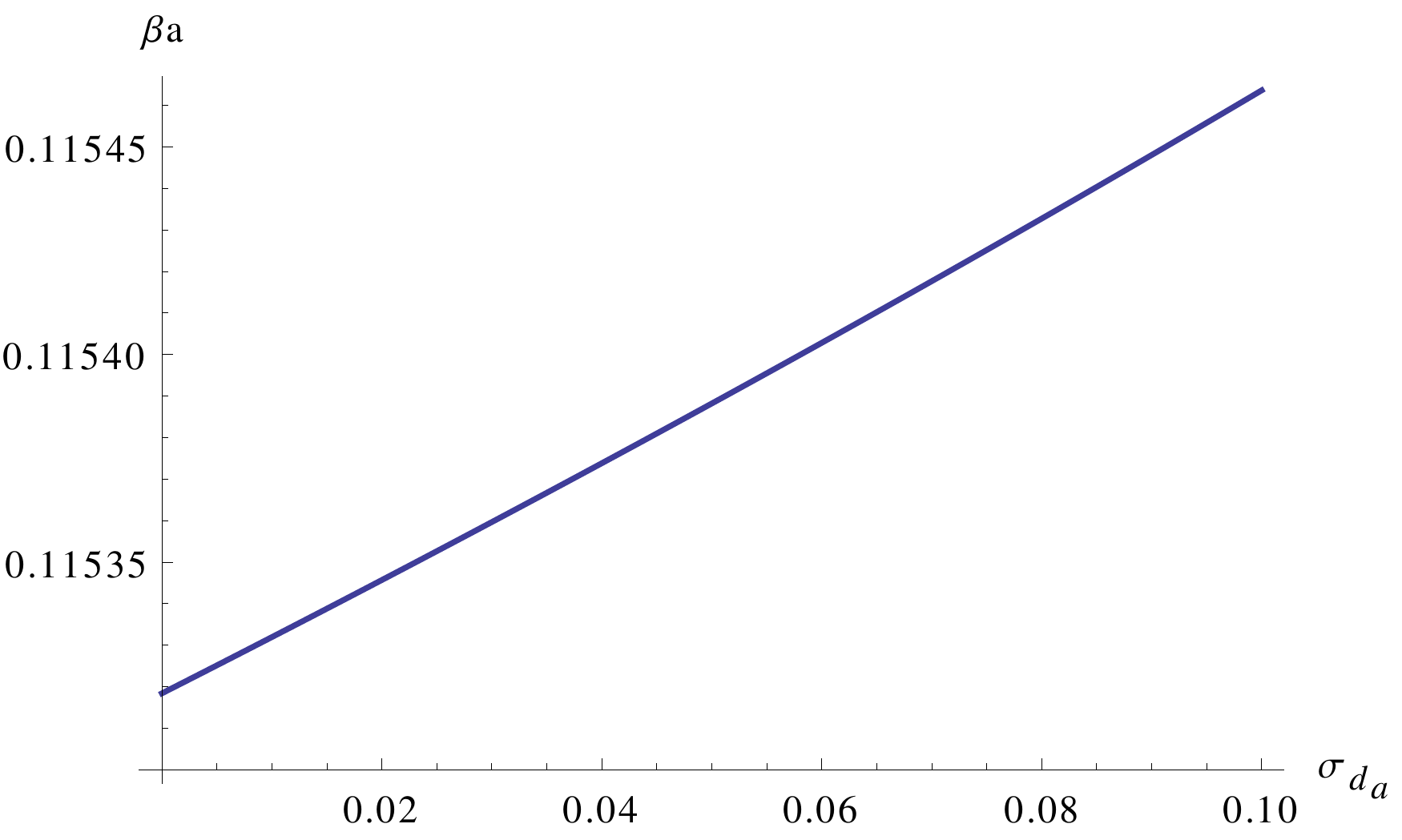}
     \caption{{\small \emph{The local toroidal beta increases a bit with the increase of pressure anisotropy for non-parallel flows, $\lambda=0.5$.}}}  
\end{figure}
One finds that $\beta_{t_a}\approx 8.9\%$ for $\lambda=0$, while $\beta_{t_a}\approx 11.5\%$ for $\lambda=0.5$, which are on the same order of magnitude though a bit larger of the expected ITER ones. For the NSTX these values are $\beta_{t_a}\approx 13\%$ for $\lambda=0$, while $\beta_{t_a}\approx 16\%$ for $\lambda=0.5$, values that are closer to the actual ones for this tokamak.
\par
It is recalled that the safety factor on the magnetic axis defined by Eq. (\ref{qa u nor}) does not depend on pressure anisotropy for both static equilibria and equilibria with parallel flows due to the pertinent functional form of $\widetilde{I}$ (Eq. (\ref{Xu nor})). For NSTX equilibria with field-aligned flows it takes the value $q_a \approx 2.3$, satisfying the Kruskal-Shafranov limit. For non parallel flows there exists an extra flow term in $\widetilde{I}$, and thus, the safety factor on axis depends on the anisotropy parameter $\sigma_{d_a}$ as

\beq \label{qa1} q_a=2.24 - 0.034\sqrt{\frac{1}{1-\sigma _{d_a}}} \eeq
Thus, we observe that $q_a$ decreases with $\sigma_{d_a}$. However, similar to $\beta _{t_a}$ the effect of anisotropy on $q_a$ is very small, as shown in Fig. (3.23). For diamagnetic ITER equilibria we find $q_a\approx1.69$ for field-aligned flows, and $q_a\approx1.61$ for non-parallel flows. In general the $q$-profile for the Solovev-like equilibrium is monotonically increasing from the magnetic axis to the plasma boundary. An example for ITER is given in Fig. (3.24).
\begin{figure}
  \centering
    \includegraphics[width=3.5in]{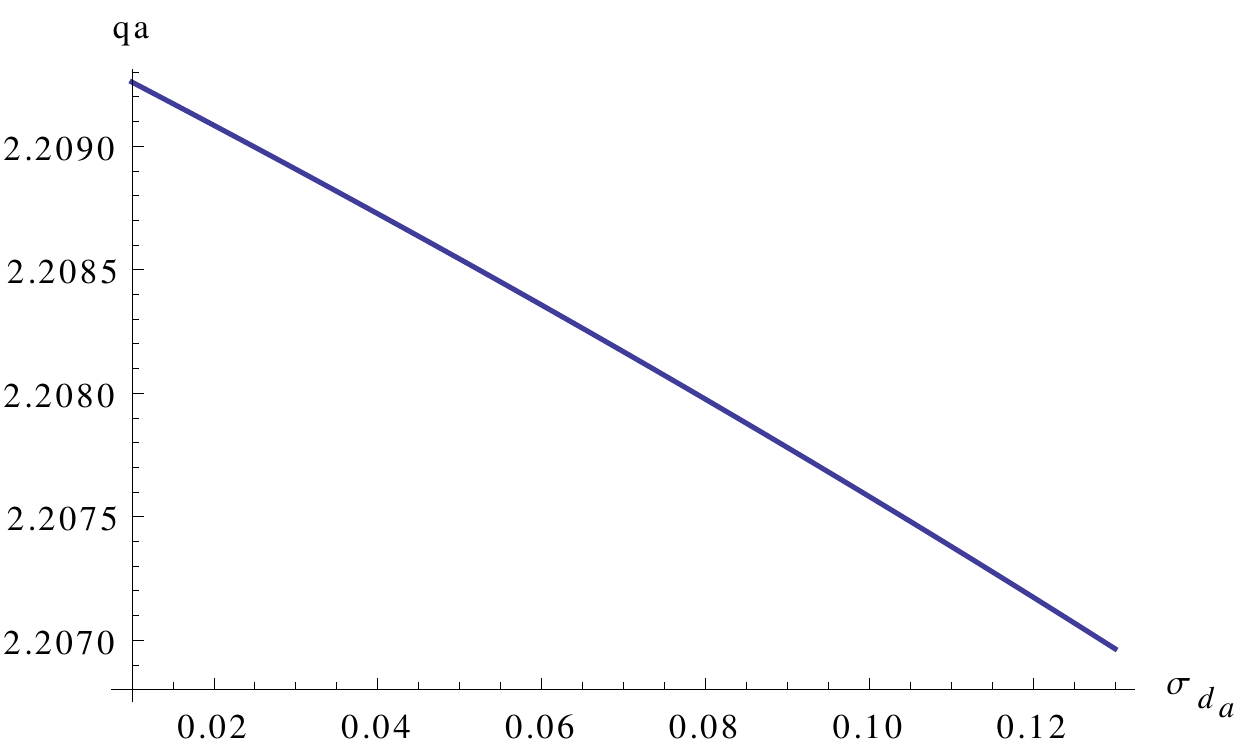}
     \caption{{\small \emph{The increase of pressure anisotropy results in a weak decrease on $q_a$ for NSTX-like diamagnetic equilibria with non-parallel flows, $\lambda =0.5$. The Kruskal-Shafranov limit is well satisfied.} }} 
\end{figure}
\begin{figure}
  \centering
    \includegraphics[width=3.5in]{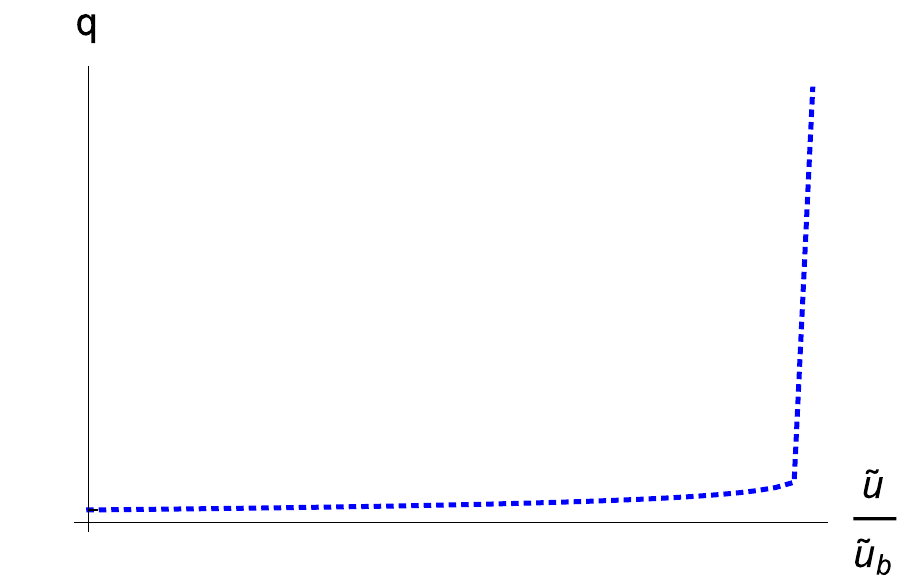}
     \caption{{\small \emph{The safety factor for ITER-like diamagnetic configuration with non-parallel flows, $\lambda=0.5$, is monotonically increasing from the magnetic axis to the plasma boundary.} }}
     \end{figure} 
     
 \section{Paramagnetic Configurations}
 
\hspace{2em}When $\epsilon<0$ the inner point of the separatrix on the midplane $\zeta=0$ is located on the axis of symmetry $\xi=0$. Thus, the flux function on the separatrix has the form
 \beq \label{us par} \widetilde{u}_s=\widetilde{u}(0,0)\Rightarrow \widetilde{u}_s=\frac{\widetilde{\overline{p}}_{s_a}(3\delta ^2+2\lambda)}{24\widetilde{u}_b(1+\delta ^2)}\eeq
 and since the plasma is assumed to extend up to the separatrix, it is
 \beq \label{ub par} \widetilde{u}_b=\left[\frac{\widetilde{\overline{p}}_{s_a}(3\delta ^2+2\lambda)}{24(1+\delta ^2)}\right]^{1/2}\eeq  
 One observes that in contrast with a diamagnetic configuration, $\widetilde{u}_b$ does not depend on $\epsilon$. Along the same lines as for the diamagnetic case, we find that the radial coordinates of the innermost and outermost points of the boundary are
 \beq \label{inner par} \xi _{in}=0\eeq
 and
 \beq \label{outter par} \xi _{out}=\left[\frac{-3\delta ^2+\sqrt{3}\sqrt{3\delta ^4+8\delta ^2\lambda+4\lambda ^2}}{2\lambda}\right]^{1/2}\eeq
In addition, we find analytical relations for the major and minor radius of a paramagnetic configuration:
 \beq \label{major par} \xi _0=\left[\frac{-3\delta ^2+\sqrt{3}\sqrt{3\delta ^4+8\delta ^2\lambda+4\lambda ^2}}{8\lambda}\right]^{1/2}\eeq
 and
 \beq \label{minor par} \widetilde{\alpha}=\frac{1}{2}\left[\frac{-3\delta ^2+\sqrt{3}\sqrt{3\delta ^4+8\delta ^2\lambda+4\lambda ^2}}{2\lambda}\right]^{1/2}\eeq
 Note that $\xi_{out}$, $\xi_0$ and $\widetilde{\alpha}$ do not depend on $\epsilon$.
 \par
 In the static limit, $\lambda\rightarrow 0$, relations (\ref{outter par})-(\ref{minor par}) reduce to
 \beq \label{ps1} \xi _{out(s)}=\sqrt{2}\eeq 
 and
 \beq \label{ps2} \xi _{0(s)}=\widetilde{\alpha}_{(s)}=\frac{1}{\sqrt{2}}\eeq
 Also, the parametric equation $\zeta _b(\xi)$ which gives the $\zeta$-coordinates of the boundary points is in the static limit
 \beq \label{ps3} \zeta _{b(s)}(\xi)=\sqrt{\frac{-\delta ^2\xi ^2(-2+\xi ^2)}{4(-\epsilon +\xi ^2)}}\eeq
 By solving the equation $\frac{\partial \zeta _{b(s)}}{\partial \xi}=0$ we find that the radial coordinate of the upper point for $\lambda=0$ is
 \beq \label{ps4} \xi _{up(s)}=\sqrt{\epsilon+\sqrt{\epsilon ^2-2\epsilon}}\eeq
 Substituting (\ref{ps4}) into (\ref{ps3}) we also find the $\zeta$-coordinate of the upper point as
 \beq \label{ps5} \zeta _{up(s)}=\frac{1}{2}(\epsilon +\sqrt{\epsilon ^2-2\epsilon})\sqrt{\frac{-\delta ^2}{\epsilon}}\eeq
 From relations (\ref{triangularity}) and (\ref{elongation}) we find for the triangularity and elongation the expressions
 \beq \label{elo static par} \kappa _{(s)}=(\epsilon +\sqrt{\epsilon ^2-2\epsilon})\sqrt{\frac{-\delta ^2}{\epsilon}}\eeq
 and
 \beq \label{triang static par} t_{(s)}=1-\sqrt{2}\left(\epsilon +\sqrt{\epsilon ^2-2\epsilon}\right)^{1/2}<1\eeq
 Thus, by solving the system of equations [$\kappa=\kappa_{(s)}$, $R_0=R_{0(s)}$, $t=t_{(s)}$], we find the relations
 \beq \label{epsilon par} \epsilon =\frac{(t-1)^4}{4(t^2-2t-1}\eeq ,
 \beq \label{delta par} \delta =\frac{\kappa \sqrt{2}}{\sqrt{-t^2+2t+1}}\eeq
 and
 \beq \label{Ra par} R_a=\sqrt{2}R_0\eeq
 from which we can determine the values of the parameters $\epsilon$, $\delta$, and $R_a$ in terms of the elongation, triangularity and major radius. Afterwards, we can also determine the values of the characteristic parameters of the boundary presented above. All these parameters are calculated and given in Table (3.3) for the ITER and NSTX paramagnetic configurations.
 \begin{table}[ht]
\centering
 \begin{tabular}{c c c}
\hline\hline
 & ITER &NSTX  \\ [1ex] 
\hline 
$\epsilon$ & -0.032479 & -0.008929  \\
$\delta$ & 1.93039 & 2.3519 \\
$R_a\ (m)$ & 8.76812 & 1.20208  \\
$B_a\ (T)$ & 3.74767 & 0.304056  \\
$\widetilde{\overline{p}}_{s_a}$ & 0.089403 & 0.135821  \\ 
$\widetilde{u}_b$ & 0.0938664 & 0.119909  \\ 
$\xi _0$ & $1/\sqrt{2}$ & $1/\sqrt{2}$ \\
$\alpha$ & 6.2$\neq2$ & $0.85\neq0.67$ \\
$\xi _{in}$ & 0 & 0 \\
$\xi _{out}$ & $\sqrt{2}$ & $\sqrt{2}$ \\
$\xi _{up}$ & 0.473762 & 0.353553 \\
$\zeta _{up}$ & 1.20208 & 1.55563 \\[1ex]
\hline 
\end{tabular}
\caption{{\small \emph{Characteristic values of the shaping parameters for ITER and NSTX spherical tokamaks for $\lambda=0$ paramagnetic equilibria. }}}
\end{table}
The poloidal cross section with the magnetic surfaces of ITER paramagnetic configuration for the case of static or parallel flows $(\lambda=0)$ is illustrated in Figure (3.25).
\begin{figure}
  \centering
    \includegraphics[width=2.5in]{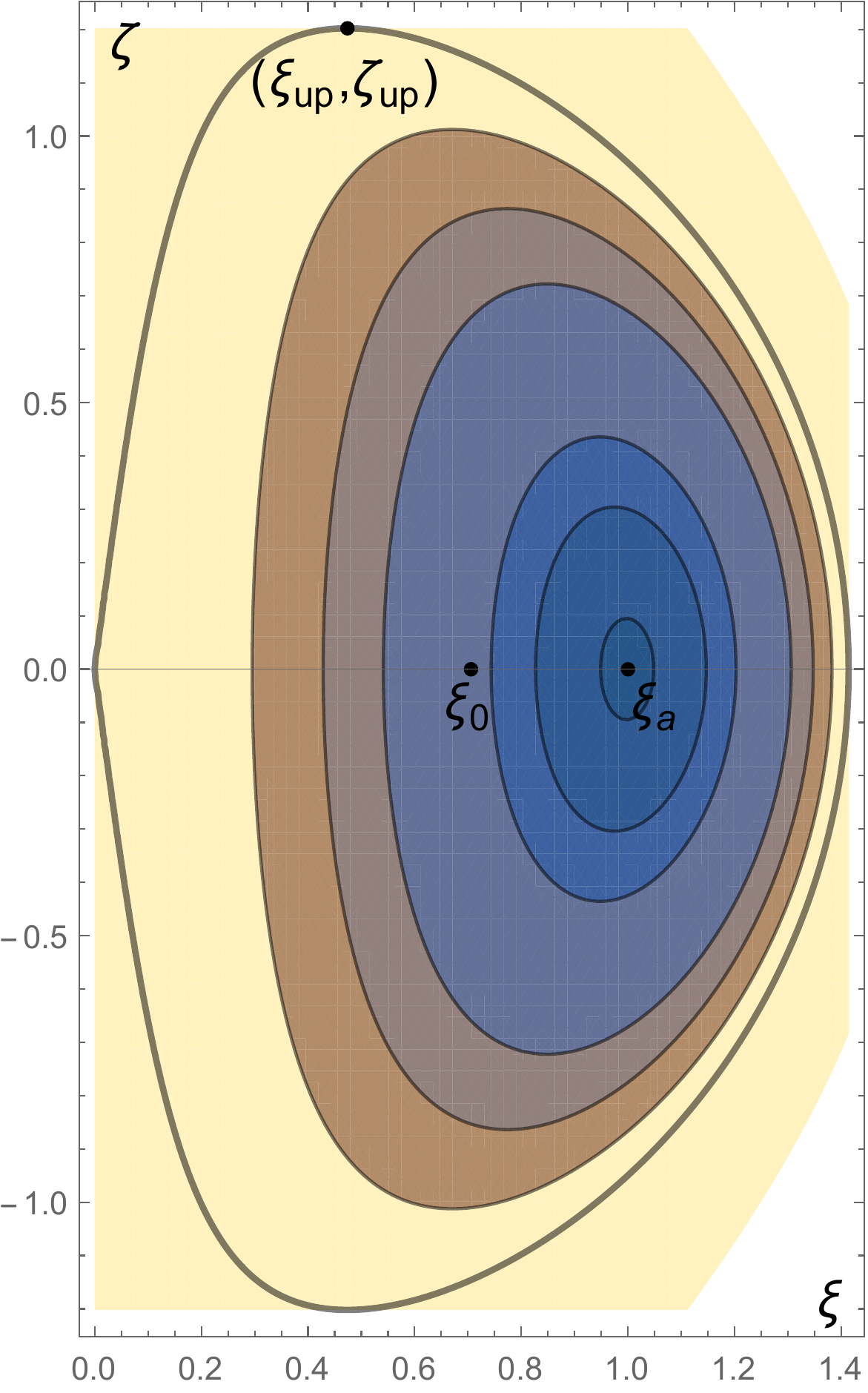}
     \caption{{\small \emph{The poloidal cross-section for ITER-like paramagnetic configuration with parallel flows.} }}
     \end{figure} 
In contrast with the diamagnetic case, in a paramagnetic configuration the plasma reaches through a corner the axis of symmetry implying values for the minor radius different from the actual ones, and thus, such a configuration is not typical for conventional tokamaks.  However,  a configuration with a similar corner was observed recently in the QUEST spherical tokamak as a self organized state \cite{mizu} (Fig. 5 therein).
\par
The influence of pressure anisotropy is not different from the diamagnetic case, since the variation of the anisotropy parameters $\sigma_{d_a}$ and $n$, affect the equilibrium quantities in a similar way (even the maximum permissible values of $\sigma_{d_a}$ in connection with pressure positivity in the whole plasma region are almost the same both for ITER- and NSTX-like configurations), except for the following two differences. First, the toroidal velocity, $\widetilde{v}_\phi$, unlikely the diamagnetic case, reverses near the axis of symmetry and then behaves as the diamagnetic one to the right of the reversal point. An example for ITER is given in Fig. (3.26). 
\par
In spherical tokamaks the reversal point is displaced closer to the magnetic axis and $\widetilde{v}_\phi$ remains positive in a larger region than in the conventional ITER-like one [see Fig. (3.27)]. Reversal of $\widetilde{v}_\phi$ during the transition to improved confinement regimes have been observed in ASDEX Upgrade \cite{McD} and in LHD \cite{Ida}.
\begin{figure}
  \centering
    \includegraphics[width=3.5in]{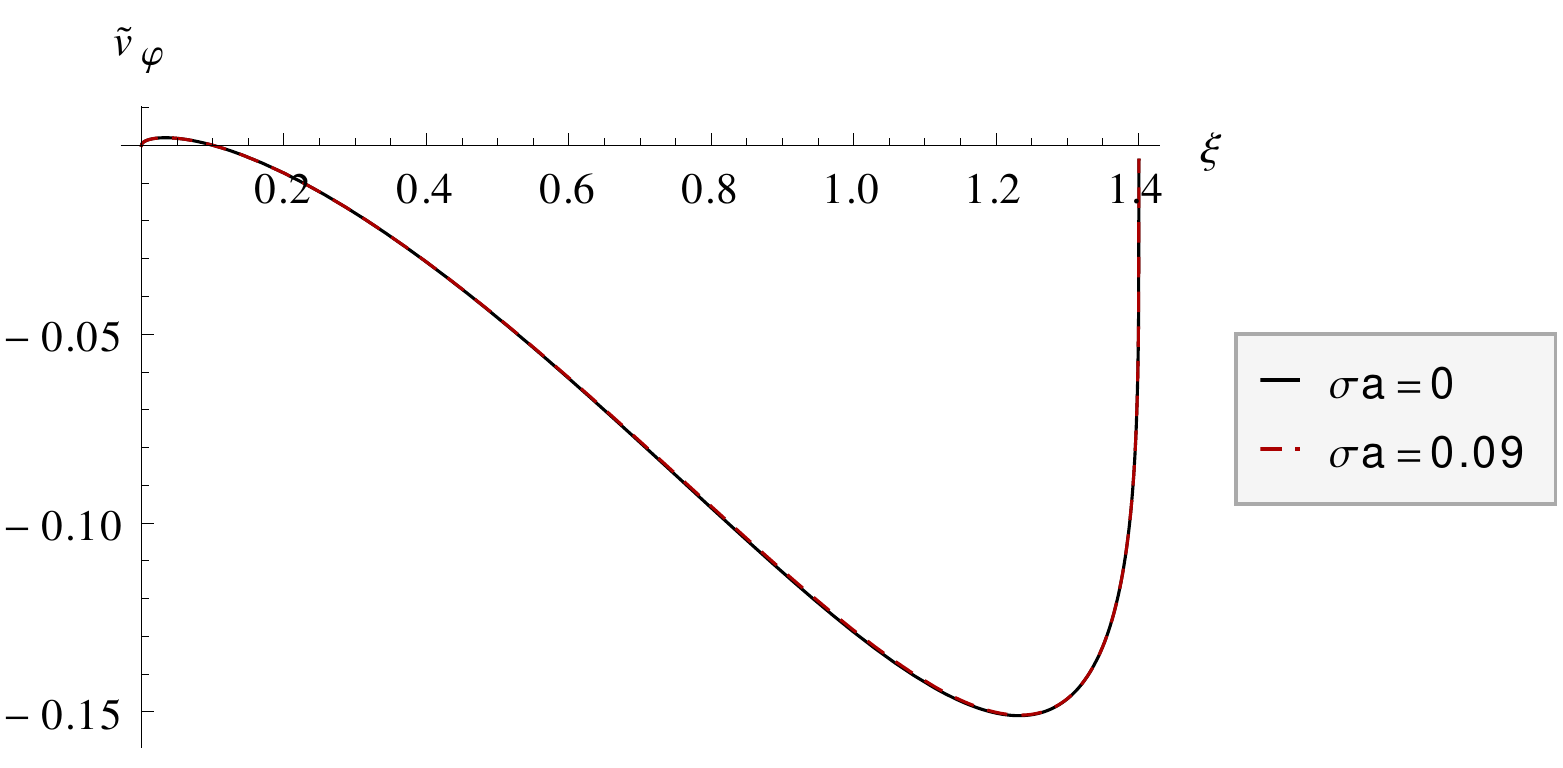}
     \caption{{\small \emph{The profile of $\widetilde{v}_\phi$ becomes slightly positive near the axis of symmetry on ITER-like paramagnetic equilibria with non-parallel flows, $\lambda=0.5$. Then it changes sign and behaves like in the diamagnetic case.} }}
     \end{figure}
     \begin{figure}
  \centering
    \includegraphics[width=3.5in]{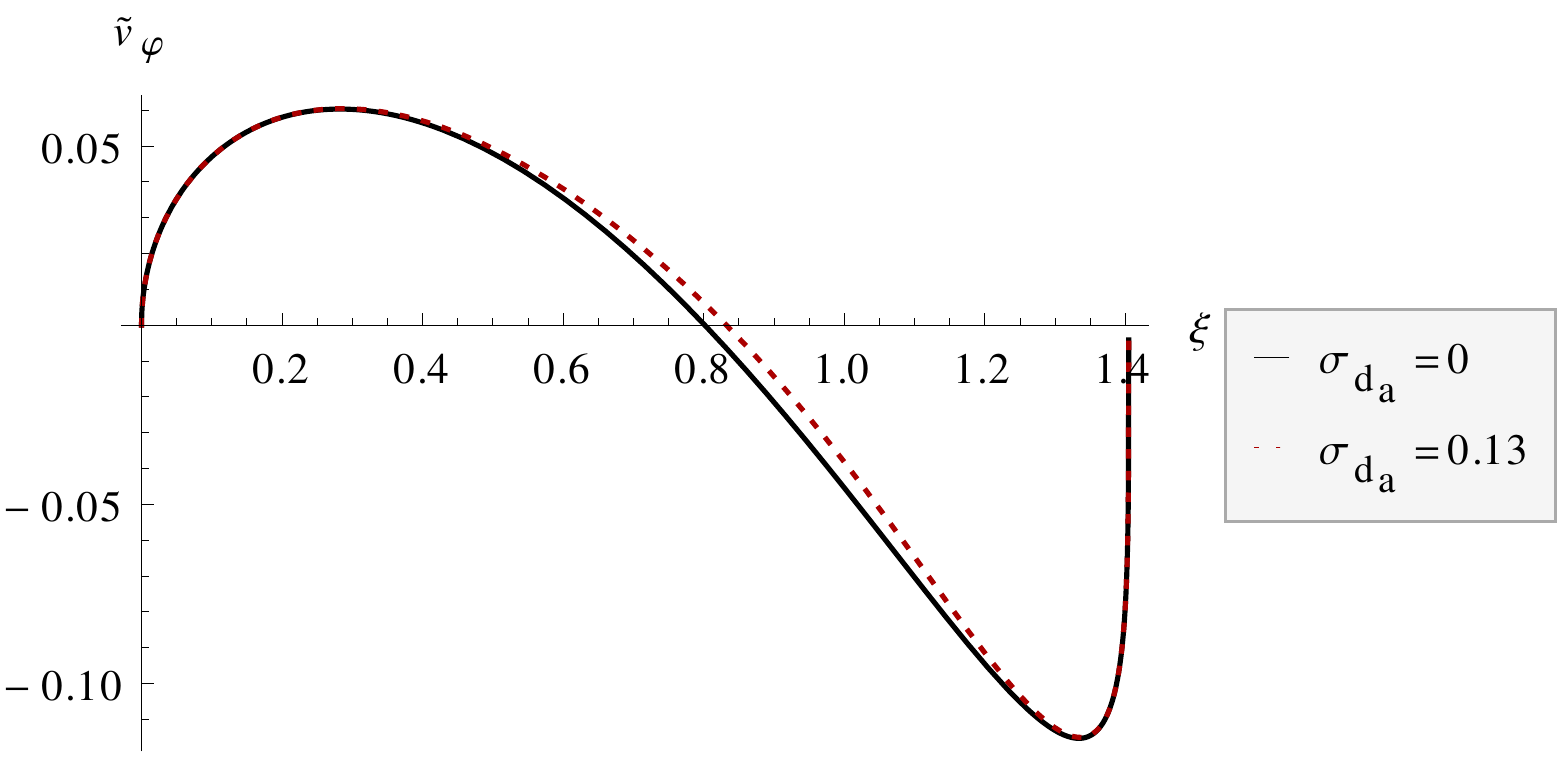}
     \caption{{\small \emph{The profile of $\widetilde{v}_\phi$ for NSTX-like paramagnetic equilibria with non-parallel flows. It has two extrema, in contrast with the diamagnetic case in which it was peaked near the magnetic axis. Its absolute values decreases a bit with the increase of $\sigma_{d_a}$ in the region near the magnetic axis.} }}
     \end{figure}
Second, the current density sharply falls off near the axis of symmetry and then increases from the high field side up to the outermost point of the boundary (Fig. 3.28).
      \begin{figure}
  \centering
    \includegraphics[width=3.5in]{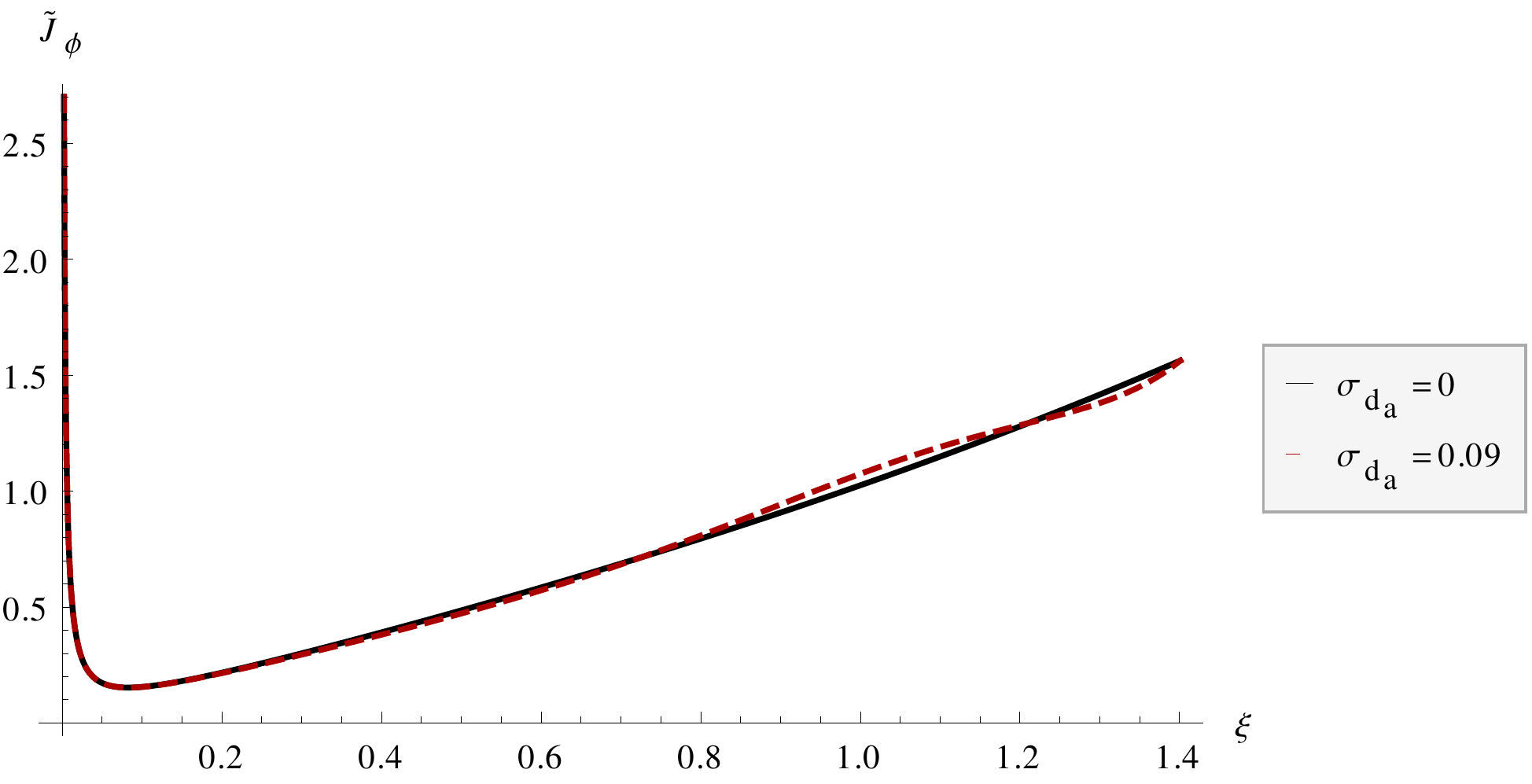}
     \caption{{\small \emph{The profile of $\widetilde{J}_\phi$ for ITER-like paramagnetic equilibria with non-parallel flows. The effect of anisotropy parameter $\sigma_{d_a}$ is similar with that in the diamagnetic case.} }}
     \end{figure}
The influence of pressure anisotropy on the confinement figures of merit for paramagnetic equilibria is similar as in the diamagnetic ones. Specifically, the local toroidal plasma beta on the magnetic axis is $\beta_{t_a}\sim 8.9\%$ for field aligned flows, and $\beta_{t_a}\sim 10\%$ for non-parallel ones for ITER paramagnetic equilibria, while $\sim 13\%$ and $\sim 15,5\%$ for NSTX ones. Thus, for field-aligned flows it takes almost the same values as in the diamagnetic case, while for flows not parallel to the magnetic field, $\beta_{t_a}$ is a little lower. In addition, the safety factor on axis is $q_a\sim 2.53$ for parallel flows, while $q_a\sim 2.45$ for non-parallel ones, for ITER paramagnetic equilibria. For NSTX the respective values are $q_a\sim2.44$, and $\sim2.35$, both higher than in diamagnetic equilibria (and both satisfying the Kruskal-Shafranov limit).

 \section{Conclusions}
 
\hspace{2em}In the present chapter, we constructed Solovev-like equilibria for linear choices of the free functions included in the GGS equation derived in chapter 2. Then we examined the influence of pressure anisotropy and plasma flow on these equilibria for peaked on axis profiles of the anisotropy and Mach functions containing a couple of free parameters (Eqs. (\ref{Mach sol})-(\ref{sigma sol})). The study includes diamagnetic and paramagnetic plasmas, flows parallel and not parallel to the magnetic field as well as ITER-like and NSTX-like tokamak configurations. The main results are summarized below.
\begin{enumerate}

\item The impact of anisotropy on most equilibrium quantities is stronger than that of the flow, with only exceptions the effective pressure and the average pressure. This is plausible because according to experimental evidence, the parameter $\sigma_{d_a}$ determining the maximum of the pressure anisotropy on the magnetic axis is larger than the respective Mach number $M_{p_a}^2$. The non-parallel flow term has a small contribution in most quantities.
\item Pressure anisotropy as well as plasma flow act paramagnetically as far as the parameters $\sigma_{d_a}$ and $M_{p_a}^2$ are concerned, since their increase raises the toroidal magnetic field from its vacuum value. On the other hand increase of $n$ acts diamagnetically.
\item Anisotropy through $\sigma_{d_a}$ has a different influence on the toroidal current density in different regions of the plasma, since near the magnetic axis it makes $\widetilde{J}_\phi$ to increase, while near the boundary it makes it lower than its isotropic value. In addition, raise of $\sigma_{d_a}$ makes the isotropic $|\widetilde{J}_{\zeta}|$ take higher values, and to present two extrema on the left and right side of the magnetic axis. The higher $n$ is the closer to axis are located both extrema.
\item The absolute values of $\widetilde{v}_\phi$, $\widetilde{v}_\zeta$, and $\widetilde{E}_\xi$ become slightly larger in the presence of anisotropy mainly through $\sigma_{d_a}$. In contrast, the anisotropy has opposite effects on plasma pressure, i.e. both the overall pressure and the average pressure decrease with $\sigma_{d_a}$. At last, the toroidal beta on axis and the safety factor thereon are rather insensitive to pressure anisotropy.
\item Comparing the diamagnetic configurations with the paramagnetic ones we found that both anisotropy and flow act in a similar way in most of the equilibrium quantities. The only differences are that in paramagnetic equilibria $\widetilde{v}_\phi$ reverses, while it is peaked on axis in diamagnetic ones, and that unlike the diamagnetic case, $\widetilde{J}_\phi$ sharply falls off near the axis of symmetry and then starts to increase from high to low field side. For field-aligned flows $\beta_{t_a}$ is almost the same both for diamagnetic and paramagnetic equilibria, while for non-parallel flows the paramagnetic beta on axis is $\sim0.5-1.0\%$ lower than the diamagnetic one, in both ITER and NSTX tokamaks. In contrast, the paramagnetic $q_a$ takes higher values than the diamagnetic one, both for parallel and not parallel flow as well as for ITER and NSTX tokamaks. This is reasonable since improving equilibrium usually has a negative impact on stability. 
\item At last by comparing the conventional ITER tokamak with the NSTX spherical one, we found that both pressure anisotropy and plasma flow act in the same way in most of their equilibrium quantities. In an NSTX, both $\sigma_{d_a}$ and $M_{p_a}^2$ take higher values than in ITER, and since in the NSTX these two parameters are comparable, the effects of the flow are more evident thereon. The main noticeable differences are that the reversal point of $\widetilde{v}_\phi$ is located closer to the magnetic axis in NSTX-like paramagnetic equilibria, and that for the same value of $\sigma_{d_a}$ the ratio $\widetilde{p}_{\parallel}/\widetilde{p}_{\bot}$ is higher on ITER than on NSTX, both diamagnetic and paramagnetic.
\end{enumerate}
\newpage

\chapter{Hernegger-Maschke-like Solution}

\hspace{2em}In this chapter the GGS equation will be solved for an alternative choice of the free functions.
Since the charged particles  move  parallel to the magnetic field free of magnetic force,  parallel flows is a plausible approximation. In particular for tokamaks this is compatible with the fact that the toroidal magnetic field is an order of magnitude larger than the poloidal one and the same scaling is valid  for the toroidal  and poloidal components of the fluid velocity. Also, for parallel flows the problem remains analytically tractable and leads to a generalised Hernegger-Maschke solution to be constructed below. The original Hernegger-Maschke solution was first introduced in \cite{Mas} for isotropic equilibria and rectangular boundary, with choices for the free functions of the GS equation to be quadratic in $\widetilde{u}$, and has also been studied in a number of papers \cite{GuFr},\cite{Oshi}. Here, respective equilibria for plasma surrounded by a fixed toroidal boundary of ITER-like cross section (presenting an X-point at the lower part) will be constructed, in relevance to the ITER and NSTX-Upgrade tokamak devices. Since the boundary is held fixed, the position of the magnetic axis of each configuration will be left free, and all quantities will be normalized with respect to the geometric center (see section (2.3)). After the generalised equilibria are constructed, we are going to examine the influence of pressure anisotropy, as well as plasma flow on their characteristics.

\section{Construction of the Solution}

\hspace{2em}In the absence of the electric field term ($\xi^4$ -term) the GGS equation (\ref{GGSu nor}) becomes
\beq \label{hmlinear}  \widetilde{\Delta ^{*}}\widetilde{u}+\frac{1}{2}\frac{d}{d\widetilde{u}}\left(\frac{\widetilde{X}^2}{1-\sigma _d -M_p ^2}\right)+\xi ^2\frac{d\widetilde{\overline{p}_s}}{d\widetilde{u}}=0\eeq
where, $\widetilde{\Delta ^{*}}\widetilde{u}=\frac{\partial^2\widetilde{u}}{\partial \xi^2}+\frac{\partial^2\widetilde{u}}{\partial \zeta^2}-\frac{1}{\xi}\frac{\partial \widetilde{u}}{\partial \xi}$.
We choose the free function terms of Eq. (\ref{hmlinear}) to be quadratic in $\widetilde{u}$ as 
\beq \label{ANS1} \widetilde{\overline{p}}_s(\widetilde{u})=\widetilde{p}_1+\widetilde{p}_2\widetilde{u}^2\eeq
and
\beq \label{ANS2} \frac{\widetilde{X}^2(\widetilde{u})}{1-\sigma _d(\widetilde{u})-M_p^2(\widetilde{u})}=\widetilde{X}_0^2+\widetilde{X}_1\widetilde{u}^2\eeq
In contrast with Solovev solution and the free boundary problem examined in chapter 3, here we want the function $\widetilde{u}$ to vanish on a fixed plasma boundary, i.e. $\widetilde{u}_b=0$. Thus, since plasma pressure must also vanish on the boundary, from (\ref{ANS1}) it should be $\widetilde{p}_1=0$. In addition, the parameter $\widetilde{X}_0^2$ is related with the vacuum toroidal magnetic field at the geometric center as $\widetilde{X}_0^2=\xi_0^2\widetilde{B}_{\phi _0}^2$, and due to the adopted normalization it is $\widetilde{X}_0^2= 1$.
Thus, the ansatz (\ref{ANS1})-(\ref{ANS2}) take the form
\beq \label{hm1} \widetilde{\overline{p}}_s(\widetilde{u})=\widetilde{p}_2\widetilde{u}^2\eeq
and 
\beq \label{hm2} \frac{\widetilde{X}^2(\widetilde{u})}{1-\sigma _d(\widetilde{u})-M_p^2(\widetilde{u})}=1+\widetilde{X}_1\widetilde{u}^2\eeq
The values of the parameters $\widetilde{p}_2$ and $\widetilde{X}_1$, will be chosen by inspection in connection with realistic shaping and values of the equilibrium figures of merit, i.e. the local toroidal beta and the safety factor on the magnetic axis.
\par
Substituting the ansatz (\ref{hm1})-(\ref{hm2}) into the Grad-Shafranov equation (\ref{hmlinear}), it reduces to
\beq \label{hm3} \frac{\partial ^2\widetilde{u}}{\partial \xi ^2}+\frac{\partial ^2\widetilde{u}}{\partial \zeta ^2}-\frac{1}{\xi}\frac{\partial \widetilde{u}}{\partial \xi}+\widetilde{X}_1\widetilde{u}+2\widetilde{p}_2\xi ^2\widetilde{u}=0\eeq
which is a linear partial differential equation.
The solution to Eq. (\ref{hm3}) is found by separation of variables,
\beq \label{separ} \widetilde{u}(\xi ,\zeta )=G(\xi)T(\zeta)\eeq
on the basis of which, it further reduces to the following form
\beq \label{hm4} \frac{1}{T(\zeta)}\frac{d^2T(\zeta)}{d\zeta ^2}=-\frac{1}{G(\xi)}\frac{d^2G(\xi)}{d\xi ^2}+\frac{1}{\xi G(\xi)}\frac{dG(\xi)}{d\xi}-\widetilde{X}_1-2\widetilde{p}_2\xi ^2=-\eta ^2\eeq  
where $\eta$ is the separation constant.
Therefore, the problem reduces to a couple of ODEs. The one for the function $T$ is
\beq \label{ode1}  \frac{d^2T(\zeta)}{d\zeta ^2}+\eta ^2T(\zeta)=0\eeq
having the general solution
\beq \label{ode1sol} T(\zeta)=a_1cos(\eta \zeta) +a_2sin(\eta \zeta)\eeq
with the coefficients $a_1$ and $a_2$ to be determined later. The second equation satisfied by function $G$ is 
\beq \label{ode2} \frac{d^2G(\xi)}{d\xi ^2}-\frac{1}{\xi }\frac{dG(\xi)}{d\xi}+(\widetilde{X}_1-\eta ^2)G(\xi)+2\widetilde{p}_2\xi ^2 G(\xi)=0\eeq
Introducing the parameters $\gamma =\widetilde{X}_1$, $\delta =2 \widetilde{p}_2$, and $\varrho =i\sqrt{\delta}\xi ^2$, so that $\frac{\partial}{\partial \xi}=2i\sqrt{\delta}\xi\frac{\partial}{\partial \varrho}$ and $\frac{\partial^2}{\partial \xi^2}=2i\sqrt{\delta}\frac{\partial}{\partial \varrho}+4i\sqrt{\delta}\varrho\frac{\partial^2}{\partial \varrho^2}$, (\ref{ode2}) becomes
\beq \label{UN1} \frac{d^2G(\varrho)}{d\varrho ^2}+\left[i\frac{\eta ^2-\gamma}{4\sqrt{\delta}}\frac{1}{\varrho}-\frac{1}{4}\right]G(\varrho)=0\eeq
Furthermore, if we set
\beq \label{UN2} \nu \equiv i\frac{\eta ^2-\gamma}{4\sqrt{\delta}}\eeq
then (\ref{UN1}) is put in the form
\newline
\beq \label{whittaker} \frac{d^2G(\varrho)}{d\varrho ^2}+\left[\frac{\nu}{\varrho}-\frac{1}{4}\right]G(\varrho)=0\eeq
which is a special case of the Whittaker's equation
 \beq \label{whittaker1} W^{''}(x)+\left[\frac{\nu}{x}-\frac{1}{4}+\frac{\frac{1}{4}-\mu ^2}{x^2}\right]W(x)=0\eeq for $\mu=\frac{1}{2}$. Thus, Eq. (\ref{whittaker}) admits the general solution
\beq \label{whisol} G(\varrho)=b_1M_{\nu ,\frac{1}{2}}(\varrho)+b_2W_{\nu ,\frac{1}{2}}(\varrho)\eeq
where $M_{\nu,\mu}$ and $W_{\nu,\mu}$ are the Whittaker functions, which are independent solutions of the Whittaker's differential equation.
\par
Consequently, a typical solution of the original equation (\ref{hm3}) is written in the form
\beq \label{typsol} \widetilde{u}(\varrho ,\zeta)=\left[b_1M_{\nu ,\frac{1}{2}}(\varrho)+b_2W_{\nu ,\frac{1}{2}}(\varrho)\right]\left[a_1cos(\eta \zeta) +a_2sin(\eta \zeta)\right]\eeq
For further treatment it is convenient to restrict the separation constant $\eta$ to positive integer values $j$. Therefore, by superposition the solution can be expressed as
\beq \label{supersol} \widetilde{u}(\varrho ,\zeta)=\sum _{j=1}^\infty \left[a_jM_{\nu _j,\frac{1}{2}}(\varrho)cos(j\zeta)+b_jM_{\nu _j,\frac{1}{2}}(\varrho)sin(j\zeta)+c_jW_{\nu _j,\frac{1}{2}}(\varrho)cos(j\zeta)+d_jW_{\nu _j,\frac{1}{2}}(\varrho)sin(j\zeta)\right]\eeq
In order to make the analysis more tractable, we factorize (\ref{supersol}) with respect to the coefficient $a_1$ so that
\begin{align} \label{supersol2}
\widetilde{u}& =a_1\left\{ M_{\nu _1,\frac{1}{2}}(\varrho)cos(\zeta)+\frac{b_1}{a_1}M_{\nu _1,\frac{1}{2}}(\varrho)sin(\zeta)+\frac{c_1}{a_1}W_{\nu _1,\frac{1}{2}}(\varrho)cos(\zeta)+\frac{d_1}{a_1}W_{\nu _1,\frac{1}{2}}(\varrho)sin(\zeta)\right. \nonumber \\
 &\left. {}+\sum _{j=2}^N \left[\frac{a_j}{a_1} M_{\nu _j,\frac{1}{2}}(\varrho)cos(j\zeta)+\frac{b_j}{a_1}M_{\nu _j,\frac{1}{2}}(\varrho)sin(j\zeta)+\frac{c_j}{a_1}W_{\nu _j,\frac{1}{2}}(\varrho)cos(j\zeta)+\frac{d_j}{a_1}W_{\nu _j,\frac{1}{2}}(\varrho)sin(j\zeta)\right]\right\}
\end{align}
Now, by setting $a_1\equiv c$, $a_j^*=\frac{a_j}{c}$, $b_j^*=\frac{b_j}{c}$, $c_j^*=\frac{c_j}{c}$, and $d_j^*=\frac{d_j}{c}$, then the solution can be expressed as
\beq \label{supersol3} \widetilde{u}(\varrho ,\zeta)=c\widetilde{u}^*(\varrho ,\zeta)\eeq
where
\beq \label{supersol4} \widetilde{u}^*(\varrho ,\zeta)=\sum _{j=1}^N \left[a_j^*M_{\nu _j,\frac{1}{2}}(\varrho)cos(j\zeta)+b_j^*M_{\nu _j,\frac{1}{2}}(\varrho)sin(j\zeta)+c_j^*W_{\nu _j,\frac{1}{2}}(\varrho)cos(j\zeta)+d_j^*W_{\nu _j,\frac{1}{2}}(\varrho)sin(j\zeta)\right]\eeq
with $a_1^*=1$.
This `trick' is done because employing the function $\widetilde{u}^*$, the system of algebraic equations that will arise from the boundary conditions that are going to be imposed in order to determine the values of the coefficients $a_j^*$, $b_j^*$, $c_j^*$, and $d_j^*$, will be inhomogeneous due to the coefficient $a_1^*=1$ of the first term of the sum in (\ref{supersol4}). So, it will be easier to solve with Mathematica, because otherwise the system becomes homogeneous and the package gets the trivial zero solution. Once the values of these coefficients are found, we can then determine the solution $\widetilde{u}$ by multiplying them with the constant $c$, see Eq. (\ref{supersol3}). The value of $c$ will be determined on the basis of the Kruskal-Shafranov stability limit, $q_a>1$.
\par
Before continuing to the presentation of the boundary conditions, we have to assign the functions $\sigma_d$, $M_p^2$ and $\widetilde{\rho}$ with respect to $\widetilde{u}$. Thus, we choose the anisotropy function, the Mach function, and the mass density profiles to be peaked on the magnetic axis of a given configuration and vanishing on the boundary:
\beq \label{sigma hm} \sigma _d(\widetilde{u})=\sigma _{d_a}\left(\frac{\widetilde{u}}{\widetilde{u}_a}\right)^n\eeq 
\beq \label{Mach hm} M_p^2(\widetilde{u})=M_{p_a}^2\left(\frac{\widetilde{u}}{\widetilde{u}_a}\right)^m\eeq 
and
\beq \label{rho hm} \widetilde{\rho} (\widetilde{u})=\widetilde{\rho} _a\left(\frac{\widetilde{u}}{\widetilde{u}_a}\right)^g\eeq
where $\widetilde{\rho}_a$ and $\widetilde{u}_a$ constant quantities.
\par
It is recalled that typical values for the poloidal Mach function is $\sim 10^{-4}$ for an ITER tokamak and $\sim 10^{-2}$ for the NSTX-U spherical one. At last, the exponents $g$ and $m$ will be fixed at the values $g=1/2$ and $m=2$, while the free parameters $\sigma_{d_a}$ and $n$, associated with the maximum value and the shape of the anisotropy function $\sigma_d$, will be let to vary through a sufficient range, in order to examine the influence of anisotropy on equilibrium. 

\section{Boundary (Shaping) Conditions}

\hspace{2em}Once all the arbitrary functions are specified, we can construct single-null diverted equilibria, e.g. ITER-like ones, by imposing appropriate boundary conditions. For an up-down asymmetric (about the midplane) magnetic surface, its shape is usually characterized by  four points which have to be fixed, namely the inner, the outer, the upper, and the lower point of the plasma boundary, which we want to be an X-point. The above characteristic boundary points are:
\begin{center}
Inner point : $(\xi_{in}=1-\frac{\alpha}{R_0},\zeta_{in}=0)$
\end{center}
\begin{center}
Outer point : $(\xi_{out}=1+\frac{\alpha}{R_0},\zeta_{out}=0)$
\end{center}
\begin{center}
Upper point : $(\xi_{up}=1-t\frac{\alpha}{R_0},\zeta_{up}=\kappa\frac{\alpha}{R_0})$
\end{center}
\begin{center}
Lower X-point : $(\xi_{x}=1+\frac{\alpha}{R_0}cos[\pi -tan^{-1}(\frac{\kappa}{t})],\zeta_{x}=-\kappa\frac{\alpha}{R_0})$
\end{center}
where $\alpha$ is the minor radius of the torus, $t$ is the triangularity, $\kappa$ the elongation. The inverse aspect ratio is $\varepsilon=\frac{\alpha}{R_0}$. Values of these parameters for ITER and NSTX-U are given in chapter 1 (see Table 1.1). In this work we assume that the upper and lower part of the separatrix have the same elongation and triangularity. It is noted however that different parameter values for these two parts are possible. The above characteristic points of an up-down asymmetric boundary are illustrated in Fig. (4.1). 
 \begin{figure}
  \centering
    \includegraphics[width=3.5in]{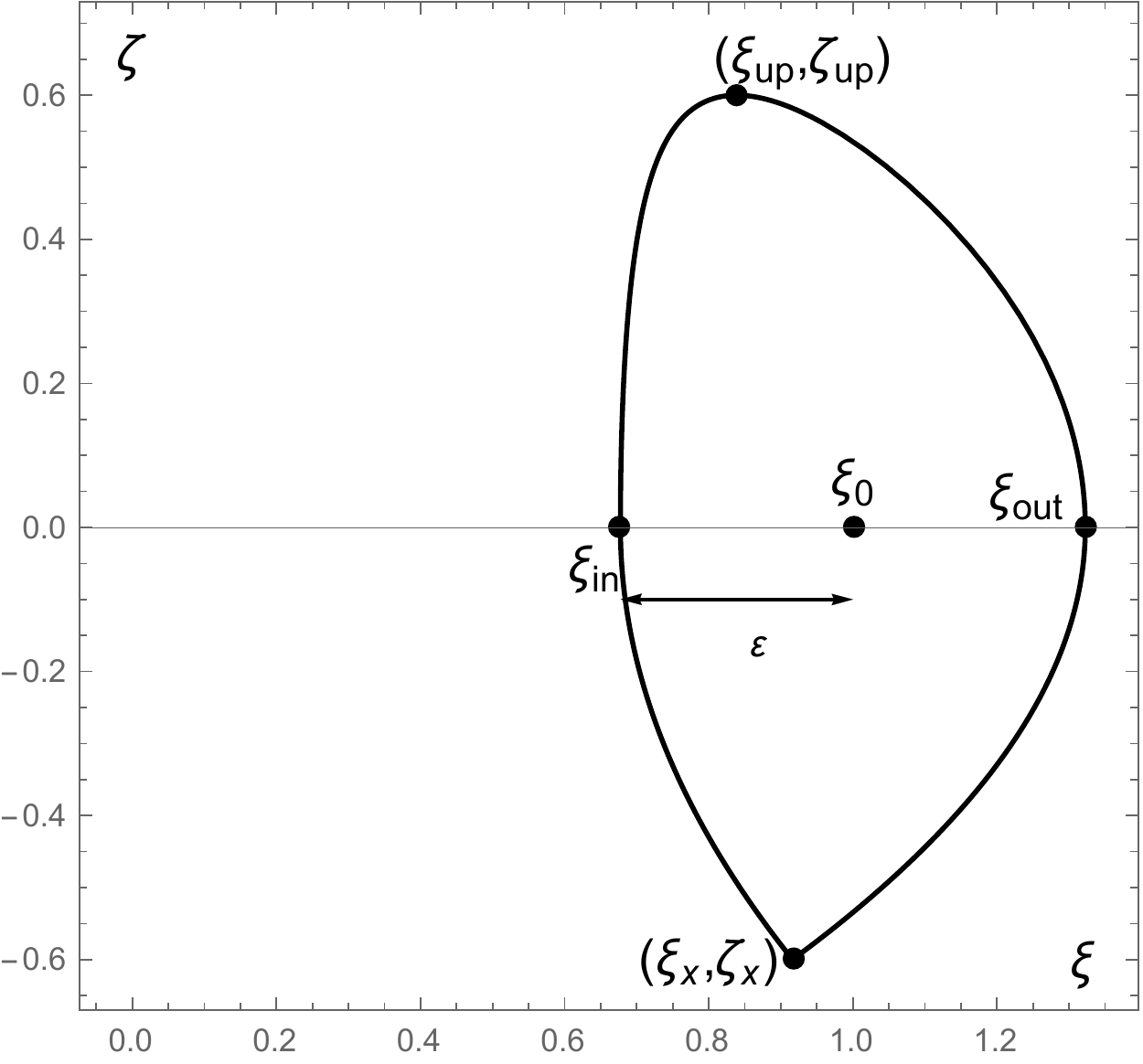}
     \caption{{\small \emph{Characteristic points determining an up-down asymmetric boundary.} }}
     \end{figure}
Such an up-down asymmetric boundary consisting of a smooth upper part and a lower part that possesses an X-point, is described by the following parametric equations introduced in \cite{Parametric}:
for the smooth upper part, where $0\leq \theta \leq \pi$,
\begin{align} \label{kui1}
\xi_u &=1+\frac{\alpha}{R_0}cos(\theta+w_1sin(\theta))\nonumber \\
&\zeta_u=\kappa\frac{\alpha}{R_0}sin(\theta)
\end{align}
for the left lower part, where $\pi\leq\theta\leq 2\pi -w_2$,
\begin{align} \label{kui2}
\xi_l &=1+\frac{\alpha}{R_0}cos(\theta)\nonumber \\
&\zeta_l=-\left[\zeta_x^2\frac{1+cos(\theta)}{1+cos(w_2)}\right]^{1/2}
\end{align}
and for the right lower part, where $2\pi-w_2\leq \theta\leq 2\pi$,
\begin{align} \label{kui3}
\xi_r &=1+\frac{\alpha}{R_0}cos(\theta)\nonumber \\
&\zeta_r=-\left[\zeta_x^2\frac{1-cos(\theta)}{1-cos(w_2)}\right]^{1/2}
\end{align}
where, $w_1=sin^{-1}(t)$, and $w_2=\pi-tan^{-1}(\frac{\kappa}{t})$.
\par
Since the value of $\widetilde{u}^*$ must equal to zero on the boundary, $\widetilde{u}^*_b=0$, then the first four boundary conditions come from the requirement that it should also be zero at the above presented characteristic points of the boundary. Function $\widetilde{u}^*$ is in general complex, and since it satisfies the GGS equation, then both its real and imaginary parts are also solutions of this equation. Here, following Ref. \cite{GuFr} we will work with the imaginary part of the flux function. So the first four conditions are:
\beq \label{c1} Im[\widetilde{u}^*(\xi _{in},\zeta _{in})]=0\eeq
\beq \label{c2} Im[\widetilde{u}^*(\xi _{out},\zeta _{out})]=0\eeq
\beq \label{c3} Im[\widetilde{u}^*(\xi _{up},\zeta _{up})]=0\eeq
\beq \label{c4} Im[\widetilde{u}^*(\xi _x,\zeta _x)]=0\eeq
In addition, boundary conditions related with the first derivative of $\widetilde{u}^*$ at these four characteristic points will also be included. These are:
\beq \label{c5} Im[\widetilde{u}_\zeta ^*(\xi _{in},\zeta _{in})]=0\eeq
\beq \label{c6} Im[\widetilde{u}_\zeta ^*(\xi _{out},\zeta _{out})]=0\eeq
\beq \label{c7} Im[\widetilde{u}_\xi ^*(\xi _{up},\zeta _{up})]=0\eeq
\beq \label{c8} Im[\widetilde{u}_\zeta ^*(\xi _x,\zeta _x)]=0\eeq
\beq \label{c9} Im[\widetilde{u}_\xi^*(\xi _x,\zeta _x)]=0\eeq
where, $\widetilde{u}_\zeta^*=\frac{\partial \widetilde{u}^*}{\partial \zeta}$, and $\widetilde{u}_\xi^*=\frac{\partial \widetilde{u}^*}{\partial \xi}$. The above conditions guarantee the curve smoothness at these characteristic points. Specifically, in the vicinity of the innermost and outermost points the desirable curve should be perpendicular and symmetrical to the midplane. 
\par
Furthermore, there exist three other conditions  introduced in \cite{Cerfon}, that involve the second derivatives of $\widetilde{u}^*$ related with the curvature of the boundary curve in these characteristic points. These are:
\beq \label{c10} Im[\widetilde{u}_{\xi \xi}^*(\xi _{up},\zeta _{up})]=\frac{\kappa}{\varepsilon cos^2w_1}Im[\widetilde{u}_\zeta ^*(\xi _{up},\zeta _{up})]\eeq
\beq \label{c11}  Im[\widetilde{u}_{\zeta \zeta}^*(\xi _{in},\zeta _{in})]=-\frac{(1-w_1)^2}{\varepsilon \kappa ^2}Im[\widetilde{u}_\xi ^*(\xi _{in},\zeta _{in})]\eeq
\beq \label{c12} Im[\widetilde{u}_{\zeta \zeta}^*(\xi _{out},\zeta _{out})]=\frac{(1+w_1)^2}{\varepsilon \kappa ^2}Im[\widetilde{u}_\xi ^*(\xi _{out},\zeta _{out})]\eeq
where, $w_1=sin^{-1}(t)$, $\widetilde{u}_{\zeta \zeta}^*=\frac{\partial ^2 \widetilde{u}^*}{\partial \zeta^2}$, and $\widetilde{u}_{\xi \xi}^*=\frac{\partial ^2 \widetilde{u}^*}{\partial \xi^2}$. Derivation of (\ref{c10})-(\ref{c12}) is presented in Appendix C.
\par
The above conditions or part of them will be employed in the next sections to completely specify the free parameters of solution (\ref{supersol4}) in order to construct ITER and NSTX-U diamagnetic configurations.

\section{ITER Diamagnetic Configuration}

\hspace{2em}In order to determine function $\widetilde{u}^*$ we choose $j_{max}=4$ in (\ref{supersol4}), and thus, we need to solve a system of fifteen algebraic equations with equal number of unknown coefficients. For this reason we take into account the conditions (\ref{c1})-(\ref{c10}), as well as the condition $\widetilde{u}^*=0$ for five more boundary points originated by the relations (\ref{kui1})-(\ref{kui3}). Setting by inspection the values of the free parameters $\widetilde{p}_2=19.5$, $\widetilde{X}_1=-0.3$, and solving the system of the above equations, we find the values for the fifteen coefficients of the solution for an ITER diamagnetic configuration, given on Table (4.1).
 \begin{table}[ht]
\centering
 \begin{tabular}{c c}
\hline\hline
 & Coefficient Value  \\ [1ex] 
\hline 
$a_2^*$ & -0.625449 \\
$a_3^*$ & 0.13783  \\
$a_4^*$ & -0.0166821  \\
$b_1^*$ & 11.2021   \\
$b_2^*$ & -4.93763   \\ 
$b_3^*$ & 1.23997  \\ 
$b_4^*$ & -0.117724  \\
$c_1^*$ & 2.2319  \\
$c_2^*$ & -1.82077 \\
$c_3^*$ & 0.52774  \\
$c_4^*$ & -0.0657547  \\
$d_1^*$ & 29.2229 \\
$d_2^*$ & -20.6201  \\
$d_3^*$ & 10.462  \\
$d_4^*$ & -2.64276  \\[1ex]
\hline 
\end{tabular}
\caption{{\small \emph{Values of the coefficients of the solution $\widetilde{u}^*$ for an ITER diamagnetic configuration. }}}
\end{table}
Once solution $\widetilde{u}^*(\rho ,\zeta)$ is fully determined, we can find the position of the magnetic axis, by solving the equations $Im[\widetilde{u}_\xi^*]=0$ and $Im[\widetilde{u}_\zeta^*]=0$ to find that the magnetic axis is located outside of the midplane $\zeta=0$ at $(\xi_a=1.05815,\zeta_a=0.0159088)$. The value of $\widetilde{u}^*$ on axis is found to be $\widetilde{u}_a^*=-0.0386774$.
\par
Since $\widetilde{u}^*$ has been determined, the next step is to find the value of coefficient $c$, and then determine function $\widetilde{u}(\rho , \zeta)$ from Eq. (\ref{supersol3}). On the basis of the ansatz (\ref{hm2}), and for field-aligned flows, Eq. (\ref{Iu}) reduces to
\beq \label{Iq} \widetilde{I}=\sqrt{\frac{1+\widetilde{X}_1\widetilde{u}^2}{1-\sigma _d-M_p^2}}\eeq
Substituting (\ref{Iq}) into (\ref{qa u nor}), the safety factor on the magnetic axis takes the following form
\beq \label{qa hm1} q_a=\frac{1}{\xi}\sqrt{1+\widetilde{X}_1\widetilde{u}^2}\left[\frac{\partial ^2\widetilde{u}}{\partial \xi ^2}\frac{\partial ^2\widetilde{u}}{\partial \zeta ^2}\right]_{\xi =\xi _a,\zeta =\zeta _a}^{-1/2}\eeq
where $\widetilde{u}=c\widetilde{u}^*$. Evaluation of (\ref{qa hm1}) gives
\beq \label{qa hm2} q_a=1.18557\sqrt{\frac{1}{c^2}-0.000448783}\eeq
Imposing the condition $q_a=1.1$, so as the Kruskal-Shafranov limit is just satisfied, we find from (\ref{qa hm2}) that $c=1.07751$. That is, solution $\widetilde{u}$ is fully determined from Eq. (\ref{supersol3}), with its value on axis to be $\widetilde{u}_a=c\widetilde{u}_a^*=-0.0416752$. Closed magnetic surfaces associated with $\widetilde{u}$-contours of the equilibrium configuration are shown in Fig. (4.2).
\begin{figure}
  \centering
    \includegraphics[width=3.5in]{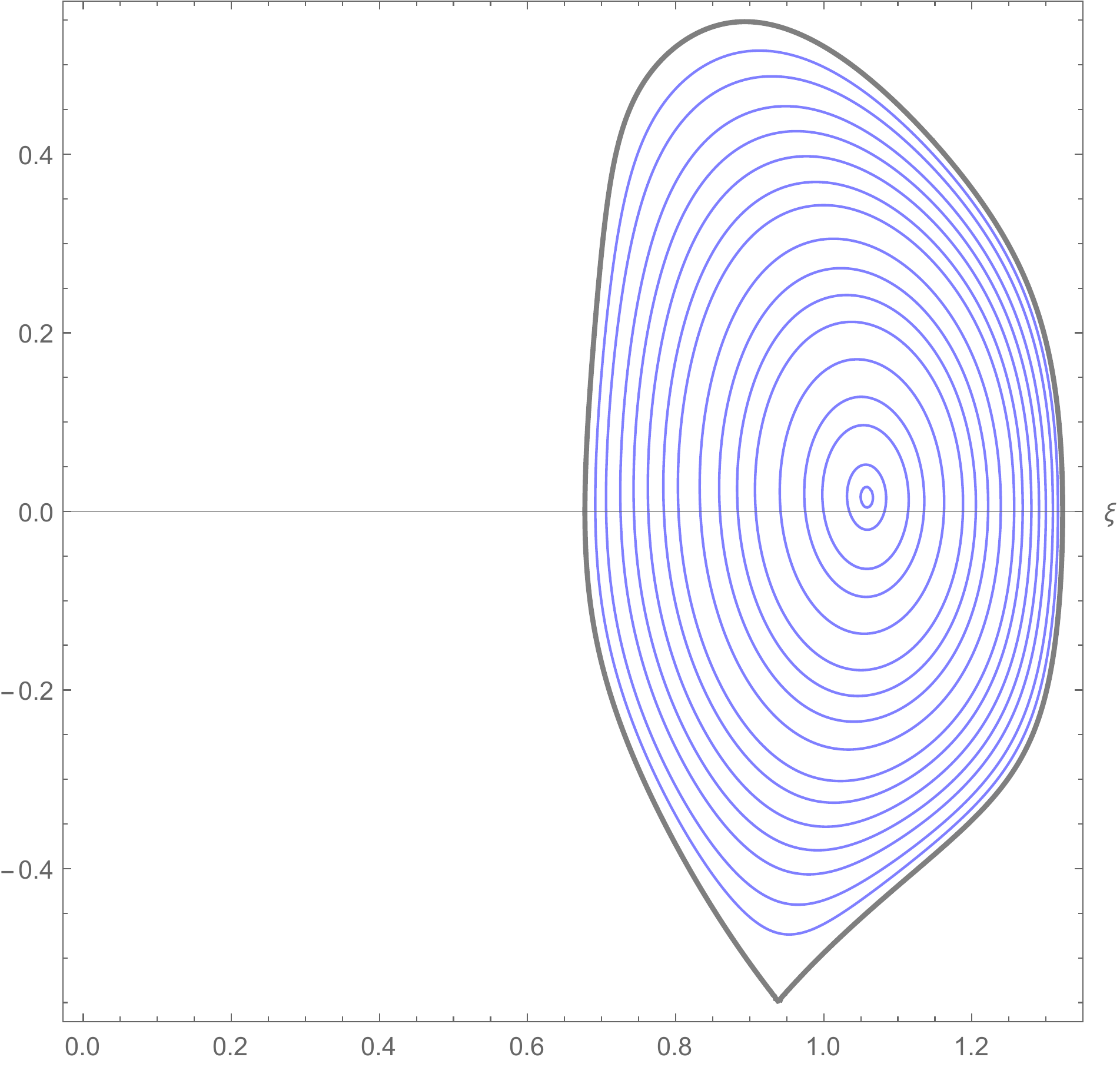}
     \caption{{\small \emph{The poloidal cross-section for an ITER-like diamagnetic equilibria with a lower X-point.} }}
     \end{figure}
     
     \subsection{Effects of Pressure Anisotropy and Flow}
     
\hspace{2em}In this subsection we are going to examine the influence of pressure anisotropy and plasma flow on the above constructed equilibria, and then compare their effects on it. The free parameters related with pressure anisotropy will be let to vary through the intervals $0\leq\sigma_{d_a}\leq 0.03$, and $2\leq n\leq 10$, so as all pressures to remain positive throughout the whole region of the plasma. Note that in this respect the maximum value of $\sigma_{d_a}$ is much less than for Solovev diamagnetic equilibria.
\par
As in Solovev-like equilibria both pressure anisotropy and plasma flow act paramagnetically. This is expected from relation (\ref{Iq}) since as $\sigma_{d_a}$ and $M_{p_a}^2$ take higher values, $\widetilde{I}$ also enhances, and as a result the magnetic field inside the plasma is increased. The paramagnetic action of the anisotropy parameter $\sigma_{d_a}$ is shown in Fig. (4.3).
  \begin{figure}
  \centering
    \includegraphics[width=3.5in]{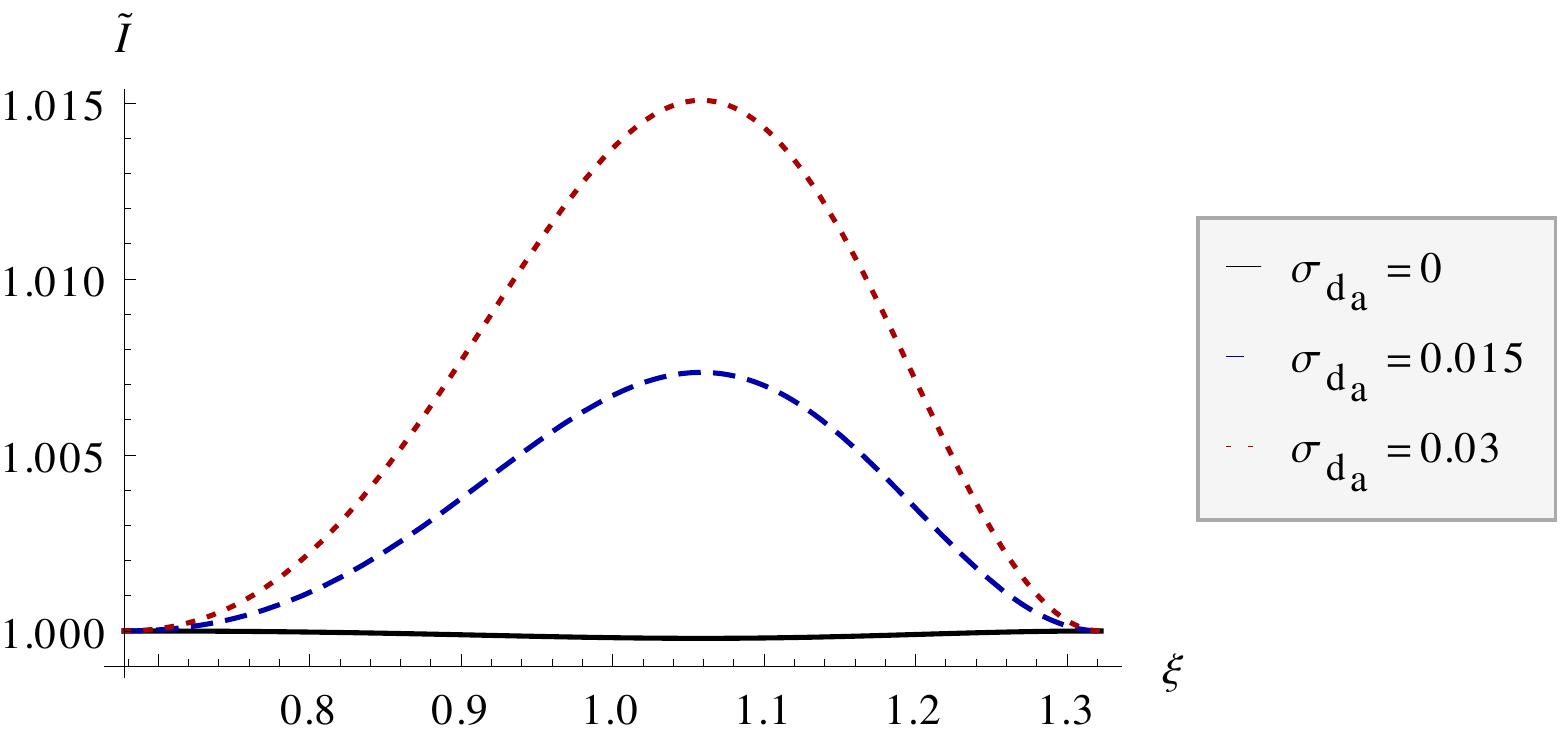}
     \caption{{\small \emph{As $\sigma_{d_a}$ increases, pressure anisotropy results to a paramagnetic action.} }}
     \end{figure}   
As we see, when the plasma is isotropic the $\widetilde{I}$ profile is hollow as expected, and as anisotropy gets larger, it results to a peaked $\widetilde{I}$ profile.
However, both $\sigma_{d_a}$ and $M_{p_a}^2$ act paramagnetically, and since $\sigma _{d_a}$ is two orders of magnitude higher than $M_{p_a}^2$ in ITER configuration, the increase of $\sigma _{d_a}$ is apparently more effective as shown in Fig. (4.4).
 \begin{figure}
  \centering
    \includegraphics[width=3.5in]{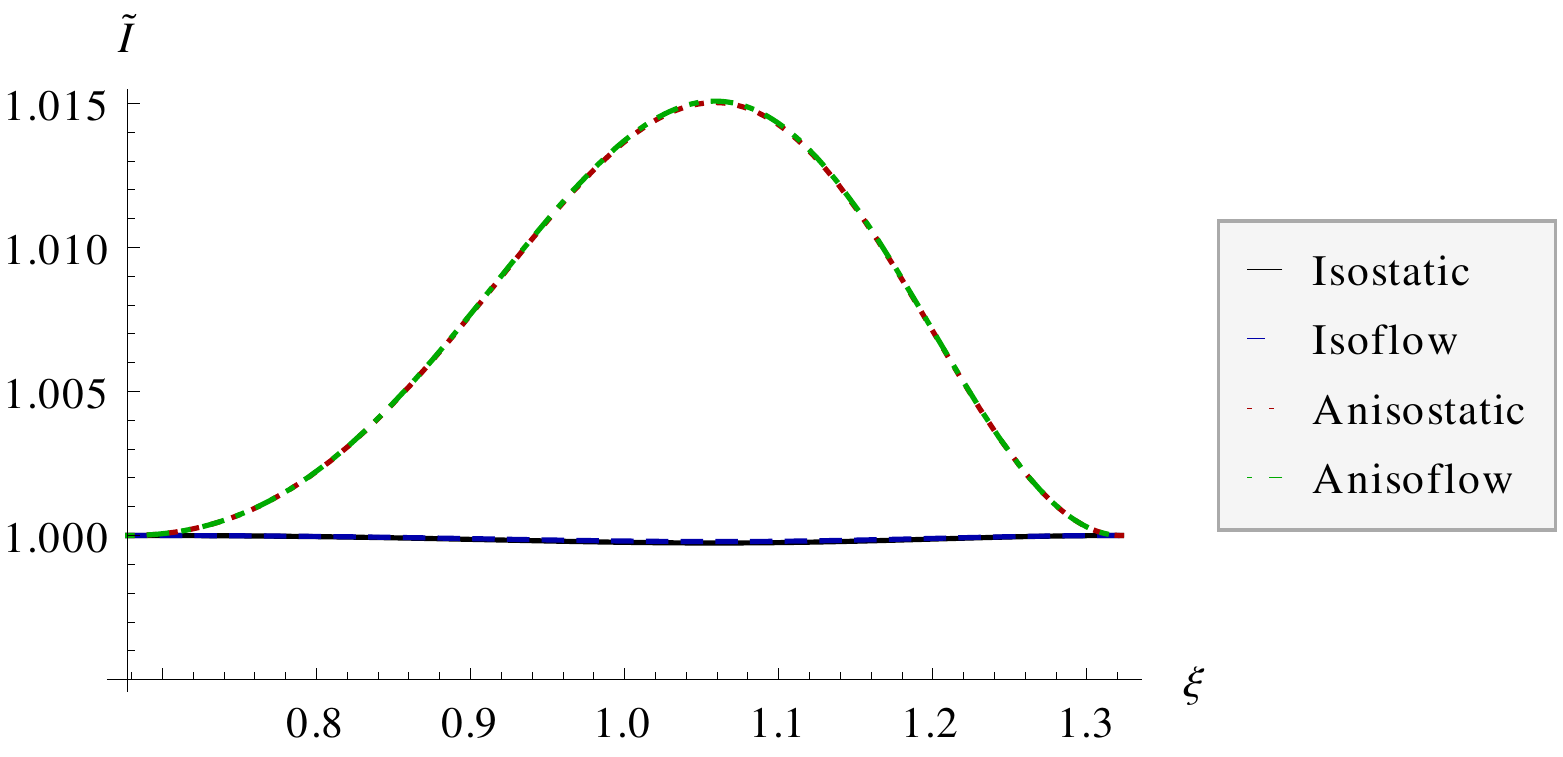}
     \caption{{\small \emph{Pressure anisotropy impact against that of the flow on $\widetilde{I}$, on $\zeta=0$ plane.} }}
     \end{figure}   
On the other hand, raising $n$ acts diamagnetically, since it makes $\widetilde{I}$ to decrease off-axis, and be localized in a shorter region as shown in Fig. (4.5).
\begin{figure}
  \centering
    \includegraphics[width=3.5in]{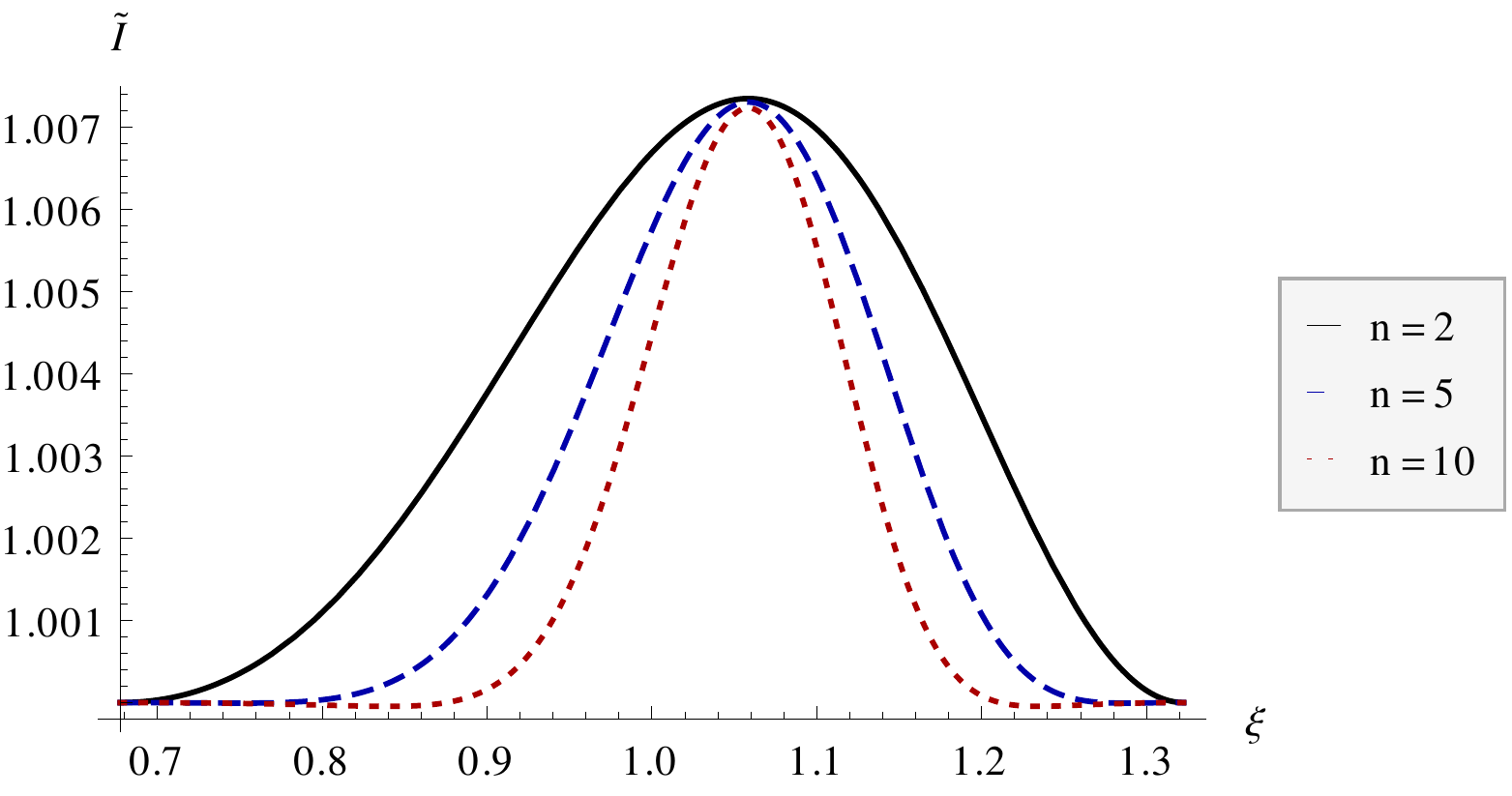}
     \caption{{\small \emph{As $n$ increases $\widetilde{I}$ decreases and pressure anisotropy results to a diamagnetic action.} }}
     \end{figure}
     \par
Pressure anisotropy has also an important influence on the $\xi$-component of the current density. Since for parallel flows the current surfaces coincide with the magnetic surfaces, the poloidal current is zero on the magnetic axis, but in the off-axis region $\widetilde{J}_\xi$ exhibits two extema and its absolute value increases with $\sigma _{d_a}$. As $\sigma_{d_a}$ takes higher values the extrema get closer to the magneic axis as shown in Fig. (4.6). 
  \begin{figure}
  \centering
    \includegraphics[width=3.5in]{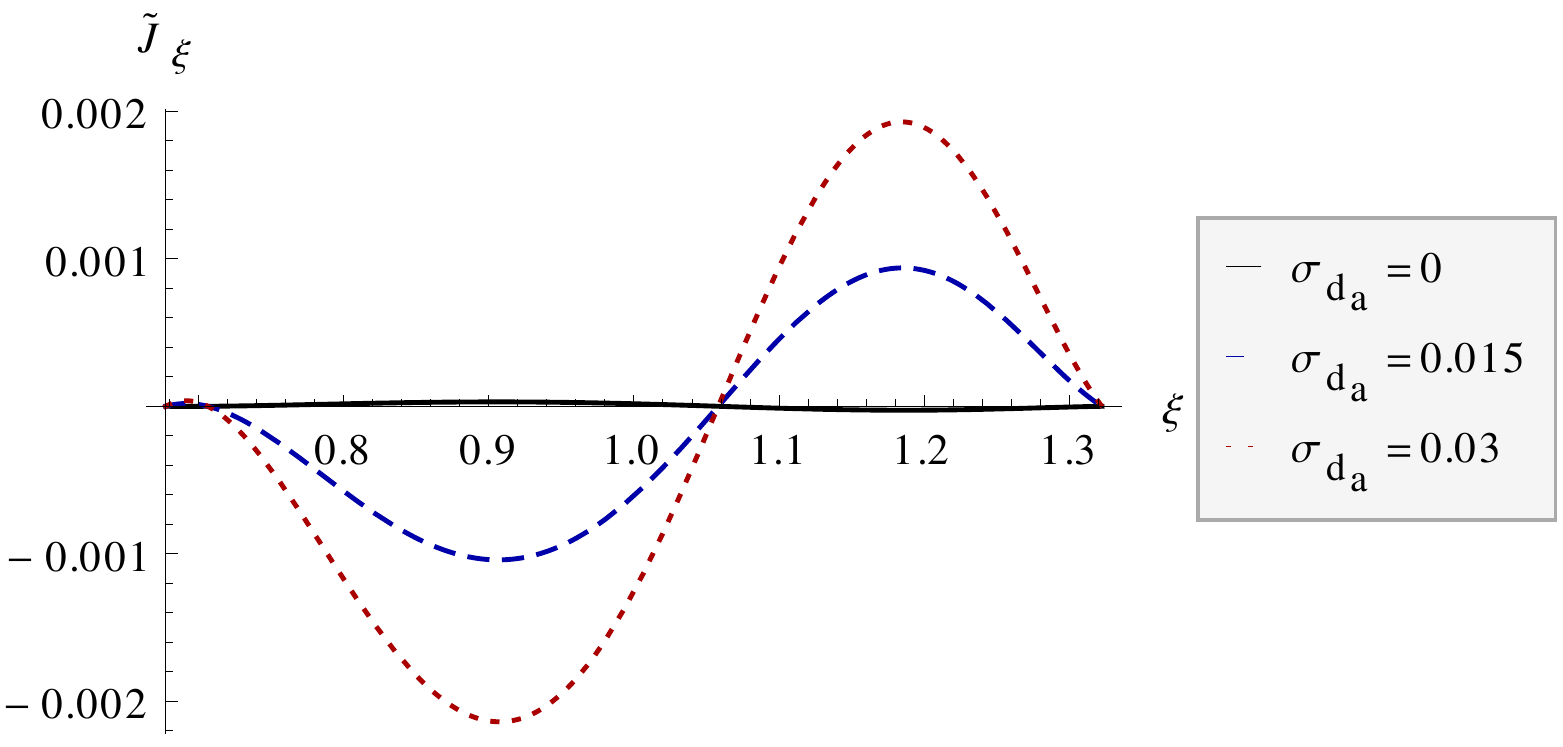}
     \caption{{\small \emph{As $\sigma_{d_a}$ increases the absolute value of $\widetilde{J}_\xi$ also increases; on the plane $\zeta =\zeta _a$ it exhibits two extrema localized near the magnetic axis.}}} 
     \end{figure}   
      Pressure anisotropy has also a more important impact than the flow on $\widetilde{J}_\xi$ as shown in Fig. (4.7), from the view of the plane $\zeta =0$.
     \begin{figure}
  \centering
    \includegraphics[width=3.5in]{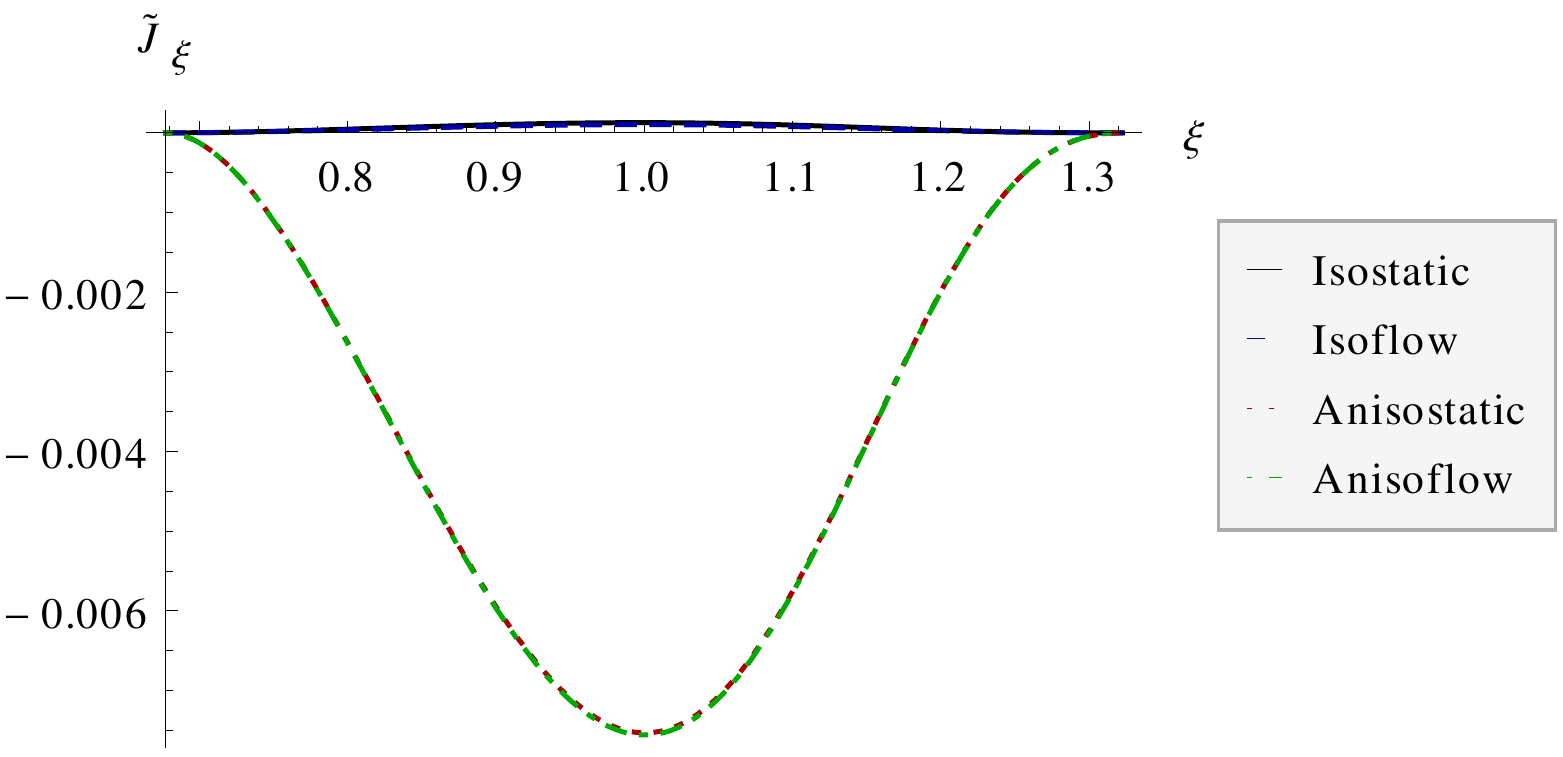}
     \caption{{\small \emph{Pressure anisotropy influence against that of the flow on $\widetilde{J}_\xi$, on the $\zeta=0$ plane. Its absolute value is peaked near the magnetic axis and increases with anisotropy.} }}
     \end{figure}   
      When $n$ gets larger values, $\widetilde{J}_\xi$ is not anymore peaked close to the axis, but its minimum gets closer to the outer point of the configuration as shown in Fig. (4.8).
 \begin{figure}
  \centering
    \includegraphics[width=3.5in]{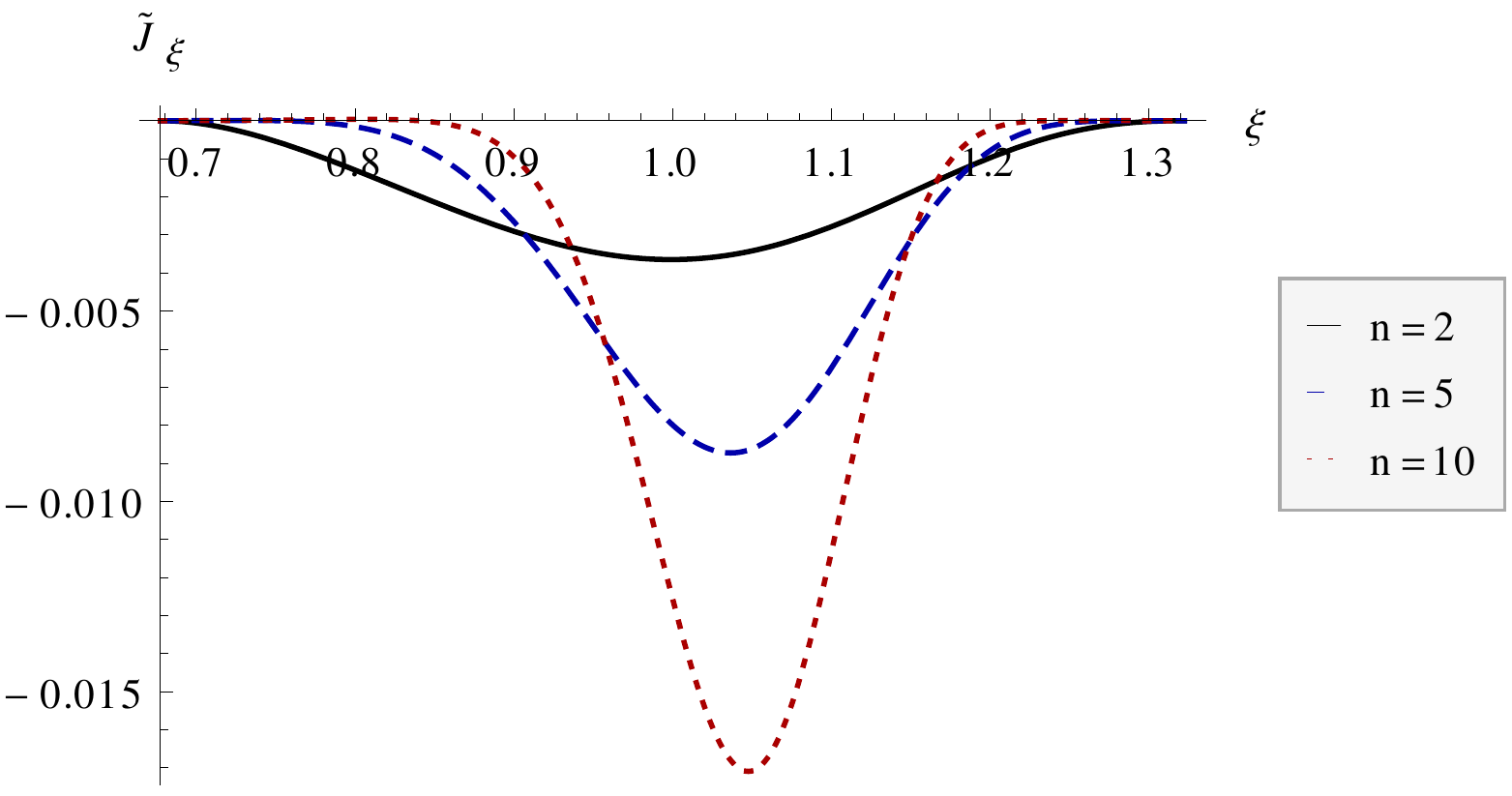}
     \caption{{\small \emph{The impact of the increase of anisotropy parameter $n$ on $\widetilde{J}_\xi$, for $\sigma_{d_a}=0.015$.}   }}   
\end{figure}
The component $\widetilde{J}_\phi$ is not monotonically increasing from $\xi_{in}$ to $\xi_{out}$ as in Solovev equilibria, but in contrast, it is peaked on the magnetic axis. In addition, $\widetilde{J}_\phi$ is not much affected by the change on anisotropy and flow -its value on axis increases a little with $\sigma _a$ and $M_{p_a}^2$.
\par
   The change on anisotropy has also an important influence on the various pressures, with the only exception of $\widetilde{\overline{p}}$. This is because the main contribution to the effective pressure is from the static one, $\widetilde{\overline{p}}_s$, which is not explicitly depended on pressure anisotropy or on plasma flow:
   \beq \label{phm}  \widetilde{\overline{p}}=\widetilde{\overline{p}}_s(\widetilde{u})-\widetilde{\rho}\frac{\widetilde{v}^2}{2}\eeq
   One can see that the only dependence on these parameters comes from the second (flow) term of relation (\ref{phm}). However, the impact of $\sigma_{d_a}$ and $M_{p_a}^2$ on $\widetilde{v}^2$ is negligible and therefore  the static pressure $\widetilde{\overline{p}}_s$ remains unaffected. On the other hand, one can expect from relations (\ref{avpress})-(\ref{pparallel}), that $\widetilde{p}_{\bot}$ and $<\widetilde{p}>$ will decrease, while $\widetilde{p}_{\parallel}$ will increase with pressure anisotropy. The variation of the average pressure with $\sigma_{d_a}$ on the midplane is given on Fig. (4.9). 
    \begin{figure}
  \centering
    \includegraphics[width=3.5in]{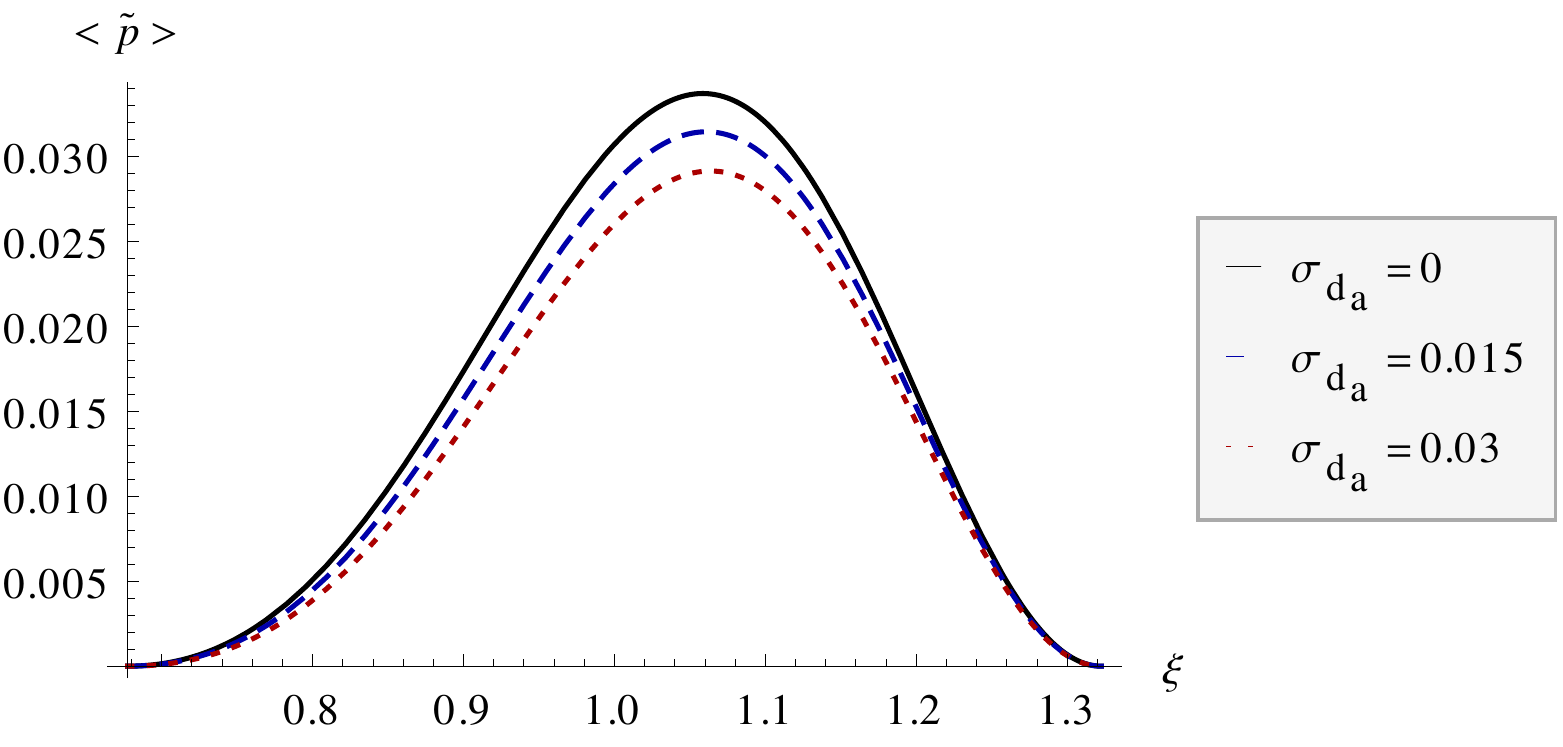}
     \caption{{\small \emph{The impact of anisotropy parameter $\sigma_{d_a}$ on $<\widetilde{p}>$ on the midplane $\zeta=0$, for $n=2$.}  }}    
\end{figure}  
In addition, pressure anisotropy has a stronger effect on $<\widetilde{p}>$ than that of the flow, as shown in Fig. (4.10).
 \begin{figure}
  \centering
    \includegraphics[width=3.5in]{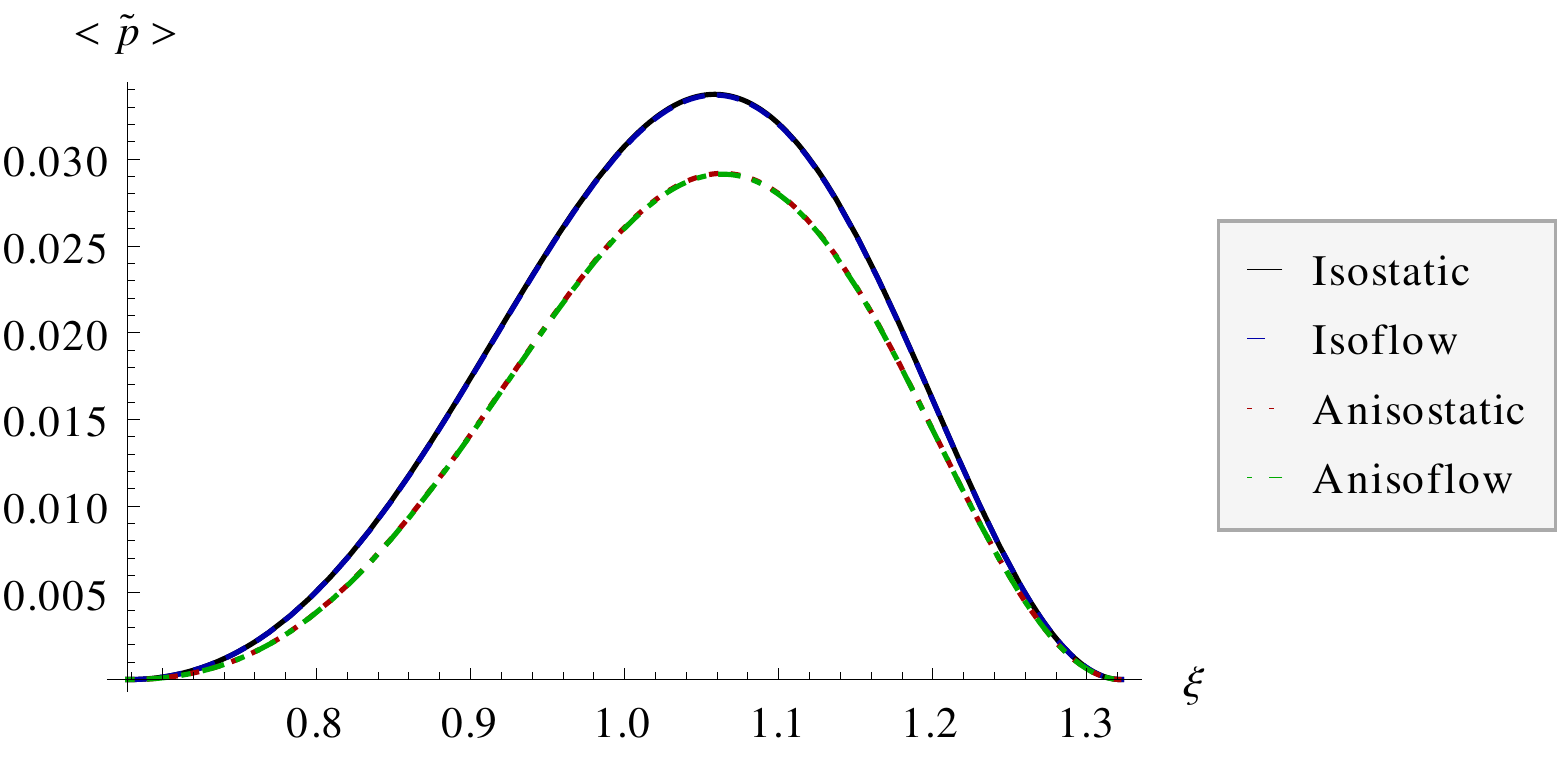}
     \caption{{\small \emph{The impact of pressure anisotropy against that of the flow on $<\widetilde{p}>$ on the midplane $\zeta=0$.} }}     
\end{figure}
Furthermore, the ratio $\frac{\widetilde{p}_{\parallel}}{\widetilde{p}_{\bot}}\approx 1.5$ is higher from that of the Solovev ITER diamagnetic equilibria. Profiles of the various pressures for the same values of $\sigma_{d_a}$ and $n$ are shown in Fig. (4.11).
The variation of flow and anisotropy parameters has no important influence on the rest quantities ($\widetilde{v}_\phi$, $\widetilde{v}_\xi$, $\widetilde{v}_\zeta$, $\widetilde{B}_\xi$, $\widetilde{B}_\zeta$, $\widetilde{J}_\zeta$).
      \begin{figure}
  \centering
    \includegraphics[width=3.5in]{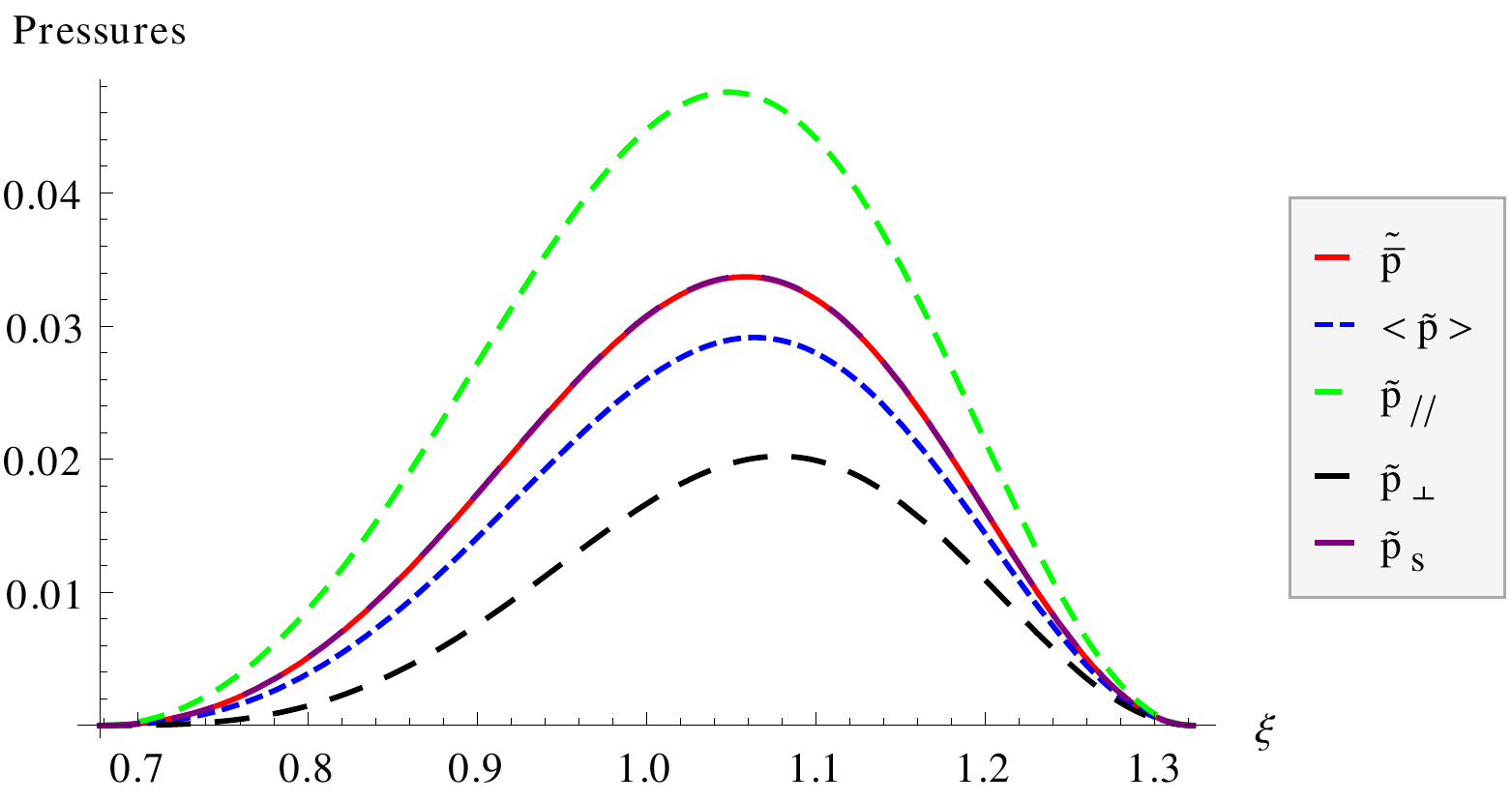}
     \caption{{\small \emph{Profiles of the various normalized pressures for ITER diamagnetic equilibria on the plane $\zeta=0$, for $\sigma_{d_a}=0.015$ and $n=2$ .}    }}  
\end{figure}
The confinement figures of merit are also weakly affected by the change on pressure anisotropy; more precisely, the local toroidal beta on the magnetic axis is
\beq \label{btorhm} \beta _{t_a}=2\widetilde{\overline{p}}_a\approx 0.0677-\frac{0.000089}{1-\sigma _{d_a}}\eeq
Thus, $\beta_{t_a}\approx 6.77\%$, almost unaffected by $\sigma_{d_a}$, while the safety factor is monotonically increasing from the magnetic axis to the plasma boundary and is not affected by the variation of $\sigma_{d_a}$, as shown in Fig. (4.12).
 \begin{figure}
  \centering
    \includegraphics[width=3.5in]{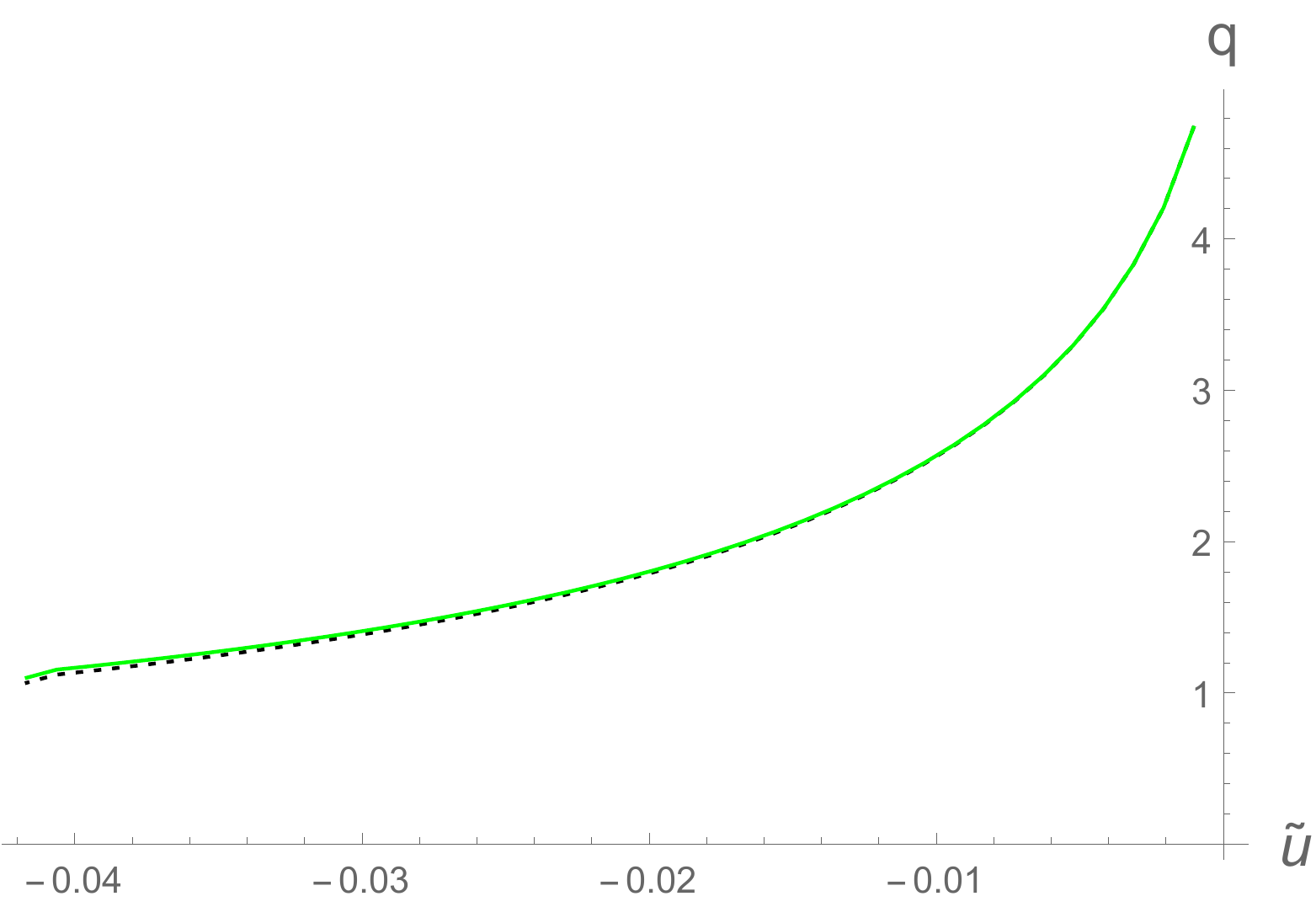}
     \caption{{\small \emph{The profile of safety factor both for $\sigma_{d_a}=0$ (dotted curve) and $\sigma_{d_a}=0.03$ (continuous curve) is monotonically increasing from axis to boundary.}  }} 
     \end{figure}
     
     \section{NSTX-Upgrade Diamagnetic Configuration}
     
\hspace{2em}In order to assign the free parameters for the function $\widetilde{u}^*$, we choose $j_{max}=6$ in the solution (\ref{supersol4}), and thus, we need to solve a system of twenty-three algebraic equations with equal number of unknown coefficients. For this reason we take into account the conditions (\ref{c1})-(\ref{c10}), as well as the condition $\widetilde{u}^*=0$ for ten more boundary points obtained by the relations (\ref{kui1})-(\ref{kui3}). Setting by inspection the values of the free parameters $\widetilde{p}_2=4.825$, $\widetilde{X}_1=-0.65$, and solving the system of the above equations, we find the values for the twenty-three coefficients of the solution for an NSTX-U diamagnetic configuration, given on Table (4.2).
    \begin{table}[ht]
\centering
 \begin{tabular}{c c}
\hline\hline
 & Coefficient Value  \\ [1ex] 
\hline 
$a_2^*$ & 0.037643 \\
$a_3^*$ & 0.014008  \\
$a_4^*$ & -0.00116133  \\
$a_5^*$ & 0.000522584  \\
$a_6^*$ & -0.0000338934  \\
$b_1^*$ & -0.369022   \\
$b_2^*$ & 0.125383   \\ 
$b_3^*$ & -0.0193287  \\ 
$b_4^*$ & 0.00040103  \\
$b_5^*$ & 0.000538235  \\ 
$b_6^*$ & -0.0000481594  \\
$c_1^*$ & 2.24895  \\
$c_2^*$ & 0.303034 \\
$c_3^*$ & -0.482659  \\
$c_4^*$ & 0.380529  \\
$c_5^*$ & 0.113029  \\
$c_6^*$ & -0.000999854  \\
$d_1^*$ & -1.23871 \\
$d_2^*$ & 0.978195  \\
$d_3^*$ & -0.92549  \\
$d_4^*$ & 0.881581  \\
$d_5^*$ & 0.355572  \\
$d_6^*$ & -0.00912552  \\[1ex]
\hline 
\end{tabular}
\caption{{\small \emph{Values of the coefficients of the solution $\widetilde{u}^*$ for an NSTX-Upgrade diamagnetic configuration. }}}
\end{table}
Then, by solving the equations $Im[\widetilde{u}_{\xi}^*]=0$ and $Im[\widetilde{u}_{\zeta}^*]=0$ we find that the magnetic axis  is located at the position $(\xi_a=1.19012,\zeta_a=0.0511745)$, while the flux function on axis takes the value $\widetilde{u}_a^*=0.948259$. Furthermore, we find the expression of the safety factor on the magnetic axis as
\beq q_a=0.161002\sqrt{\frac{1}{c^2}-0.584477}\eeq 
which for $q_a=1.1$ gives $c=0.145457$. The magnetic axis of the configuration is at $\widetilde{u}_a=0.137931$. The poloidal cross-section with a set of magnetic surfaces for such an NSTX-U diamagnetic configuration is illustrated in Fig. (4.13).
\begin{figure}
  \centering
    \includegraphics[width=3.5in]{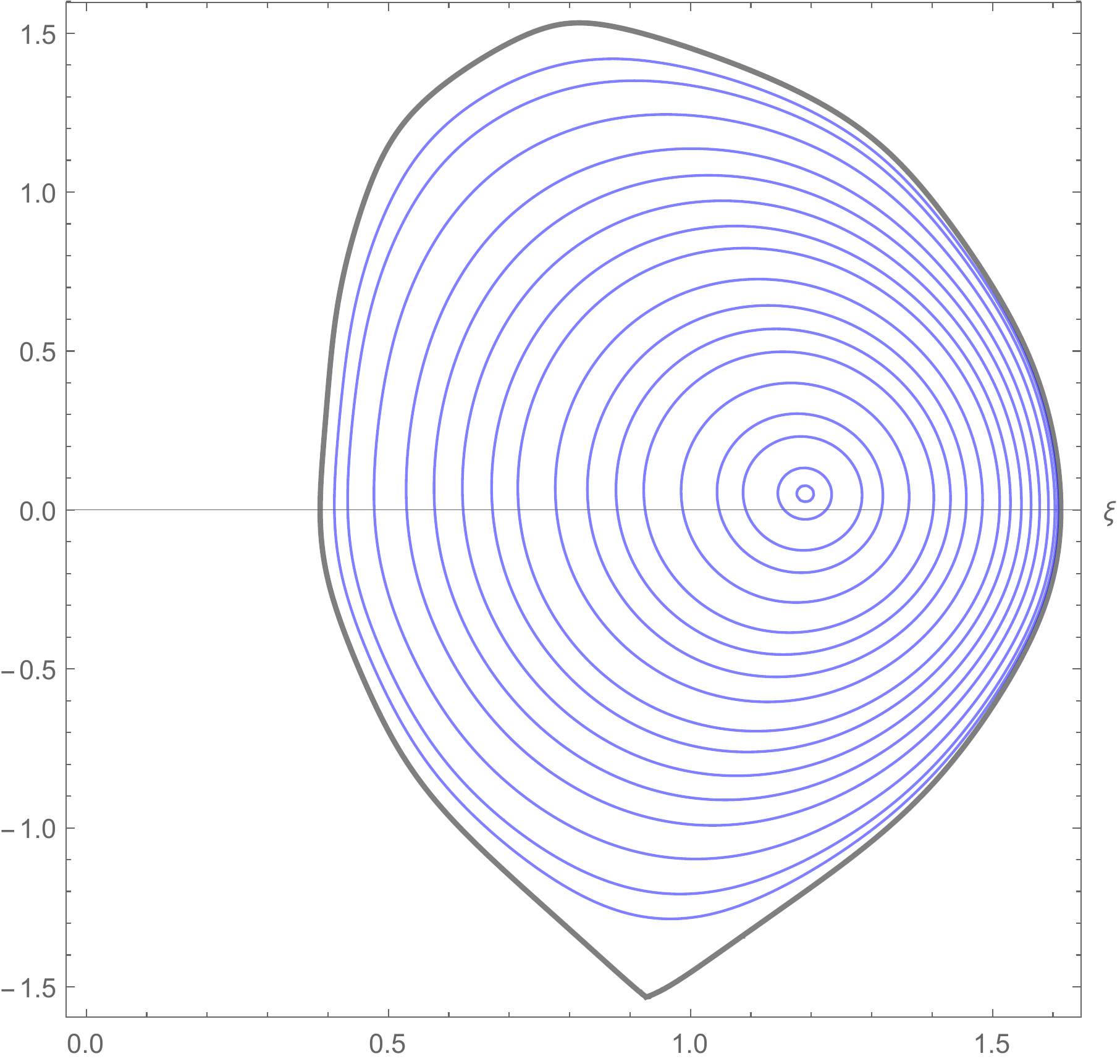}
     \caption{{\small \emph{The poloidal cross-section with closed magnetic surfaces for an NSTX-U diamagnetic configuration. The outermost black-colored curve corresponds to the separatrix}}}   
     \end{figure}
     
  \subsection{Effects of Pressure Anisotropy and Flow}
  
\hspace{2em}In order for the pressure to remain positive throughout the whole region of the plasma, we find that the maximum attainable value of the anisotropy parameter on axis is $\sigma_{d_a}=0.017$. This value is lower than the corresponding one for ITER diamagnetic equilibria in contrast to the order of the respective maximum permissible Solovev-equilibria values of $\sigma_{d_a}$ for the two tokamaks. Thus, the free parameters of pressure anisotropy will be let to vary in the intervals $0\leq \sigma_{d_a} \leq 0.017$, and $2\leq n \leq 10$.
 \par
 The main observation is that both pressure anisotropy and flow affect the equilibrium in the same way as on the ITER diamagnetic configuration, but since $M_{p_a}^2$ is two orders of magnitude  larger than on ITER, plasma flow has a greater influence on the NSTX-U, and its effects are more noticeable
 thereon. This can be seen for example from the effective pressure profile in Fig. (4.14).
 \begin{figure}
  \centering
    \includegraphics[width=3.5in]{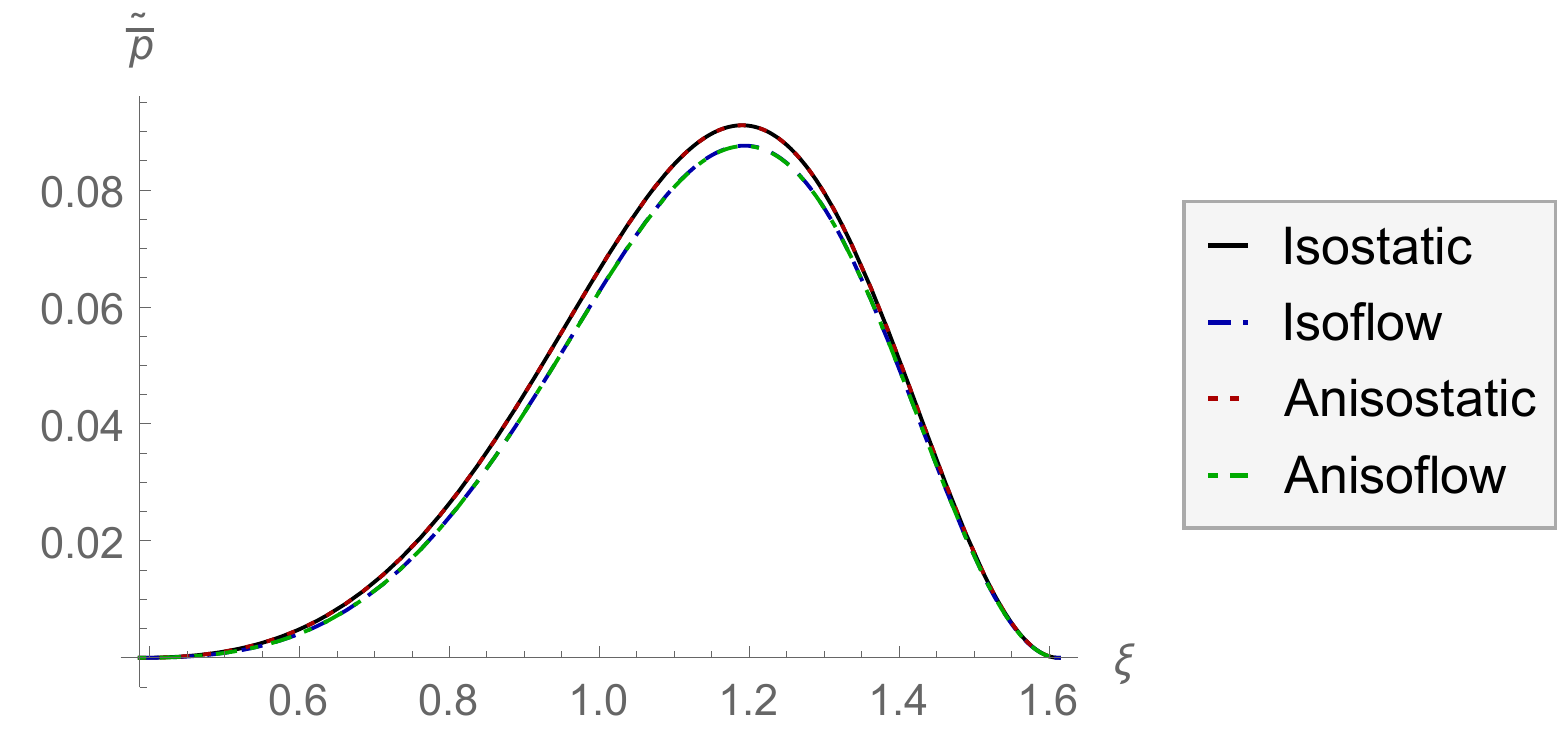}
     \caption{{\small \emph{The impacts of pressure anisotropy and plasma flow on $\widetilde{\overline{p}}$ on the midplane $\zeta=0$.}  }} 
     \end{figure}
     As we can see, the plots are separated by the flow/no-flow criterion, and this is because $M_{p_a}^2$ has some impact on $\widetilde{v}^2$, while the respective impact of $\sigma_{d_a}$ is rather negligible.
     \par
     Additionally, both anisotropy and flow have the same influence on the quantities $\widetilde{I}$ and $\widetilde{J}_\xi$ as on ITER diamagnetic equilibria. When they are both present, they increase the magnetic field inside the plasma resulting to a paramagnetic action, while they also make $|\widetilde{J}_\xi|$ enhance, as shown in Figures (4.15) and (4.16). In general, the effects of the variation of $\sigma_{d_a}$ and $M_{p_a}^2$ are similar to the respective ones in the Solovev-like solution.
  \begin{figure}
  \centering
    \includegraphics[width=3.5in]{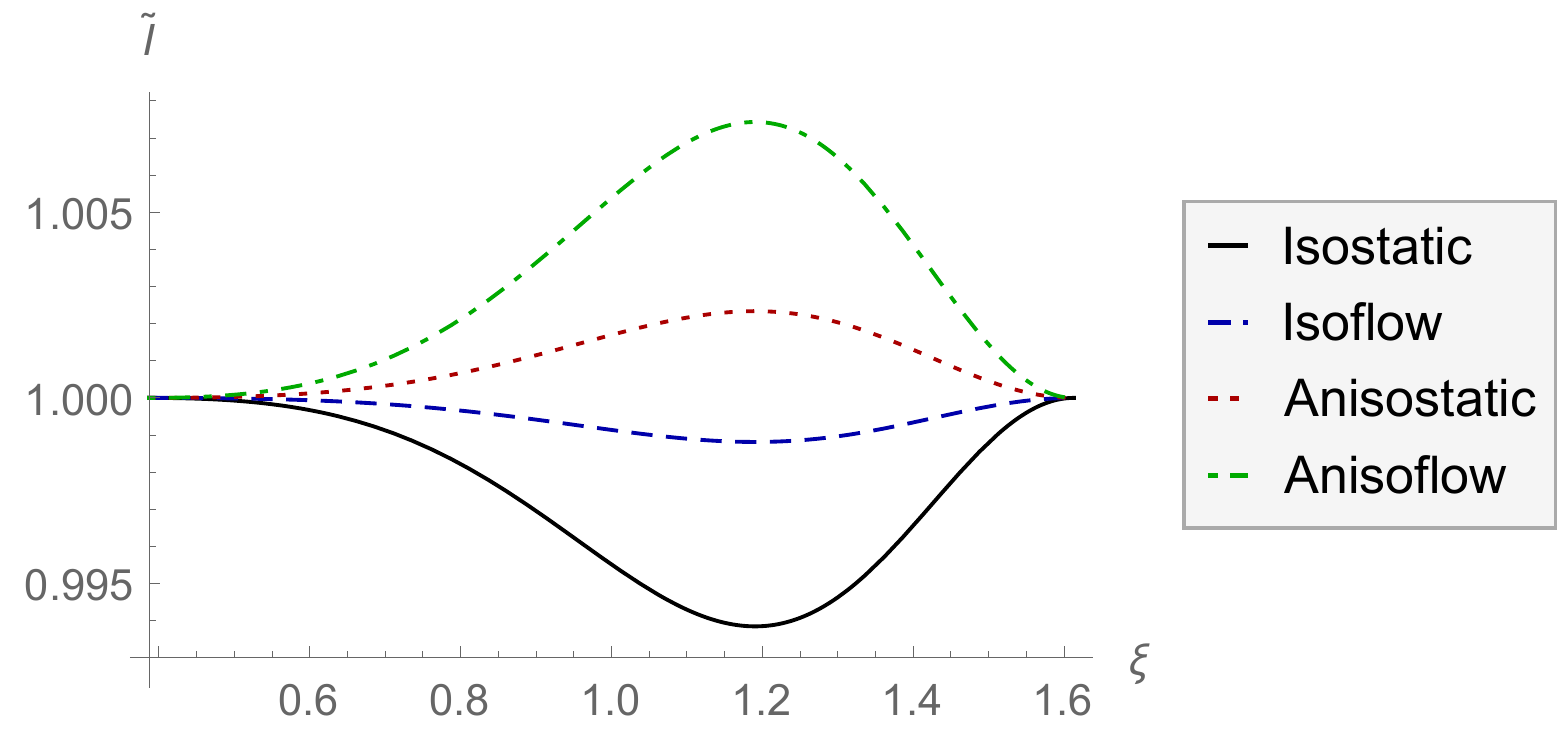}
     \caption{{\small \emph{The paramagnetic action of pressure anisotropy and plasma flow on $\widetilde{I}$ on the midplane $\zeta=0$.}}}   
     \end{figure}
      \begin{figure}
  \centering
    \includegraphics[width=3.5in]{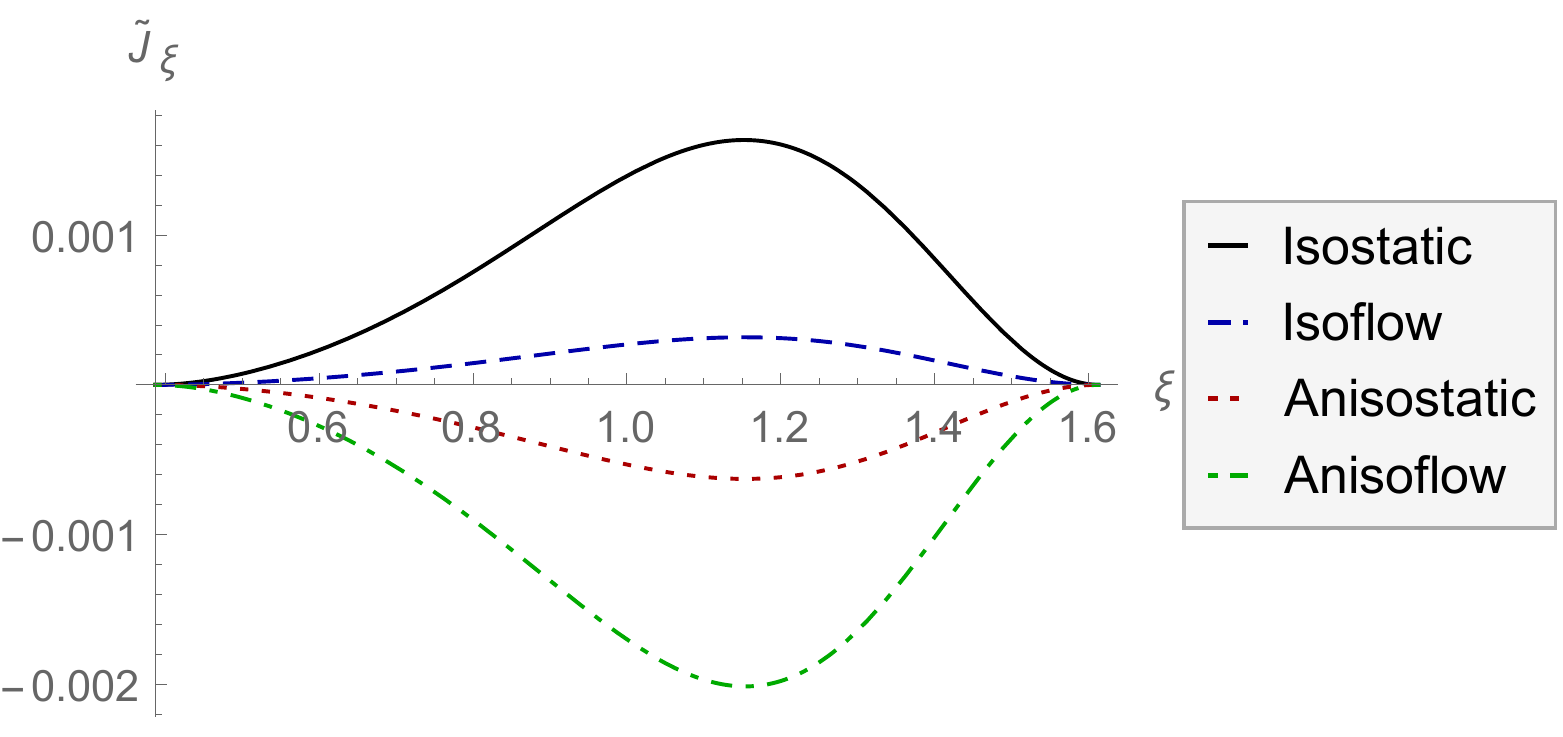}
     \caption{{\small \emph{The impacts of pressure anisotropy and plasma flow on $\widetilde{J}_\xi$ on the midplane $\zeta=0$. As $\sigma_{d_a}$ and $M_{p_a}^2$ increase, $\widetilde{J}_\xi$ changes sign and its absolute value also increases.  }}}   
     \end{figure}
Anisotropy through $\sigma_{d_a}$ has some influence on the parallel and perpendicular pressures, as the first increases, while the second one decreases when $\sigma _{d_a}$ takes larger values.
     The ratio of these two pressures for the same value of $\sigma_{d_a}$ is $\frac{\widetilde{p}_{\parallel}}{\widetilde{p}_{\bot}}\approx 1.08$, which is lower than on ITER and similar to Solovev diamagnetic equilibria for the two kinds of tokamaks. In contrast, the variation of anisotropy parameter $n$ does not have an important influence on these quantities. Furthermore, the ratio of the maximum average pressure for the two configurations is $\frac{<\widetilde{p}>_{NSTX-U}}{<\widetilde{p}>_{ITER}}\approx 2.73$, a value which is near to the respective Solovev one. \par
     The local toroidal beta on the magnetic axis depends on $\sigma_{d_a}$ as 
     \beq \label{bNSTXU} \beta _{t_a}=0.1836-\frac{0.00697}{1-\sigma _{d_a}}\eeq
     so that $\beta _{t_a}\approx 18.4\% $ for NSTX-U diamagnetic equilibria, in agreement with experimental results. Also, the safety factor is monotonically increasing from the magnetic axis to the plasma boundary as shown in Fig. (4.17).
    \begin{figure}
  \centering
    \includegraphics[width=3.5in]{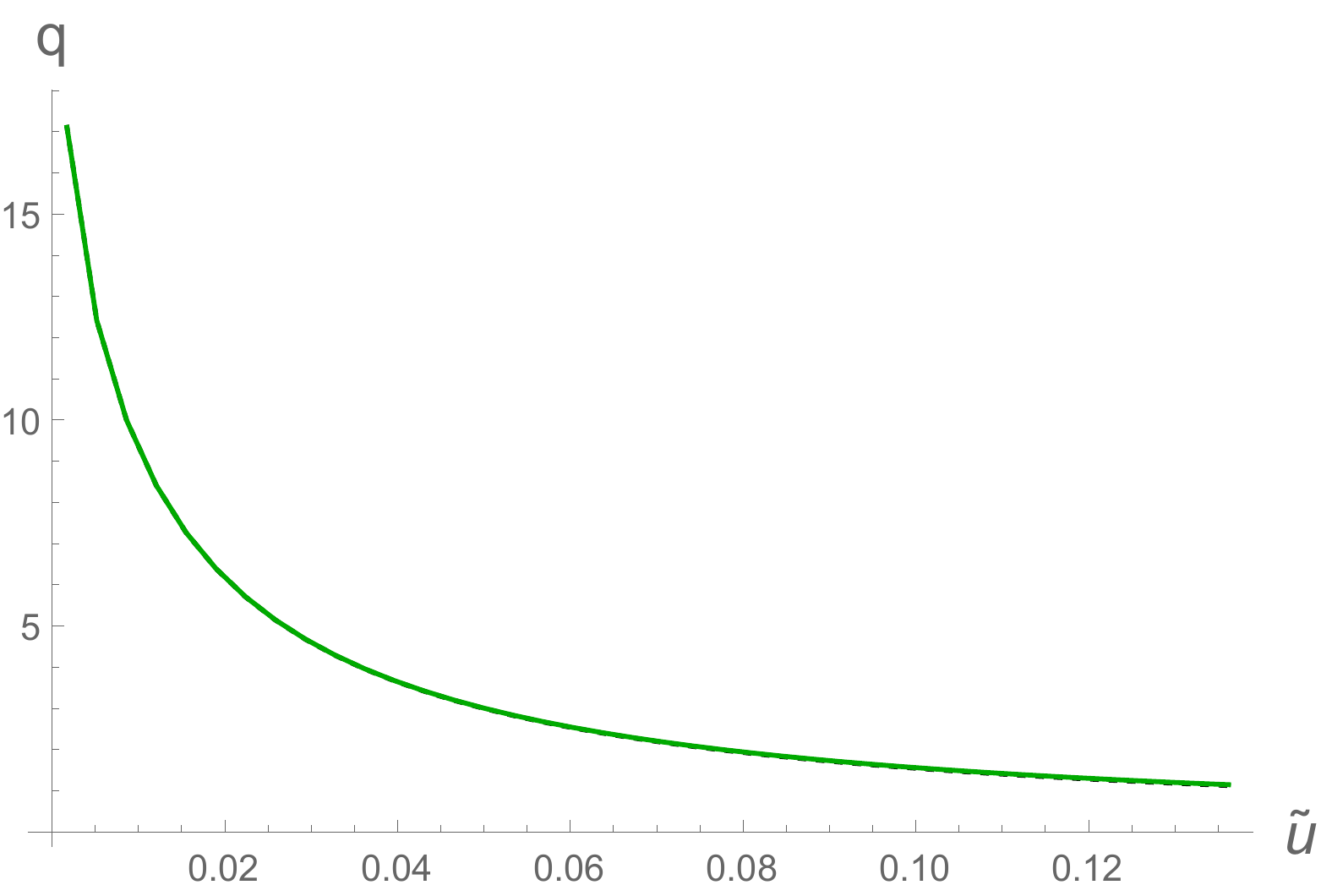}
     \caption{{\small \emph{The safety factor profile is monotonically increasing from the magnetic axis to the plasma boundary and is not affected by the change on pressure anisotropy.  }}}   
     \end{figure}  
   Both pressure anisotropy and plasma flow have a minimal effect on the rest equilibrium quantities examined.
   \newpage  
   
 \section{Conclusions}  
 
\hspace{2em}In the present chapter Hernegger-Maschke-like equilibria were constructed for quadratic in $\widetilde{u}$ choices of the free functions involved in the linearised GGS equation (\ref{hmlinear}). Appropriate boundary conditions were imposed to construct fixed boundary equilibria with a diverted boundary possessing an X-point. On the basis of the solution obtained we examined the influence both of pressure anisotropy and plasma flow on equilibrium quantities and figures of merit, for ITER and NSTX-Upgrade diamagnetic configurations and for field-aligned flow. The main conclusions are summarized below.
 
 \begin{enumerate}
 \item In general, pressure anisotropy has a stronger impact on equilibrium than the flow, mainly because the pertinent maximum permissible parametric values are higher than the ones for the flow. Specifically for an ITER tokamak the maximum attainable value for the anisotropy parameter $\sigma_{d_a}$ is approximately 0.03, which is two orders of magnitude higher than the Mach number $M_{p_a}^2$. For the NSTX-U though the respective values of $\sigma_{d_a}$ and $M_{p_a}^2$ are comparable, the effect of $\sigma_{d_a}$ on equilibrium remains stronger. However, the flow effects are more pronounced than on ITER because of the higher NSTX-U values of $M_{p_a}^2$. The only exceptions are the effective pressure and the components of the velocity field, in which the flow has a stronger influence than anisotropy. A peculiarity of the Hernegger-Maschke solution is that the maximum attainable value of parameter $\sigma_{d_a}$ for the spherical tokamak is lower than that of the conventional one.
 \item Similar to the case of Solovev-like equilibria $\sigma_{d_a}$ and $M_{p_a}^2$ have an additive paramagnetic action, increasing the magnetic field inside the plasma, while the parameter $n$ associated with the peakedness of the anisotropy function profile acts diamagnetically.
 \item The equilibrium quantities mostly affected by pressure anisotropy are the $\xi$-component of current density and the various pressures. Specifically, the isotropic $\widetilde{J}_\xi$ being positive- with smaller ITER values than NSTX-U ones- in the presence of anisotropy changes sign. The larger the maximum $|\widetilde{J}_\xi|$ is the higher $\sigma_{d_a}$. In addition, increase of $\sigma_{d_a}$ makes $\widetilde{p}_{\parallel}$ to take higher values but $\widetilde{p}_{\bot}$ and $<\widetilde{p}>$ lower ones. At last, as in the diamagnetic Solovev-like equilibria, the ratio $\frac{\widetilde{p}_{\parallel}}{\widetilde{p}_{\bot}}$ is higher for ITER equilibria than that for NSTX-U ones.
 \end{enumerate}
 \newpage\null\newpage
 
 \chapter{General Conclusions and Future Prospects}
 \section{Conclusions}
 \hspace{2em}In the present thesis we studied the equilibrium properties of a magnetically confined axisymmetric toroidal plasma with pressure anisotropy and incompressible flow of arbitrary direction. Also, we examined the impact of both anisotropy and flow on ITER-like configurations as well as on NSTX and NSTX-Upgrade spherical tokamak ones, both paramagnetic and diamagnetic.
 \begin{enumerate}
 \item In the first part of the study we derived a generalised Grad-Shafranov equation [Eq. (\ref{GGS psi})] governing axisymmetric plasma equilibria in the presence of pressure anisotropy and flow. To this end we adopted a diagonal pressure tensor with one element parallel to the magnetic field, $p_{\parallel}$, and two equal perpendicular ones, $p_{\bot}$. As a measure of the pressure anisotropy we introduced the function $\sigma_d=\mu_0\frac{p_{\parallel}-p_{\perp}}{B^2}$, assumed to be uniform on magnetic surfaces,
 while the flow is expressed by the poloidal Alfv\' enic Mach function $M_{p}=\frac{v_{pol}}{v_{A_{pol}}}$, where $v_{A_{pol}}$ is the Alfv\' en velocity. This equation recovers known GS-like equations governing static anisotropic equilibria and isotropic equilibria with plasma flow. Also for static isotropic equilibria the equation is reduced to the usual well known GS equation.
The form of the equation containing the sum $M_{p}^2+\sigma_d$ indicates that pressure anisotropy and flow act additively with the only exception the electric field term. In addition we derived a generalised Bernoulli equation [Eq. (\ref{Bernoullipsi})] involving the function $\overline{p}=\frac{p_{\parallel}+p_{\bot}}{2}$ which may be interpreted as an effective isotropic pressure.
 \item In the second part of the study two parametric linear equilibrium solutions of the GGS equation were derived for appropriate choices of the free functions appearing in the equation. They consist extensions of the most important static isotropic solutions widely employed for tokamak studies. Specifically, we derived an extended Solovev solution describing configurations with a non-predefined boundary, and an extended Hernegger-Maschke solution with a fixed boundary possessing an X-point imposed by appropriate boundary conditions. On the basis of these solutions we examined ITER, NSTX and NSTX-U equilibria for arbitrary flows, both diamagnetic and paramagnetic. Furthermore, we examined the impact of anisotropy -through the parameters $\sigma_{d_a}$ and $n$, defining the maximum value and the shape of the function $\sigma_d$- and flow -through the Alfv\' enic Mach number $M_{p_a}^2$ defining the maximum of the function $M_p^2$- on their characteristics and came to the following conclusions. 
 \begin{enumerate}
 \item Pressure anisotropy has a stronger impact on equilibrium than that of the flow. This is because the maximum permissible values of $\sigma_{d_a}$ are in general higher than the respective $M_{p_a}^2$ ones. The effects of the flow are more noticeable in the STs since the values of $M_{p_a}^2$ are comparable with the $\sigma_{d_a}$ ones, and larger than the respective ITER ones. In general, anisotropy in connection with the requirement of positiveness of the pressure within the plasma region, is higher in spherical tokamaks than in ITER configurations -both for diamagnetic and paramagnetic Solovev equilibria-, and vice versa for the Hernegger-Maschke diamagnetic solution. This may be a peculiarity of the Hernegger-Maschke solution since anisotropy is believed to play a more important role in spherical tokamaks than in conventional ones.
 \item In addition, we found that anisotropy and flow through the parameters $\sigma_{d_a}$ and $M_{p_a}^2$ have an additive paramagnetic impact on equilibrium, while anisotropy through $n$ acts diamagnetically. The paramagnetic effects of anisotropy are stronger on the spherical tokamaks.
 \item Pressure anisotropy has an appreciable impact on equilibrium quantities such as the current density and the various pressures. Specifically, the toroidal current density for Solovev equilibria, monotonically increasing from the high to low field side, presents different behaviour with $\sigma_{d_a}$ in different regions of the plasma, while for a Hernegger-Maschke solution $J_\phi$ is peaked on the magnetic axis and slightly increases with anisotropy. Also, the anisotropic poloidal components of the current density ($J_z$ in Solovev and $J_R$ in Hernegger-Maschke equilibria) present two extrema with their absolute values increasing with $\sigma_{d_a}$. In addition, $p_{\perp}$ and $<p>$ decrease, and $p_{\parallel}$ increase with $\sigma_{d_a}$ in comparison with isotropic pressure, while $\overline{p}$ is slightly affected by pressure anisotropy and more by plasma flow.
 \item In diamagnetic equilibria the toroidal velocity is peaked on the magnetic axis and slightly increase with $\sigma_{d_a}$, while the paramagnetic $v_\phi$ reverses near the axis of symmetry and then behaves as the diamagnetic one to the right of the reversal point. In spherical tokamaks the reversal point is displaced closer to the magnetic axis and $v_\phi$ remains positive in a larger region than in the conventional ITER one. At last, the rest of the equilibrium quantities and confinement figures of merit as the local beta on axis and the safety factor are almost insensitive to anisotropy.
 \end{enumerate}

 \end{enumerate}
 
\par
Let us finally note that complete understanding of the equilibrium with plasma flow and pressure anisotropy requires substantial additional work in connection with compressibility, alternative potentially more  pertinent  physical assumptions on the functional dependence of the anisotropy function $\sigma_d$ and more realistic numerical solutions. However, in these cases   the  reduced equilibrium equations are expected to  be much more complicated compared with the relative simple GGS derived in the present study which contributes to understanding  the underlying physics. 
 
 \section{Future Prospects}

 \hspace{2em}It is interesting to extend the present thesis in connection to the following projects:
 \begin{enumerate}
 \item Further generalisation of the Solovev-like solution for diverted boundaries by introducing additional terms with an arbitrary number of free parameters \cite{Sri},\cite{Cerfon},\cite{TROUM}.
 \item Study of ``compressible'' equilibria with pressure anisotropy, toroidal flow and density varying on magnetic surfaces.
 \item On the basis of the GGS equation obtained, isotropic equilibrium codes as HELENA \cite{HELENA} can be extended or novel codes can be developed. The analytic solutions constructed can then be employed for code benchmarking. 
 \item Possible generalisation of the sufficient condition for linear stability of Ref. \cite{stability} in the presence of pressure anisotropy and plasma flow. Then for parallel flow the extended condition could be applied to the analytic equilibria constructed here.
 \item Generalisation of the GGS equation obtained here to the more generic class of helically symmetric equilibria.
 \item Possible extension of the papers on static equilibria with reversed current density \cite{Wang}-\cite{Martins} in the presence of incompressible flow and pressure anisotropy.

 \end{enumerate}

 \appendix
\newpage\null\newpage
 \chapter{Safety factor profile}
 
 \hspace{2em}The general expression for the safety factor is
 \beq \label{A1} q=\frac{1}{2\pi}\left(\int_{0}^{2\pi}\frac{\widetilde{X}\sqrt{\widetilde{r}^2+\left(\frac{\widetilde{u}_\theta}{\widetilde{u}_r}\right)^2}}{\xi (1-\sigma _d-M_p^2)^{1/2}|\widetilde{\vec{\nabla}}\widetilde{u}|}d\theta -\int_{0}^{2\pi}\frac{\xi \left(\frac{d\widetilde{F}}{d\widetilde{u}}\right)\left(\frac{d\widetilde{\Phi}}{d\widetilde{u}}\right)\sqrt{\widetilde{r}^2+\left(\frac{\widetilde{u}_\theta}{\widetilde{u}_r}\right)^2}}{(1-\sigma _d-M_p^2)^{-1/2}|\widetilde{\vec{\nabla}}\widetilde{u}|}d\theta \right)\eeq
 This equation comes from the substitution of (\ref{Xu nor}) into (\ref{qu nor}), and will be used in order to numerically calculate the profile of the safety factor with Wolfram Mathematica suite, by developing a simple ``do loop'' programme.
 \par
 At first, we change variables with respect to the Shafranov coordinates $\xi=1+\widetilde{r}cos\theta$, $\zeta=\widetilde{r}sin\theta$, and then in connection with (\ref{A1}) we introduce the functions
 \begin{eqnarray} \label{A2}
 &&f_1=\frac{\sqrt{\widetilde{r}^2+\left(\frac{\widetilde{u}_\theta}{\widetilde{u}_r}\right)}}{\xi |\widetilde{\vec{\nabla}}\widetilde{u}|} \nonumber \\
 &&f_2=\frac{\xi \sqrt{\widetilde{r}^2+\left(\frac{\widetilde{u}_\theta}{\widetilde{u}_r}\right)}}{|\widetilde{\vec{\nabla}}\widetilde{u}|}
 \end{eqnarray}
 Next by using the analytical form of the solution $\widetilde{u}$ we find a number of points $(\xi_n,\zeta_n)$ corresponding to $(\widetilde{r}_n,\theta_n)$, and then we calculate the functions $f_{1_n}$, $f_{2_n}$ for each pair. The created numbers are put in a list, and after being interpolated we find the corresponding values of the safety factor $q_n$ from Eq. (\ref{A1}). Putting the pairs of $q_n$ and $\widetilde{u}_n$ in a list, and plotting them, we finally obtain the profile of the safety factor $q(\widetilde{u})$. The above described procedure is applied in the math programme given below.
 \begin{center}
 The math programme
 \end{center}

\begin{flushleft}

$\text{u}\xi=\frac{\partial u}{\partial \xi };$ $\text{u}\zeta=\frac{\partial u}{\partial \zeta };$ $\text{u}\xi \xi =\frac{\partial ^2u}{\partial \xi \partial \xi };$ $\text{u}\zeta \zeta=\frac{\partial ^2u}{\partial \zeta \partial \zeta };$
\newline
$\text{up}=\sqrt{\text{u}\zeta ^2+\text{u}\xi ^2};$
\newline

$\xi =r \cos (\theta )+1; \, \zeta =r \sin (\theta );$
\newline
$\text{ur}=\frac{\partial u}{\partial r}; \, \text{u}\theta =\frac{\partial u}{\partial \theta };$
\newline
$\text{f1}=\frac{\sqrt{r^2+\left(\frac{\text{u}\theta }{\text{ur}}\right)^2}}{\xi  \text{up}}; \, \text{f2}=\frac{\xi  \sqrt{r^2+\left(\frac{\text{u}\theta}{\text{ur}}\right)^2}}{\text{up}};$
\newline

$\text{Do}[\text{um}=N\left[\frac{c (m-1) \text{ub}}{\text{mmax}-1}\right];  \text{Do}[\theta n=N\left[\frac{2 \pi  (n-1)}{\text{nmax}-1}\right];\text{u}\theta =u\text{/.}\, \{\theta \to \theta n\};\text{a0}=a-\frac{a (\text{mmax}-m)}{\text{mmax}}; 
\text{sol}=\text{FindRoot}[\text{u}\theta =\text{um},\{r,\text{a0}\}];$
\newline
$\text{rn}=r\text{/.}\, \text{sol}[[1]];$
\newline
$\text{fn1}=\text{f1}\text{/.}\, \{r\to \text{rn},\theta \to \theta n\};  \text{PutAppend}[\theta n,\text{fn1},\text{fn1.txt}];  \text{fn2}=\text{f2}\text{/.}\, \{r\to \text{rn},\theta \to \theta n\};  \text{PutAppend}[\theta n,\text{fn2},\text{fn2.txt}];$
\newline
$\xi n=\xi \text{/.}\, \{r\to \text{rn},\theta \to \theta n\};\zeta n=\zeta \text{/.}\, \{r\to \text{rn},\theta \to \theta n\};
\text{PutAppend}[\xi n,\zeta ,\text{dd2.txt}],\{n,1,\text{nmax}\}];  \text{lfn1}=\text{ReadList}[\text{fn1.txt},\{\text{Number},\text{Number}\}];$
\newline
$\text{infn1}=\text{Interpolation}[\text{lfn1}];\text{intfn1}=\text{NIntegrate}[\text{infn1}(\text{th}),\{\text{th},0,2 \pi \}]
;\text{lfn2}=\text{ReadList}[\text{fn2.txt},\{\text{Number},\text{Number}\}];$
\newline
$\text{infn2}=\text{Interpolation}[\text{lfn2}];  \text{intfn2}=\text{NIntegrate}\left[\text{infn2}(\text{th}),\left\{\text{th},\frac{1}{10^6},2 \pi -\frac{1}{10^6}\right\}\right];$
\newline
$\text{q1}=\frac{\text{intfn1} (X\text{/.}\, \{\text{u1}\to \text{um},\sigma a\to \sigma a1,w\to \text{w1}\})}{2 \pi  \sqrt{G\text{/.}\, \{\text{u1}\to \text{um},\sigma a\to \sigma a1,w\to \text{w1}\}}}-\frac{\text{intfn2} (W\text{/.}\, \{\text{u1}\to \text{um},\sigma a\to \sigma a1,w\to \text{w1}\})}{2 \pi  \frac{1}{\sqrt{G\text{/.}\, \{\text{u1}\to \text{um},\sigma a\to \sigma a1,w\to \text{w1}\}}}};$
\newline
$\text{PutAppend}\left[\frac{\text{um}}{\text{ub}},\text{q1},\text{qd3.txt}\right];$
\newline
$\text{DeleteFile}[\text{fn1.txt}];\text{DeleteFile}[\text{fn2.txt}],\{m,1,\text{mmax}\}]$
\newline

$\text{lq1}=\text{ReadList}[\text{qd3.txt},\{\text{Number},\text{Number}\}];$
\newline
$\text{pq1}=\text{ListPlot}[\text{lq1},\text{Joined}\to \text{True},\text{PlotRange}\to \text{All},\text{AxesOrigin}\to \{0,0\},\text{PlotStyle}\to \{\text{Blue},\text{Dotted},\text{Thick}\}];$
\newline
$\text{g1}=\text{Show}[\text{pq1},\text{AxesLabel}\to \{\frac{\tilde{u}}{\tilde{u}_b},\text{q}\}]$
 \begin{figure}
  \centering
    \includegraphics[width=3.5in]{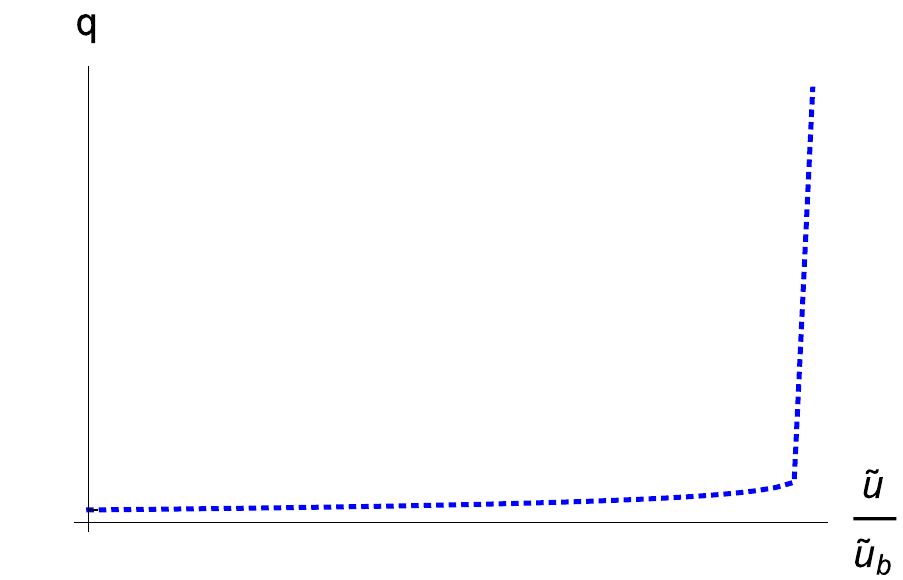}
     \caption{{\small \emph{Safety factor profile which is monotonically increasing from axis to boundary.  }}}   
     \end{figure}  
  
\end{flushleft}  
     
 \chapter{Calculation of the safety factor on the magnetic axis}
 
 \hspace{2em}Assume a static or stationary equilibrium plasma state described by the poloidal flux function $\psi(R,z)$, where $(R,z,\phi)$ are the cylindrical coordinates. By expanding $\psi$ in the vicinity of the magnetic axis $(R=R_m,z=z_m)$ on which $\psi$ has an extremum, we get
 \beq \label{B1} \psi (R,z)=\psi _m+\frac{1}{2}[A(R-R_m)^2+B(z-z_m)^2]+...\eeq
 where $A=[\frac{\partial^2\psi}{\partial R^2}]_{R=R_m,z=z_m}$, $B=[\frac{\partial^2\psi}{\partial z^2}]_{R=R_m,z=z_m}$, and $\psi_m=[\psi(R,z)]_{R=R_m,z=z_m}$.
 \par
 For a magnetic surface ($\psi=$const.) very close to the magnetic axis we take from (\ref{B1})
 \beq \label{B2} \frac{(R-R_m)^2}{a^2}+\frac{(z-z_m)^2}{b^2}=1\eeq
 where $a^2=\frac{2(\psi-\psi_m)}{A}$ and $b^2=\frac{2(\psi-\psi_m)}{B}$. Eq. \ref{B2}) implies that the intersection of a magnetic surface with the poloidal plane is an ellipse. Also, from the definition of $a^2$, and $b^2$ we observe that
 \beq \label{B3} Aa^2=Bb^2\eeq
 In order to calculate $q_a$ we will make use of Eq. (2.68) for the safety factor
 \beq \label{B4} q=\frac{1}{2\pi}\oint \frac{Idl}{R|\vec{\nabla}\psi|}\eeq
 where $\oint dl$ is the line integral along the ellipse, with $dl=dR\left[1+\left(\frac{dR}{dz}\right)^2\right]^{1/2}$. 
 \par
 If we define $f(R)=1-\frac{(R-R_m)^2}{a^2}$, then from Eq. (\ref{B1}) we find
 \beq \label{B5} \frac{b^2}{a^4}(R-R_m)^2[f(R)]^{-1}=\left(\frac{a}{b}\right)^4\frac{1}{(dz/dR)^2}\eeq
 which on the basis of Eq. (\ref{B3}) becomes
 \beq \label{B6} \frac{b^2}{a^4}(R-R_m)^2[f(R)]^{-1}=\left(\frac{A}{B}\right)^2\frac{1}{(dz/dR)^2}\eeq
 From the definitions of $A$, and $B$ we also have
 \beq \label{B7} \frac{A}{B}=\frac{\partial ^2\psi /\partial R^2}{\partial ^2\psi /\partial z^2}\Rightarrow \left(\frac{A}{B}\right)^2=\left(\frac{dz}{dR}\right)^4\eeq
 Substituting (\ref{B7}) into (\ref{B6}), the former becomes
 \beq \label{B8} \frac{b^2}{a^4}(R-R_m)^2[f(R)]^{-1}=\left(\frac{dz}{dR}\right)^2\eeq
 by means of which $dl$ takes the following form
 \beq \label{B9} dl=dR\left[1+\frac{b^2}{a^4}(R-R_m)^2[f(R)]^{-1}\right]^{1/2}\eeq
 Additionally, from Eq. (\ref{B1}) one obtains
 \beq \label{B10} |\vec{\nabla}\psi |=\left[A^2(R-R_m)^2+B^2(z-z_m)^2\right]^{1/2}\eeq
 On the basis of Eqs. (\ref{B9}) and (\ref{B10}), (\ref{B4}) becomes
 \beq \label{B11} q=\frac{1}{2\pi}\oint \frac{IdR}{R}\left[\frac{1+\frac{b^2}{a^4}(R-R_m)^2[f(R)]^{-1}}{A^2(R-R_m)^2+B^2(z-z_m)^2}\right]^{1/2}\eeq
 which after some algebraic manipulations reduces to
 \beq \label{B12} q=\frac{1}{2\pi}\int \frac{IdR}{RBb[f(R)]^{1/2}}\eeq
 In order to make the analysis tractable, we change the integration variable as $R-R_m=asin\theta$, so that $[f(R)]^{1/2}=cos\theta$, and the safety factor takes the following form
 \beq \label{B13} q=\frac{1}{2\pi}\int_{0}^{2\pi} \frac{Ia}{RBb}d\theta\eeq
 In addition, from the definitions of $a,b,A$, and $B$, we have 
 \beq \label{B14} \frac{a}{Bb}=\left(\frac{\partial ^2\psi}{\partial R^2}\frac{\partial ^2\psi}{\partial z^2}\right)^{-1/2}\eeq
 Thus, Eq. (\ref{B13}) becomes
 \beq \label{B15} q=\frac{1}{2\pi}\left(\frac{\partial ^2\psi}{\partial R^2}\frac{\partial ^2\psi}{\partial z^2}\right)^{-1/2}\int_{0}^{2\pi} \frac{I}{R}d\theta\eeq
 The last step is the calculation of the integral of equation (\ref{B15}). By the use of Eq. (\ref{Ipsi}) for the functional form of $I$; this is
 \begin{eqnarray} \label{B16}
 \int_{0}^{2\pi} \frac{I}{R}d\theta & =&\frac{X}{1-\sigma _d-M_p^2}\int_{0}^{2\pi} \frac{1}{R}d\theta -\frac{\mu _0F^{'}\Phi ^{'}}{1-\sigma _d-M_p^2}\int_{0}^{2\pi}Rd\theta \nonumber \\
 &=& 2\pi \left(\frac{X}{1-\sigma _d-M_p^2}\frac{1}{\sqrt{R_m^2-a^2}}-\frac{\mu _0R_mF^{'}\Phi ^{'}}{1-\sigma _d-M_p^2}\right)
 \end{eqnarray}
 Substituting (\ref{B16}) into Eq. (\ref{B15}) we find that the general expression for the safety factor is
 \beq \label{B17} q=\left(\frac{X}{1-\sigma _d-M_p^2}\frac{1}{\sqrt{R_m^2-a^2}}-\frac{\mu _0R_mF^{'}\Phi ^{'}}{1-\sigma _d-M_p^2}\right)\left(\frac{\partial ^2\psi}{\partial R^2}\frac{\partial ^2\psi}{\partial z^2}\right)^{-1/2}\eeq
 Especially, on the magnetic axis denoted by $m$, where $a^2|_m=0$, it reduces into the following form
 \beq \label{B18} q_m=\frac{X-\mu _0R^2F^{'}\Phi ^{'}}{R(1-\sigma _d-M_p^2)}\left(\frac{\partial ^2\psi}{\partial R^2}\frac{\partial ^2\psi}{\partial z^2}\right)^{-1/2}_{\arrowvert _m}=\frac{I}{R}\left(\frac{\partial ^2\psi}{\partial R^2}\frac{\partial ^2\psi}{\partial z^2}\right)^{-1/2}_{\arrowvert _m}\eeq
 
 \chapter{Proof of the relations (\ref{c10})-(\ref{c12})}
 
 \hspace{2em}The differential of $\widetilde{u}(\xi,\zeta)$ is
 \begin{center}
 $d\widetilde{u}=\frac{\partial \widetilde{u} }{\partial \xi}d\xi+\frac{\partial \widetilde{u} }{\partial \zeta}d\zeta$
 \end{center}
On a given magnetic surface it is $d\widetilde{u}=0$ so that 
\beq \label{C1} \frac{\partial ^2 \widetilde{u}}{\partial \zeta ^2}=-\frac{\partial \widetilde{u}}{\partial \xi}\frac{d^2\xi}{d\zeta ^2}\eeq
and
\beq \label{C2} \frac{\partial ^2 \widetilde{u}}{\partial \xi ^2}=-\frac{\partial \widetilde{u}}{\partial \zeta}\frac{d^2\zeta}{d\xi ^2}\eeq
The parametric equations for the smooth upper part of the curve presented in Fig. (4.1) are
\beq \label{C3} \xi(\theta) =1+\epsilon cos(\theta +w_1sin\theta )\eeq
and
\beq \label{C4} \zeta(\theta)=\kappa \epsilon sin\theta \eeq
where $\epsilon=\frac{\alpha}{R_0}$, and $w_1=sin^{-1}(t)$.
Taking the $\theta$-derivative of (\ref{C3}) and (\ref{C4}) we find the following relations
\beq \label{C5} \frac{d^2\xi}{d\zeta ^2}=\frac{d^2\xi /d\theta ^2}{d\zeta ^2/d\theta ^2}=\frac{w_1sin\theta sin(\theta +w_1sin\theta )-(1+w_1cos\theta)^2cos(\theta +w_1sin\theta)}{\epsilon \kappa ^2 cos^2\theta}\eeq
and
\beq \label{C6} \frac{d^2\zeta}{d\xi ^2}=\frac{d^2\zeta /d\theta ^2}{d\xi ^2 /d\theta ^2}=-\frac{\kappa sin\theta}{\epsilon (1+w_1cos\theta)^2sin^2(\theta +w_1sin\theta)}\eeq
For $\theta=0$, corresponding to the outermost point, Eq. (\ref{C5}) gives
\beq\label{C7} \left[\frac{d^2\xi}{d\zeta ^2}\right]_{(\xi _{out},\zeta _{out})}=-\frac{(1+w_1)^2}{\epsilon \kappa ^2}\eeq
Thus, by substituting (\ref{C7}) into (\ref{C1}) we get
\beq \label{C8} \widetilde{u}_{\zeta ,\zeta}(\xi _{out},\zeta _{out})=\frac{(1+w_1)^2}{\epsilon \kappa ^2}\widetilde{u}_\xi (\xi _{out},\zeta _{out})\eeq
In addition, for $\theta=\pi$, corresponding to the innermost point, Eq. (\ref{C5}) gives
\beq \label{C9} \left[\frac{d^2\xi}{d\zeta ^2}\right]_{(\xi _{in},\zeta _{in})}=\frac{(1-w_1)^2}{\epsilon \kappa ^2}\eeq
and from (\ref{C1}) we take the relation
\beq \label{C10} \widetilde{u}_{\zeta ,\zeta}(\xi _{in},\zeta _{in})=-\frac{(1-w_1)^2}{\epsilon \kappa ^2}\widetilde{u}_\xi (\xi _{in},\zeta _{in})\eeq
At last, for $\theta=\frac{\pi}{2}$, corresponding to the upper point, Eq. (\ref{C6}) gives
\beq \label{C11} \left[\frac{d^2\zeta}{d\xi ^2}\right]_{(\xi _{up},\zeta _{up})}=-\frac{\kappa}{\epsilon cos^2w_1}\eeq
which substituted into (\ref{C2}) yields
\beq \label{C12} \widetilde{u}_{\xi ,\xi}(\xi _{up},\zeta _{up})=\frac{\kappa}{\epsilon cos^2w_1}\widetilde{u}_\zeta (\xi _{up},\zeta _{up})\eeq
In the above derivation, the imaginary parts of relations (\ref{C8}), (\ref{C10}) and (\ref{C12}) were used.

\newpage
\centering

\vspace{25mm}
\begin{center}
Part of this thesis consist the object of publication:
 ``A. Evangelias and G. N. Throumoulopoulos, Plasma Phys. Control. Fusion 58 (2016) 045022''.
\end{center}

\newpage\null\newpage
\centering

\begin{minipage}{0.8\textwidth}
	\vspace{110mm}
This work has been carried out within the framework of the EUROfusion Consortium and has received funding from  (a) the National Programme for the Controlled Thermonuclear Fusion, Hellenic Republic, (b) Euratom research and training programme 2014-2018 under grant agreement No 633053. The views and opinions expressed herein do not necessarily reflect those of the European Commission.

\end{minipage}

\end{document}